\documentclass[opre,nonblindrev]{informs3}

\OneAndAHalfSpacedXI 

\usepackage{endnotes}
\let\footnote=\endnote
\let\enotesize=\normalsize
\def\notesname{Endnotes}%
\def\makeenmark{\hbox to1.275em{\theenmark.\enskip\hss}}
\def\enoteformat{\rightskip0pt\leftskip0pt\parindent=1.275em
  \leavevmode\llap{\makeenmark}}

\usepackage{multirow}
\usepackage{threeparttable}
\usepackage{mathrsfs}
\usepackage{calligra}
\DeclareMathAlphabet{\mathcalligra}{T1}{calligra}{m}{n}

\usepackage{booktabs}
\usepackage{enumitem}
\usepackage{amsmath}
\usepackage{bm} 
\usepackage{framed}
\usepackage[outdir=./]{epstopdf}

\usepackage{graphicx}
\usepackage{subfig}
\usepackage{xcolor,cancel}
\usepackage{tabularx}

\newcolumntype{C}[1]{>{\centering\arraybackslash}p{#1}}

\usepackage{natbib}
 \bibpunct[, ]{(}{)}{,}{a}{}{,}%
 \def\bibfont{\small}%
 \def\bibsep{\smallskipamount}%
 \def\bibhang{24pt}%
\TheoremsNumberedThrough     
\ECRepeatTheorems

\EquationsNumberedThrough    

\begin{document}



\RUNTITLE{Optimal Sequential Multi-class Diagnosis}

\TITLE{Optimal Sequential Multi-class Diagnosis}

\ARTICLEAUTHORS{%
\AUTHOR{ (forthcoming in \emph{Operations Research}) }
\AUTHOR{Jue Wang}
\AFF{Smith School of Business, Queen's University, Kingston, Ontario, Canada, K7L 3N6, \EMAIL{jw171@queensu.ca}}  
} 

\ABSTRACT{%
Sequential multi-class diagnosis, also known as multi-hypothesis testing, is a classical sequential decision problem with broad applications. However, the optimal solution remains, in general, unknown as the dynamic program suffers from the curse of dimensionality in the posterior belief space. We consider a class of practical problems in which the observation distributions associated with different classes are related through exponential tilting, and show that the reachable beliefs could be restricted on, or near, a set of low-dimensional, time-dependent manifolds with closed-form expressions. This sparsity is driven by the low dimensionality of the observation distributions (which is intuitive) as well as by specific structural interrelations among them (which is less intuitive). We use a matrix factorization approach to uncover the potential low dimensionality hidden in high-dimensional beliefs and reconstruct the beliefs using a diagnostic statistic in lower dimension. For common univariate distributions, e.g., normal, binomial, and Poisson, the belief reconstruction is exact, and the optimal policies can be efficiently computed for a large number of classes. We also characterize the structure of the optimal policy in the reduced dimension. For multivariate distributions, we propose a low-rank matrix approximation scheme that works well when the beliefs are near the low-dimensional manifolds. The optimal policy significantly outperforms the state-of-the-art heuristic policy in quick diagnosis with noisy data.

}%


\maketitle

%

 
\section{Introduction}  
In multi-class classification, challenges often arise when the available data are inconclusive, e.g., when they fall near the decision boundary, in which case the diagnostic uncertainty is high, and a small perturbation of the data may lead to dramatically different outcomes. A common remedy is to reduce the uncertainty of inference by taking more samples. For example, physicians often repeat a test after receiving an inconclusive result. However, additional testing invariably delays the diagnosis and incurs extra cost. There is an urgent need to know how to reach a diagnosis with the desired accuracy using the minimum number of samples. Achieving this goal often requires a sequential procedure in which, after receiving each sample, it must be decided whether to collect an additional sample. This kind of decision arises in a wide range of contexts:

\begin{itemize}
\item \textbf{Personalized medicine.} Increasing evidence suggests that some chronic diseases, such as Parkinson's disease (PD), are heterogeneous rather than unitary, encompassing divergent symptom profiles, natural progressions, and responses to treatment. Early diagnosis of the underlying subtype is therefore critical for ensuring that the correct management program is initiated as soon as possible. The current diagnosis of PD,  based mainly on symptoms, has relatively low discriminatory power early in the disease, so the subtyping decision is rarely based on a single assessment, requiring instead an observation process in which the patient is reassessed periodically. Repeat assessment improves the diagnostic accuracy, but it also delays personalized treatment and reduces its effectiveness. This puts the onus on the physician to decide when to stop following up, and which treatment to initiate.  

\item \textbf{Servicing smart products.} Advances in digital technology have created various smart, connected products, which comprise sensors, microprocessors, software, and connectivity~\citep{Porter2014}. Rolls-Royce, for example, monitors aircraft engines in flight for signs of malfunction using embedded sensors. Sensor data are analyzed in real time to identify the health condition of an engine and to support triage decisions, such as whether the engine needs to be serviced, and which type of service is needed. Since sensor data are often imperfect indicators of the true condition of an engine, it may be prudent to wait and closely monitor the situation before initiating expensive interventions such as an emergency landing, but delaying the intervention could lead to catastrophic consequences if the underlying failure is critical.  
\end{itemize}

The problem of sequential multi-class diagnosis has been extensively studied in the literature~\citep{Tartakovsky2014}. Existing methods provide performance results in asymptotic regimes, namely, when a large number of samples can be collected before diagnosis. However, in safety-critical applications such as fault diagnosis in aircraft engines, it is crucial to make quick diagnoses which often lie outside the asymptotic regime, so a more sample-efficient approach is required. This paper focuses on optimal and near-optimal policies for cases with short horizons, where a diagnosis needs to be made quickly. 

\subsection*{Relevant Literature}

The framework for balancing the trade-off between information and decision was established in the seminal work of~\cite{Wald1945}, in which a dynamic programming model was formulated to differentiate between two simple hypotheses or classes (e.g., healthy or unhealthy) as imperfect observations accumulate. Observations are imperfect in the sense that they follow distinct statistical distributions in different classes. This model is known as sequential hypothesis testing, or the sequential probability ratio test (SPRT). The optimal policy is characterized by two thresholds on the posterior probability of the unhealthy class, or equivalently, the probability ratio between the two classes. The optimal solution is to wait if this probability stays between the thresholds, to make an unhealthy diagnosis when it exceeds the upper threshold, and to make a healthy diagnosis when it falls below the lower threshold. Although SPRT has its origins in statistics, it has served as a classical application of stochastic dynamic programming~\citep{Bersekas1976, Ross1983} and has inspired important works in  operations management~\citep{wang2010design, alizamir2013diagnostic}. Since the true state is not directly observable, SPRT can be viewed as a partially observable Markov decision process (POMDP) model with static hidden states~\citep{Monahan1982}. 

While SPRT is optimal for the diagnosis of {two} classes, in practice the number of classes can be large. For example, a smart jet engine is subject to multiple types of failure, including mechanical, hydraulic, electrical, and software failure, each requiring a distinct response. Software failures are easily fixable using over-the-air updates, whereas mechanical failures may require grounding the aircraft. The multi-class generalization of SPRT, also known as sequential multi-hypothesis testing, is considerably more complex; it requires keeping track of the posterior probabilities of all classes, and hence the state of the dynamic program (i.e., the belief state) becomes a probability vector. In this case, the computation of the optimal policy suffers from the curse of dimensionality and becomes prohibitive to implement~\citep{Dayanik2008, Tartakovsky2014}. 

Over the past several decades, developments in sequential multi-class diagnosis have diverged from dynamic programming and been directed toward suboptimal procedures. There is a large body of research in the statistics literature that is based on decoupling multiple classes to pairwise SPRTs operated in parallel~\citep{Sobel1949, Armitage1950, Paulson1962, Simons1967, Lorden1977, Eisenberg1991}. In particular, \cite{Baum1994} developed an M-ary sequential probability ratio test (MSPRT) that sets an individual threshold on the posterior probability of each class. The threshold of a particular class depends only on that class and the class that is nearest in terms of the Kullback-Leibler divergence, ignoring all other classes. \cite{Veeravalli1995} showed that this approach is asymptotically optimal as the delay cost approaches zero, that is, when it is possible to collect a large number of samples. However, although the asymptotic approaches are mathematically tractable and yield elegant performance bounds, their performance may deteriorate in non-asymptotic regimes, such as those encountered in quick diagnosis.

There have been advances concerning the multi-class generalization of Chernoff's classical problem of sequential diagnosis with sampling control~\citep{Chernoff1959}. For example, \cite{Nitinawarat2013} provided heuristic policies that are asymptotically optimal. \cite{Naghshvar2013} provided non-asymptotic bounds on the value of information using a dynamic programming argument without solving the dynamic program. \cite{gurevich2019sequential} proposed an asymptotically optimal algorithm for the sequential search of multiple processes, in which each process is either normal or abnormal.

After the literature of sequential hypothesis testing diverged from dynamic programming, much progress has been made in the latter. In particular, a stream of research on belief-compressed POMDP has been developed in the artificial intelligence literature~\citep{roy2005finding} which seeks the low-dimensional approximation of the reachable state space through simulation. In this paper, we return to the dynamic programming framework and integrate recent advances in POMDP as well as new ideas from machine learning into sequential multi-hypothesis testing. 

The optimal diagnosis policy is also studied in the machine maintenance literature, in which the goal is to locate the failed component(s) in a multi-component system~\citep{Gluss1959, Butterworth1972, Cho1991}. The focus in these works is on optimizing the order in which components are inspected, under the assumption that inspection reveals the true state of the component. In contrast, we consider situations in which the data are imperfect at revealing any state. Another related subject is the repetitive testing in quality assurance~\citep{greenberg1995repetitive, ding1998repetitive}, but existing models consider binary classes (conforming/nonconforming) rather than multiple classes.  In the medical decision making literature, \cite{skandaripatient} formulated a POMDP model for treatment planning that takes into account patient heterogeneity. In their model, the state of the dynamic program is the posterior belief about the patient type. The present work uses a lower-dimensional diagnostic statistic to reconstruct the higher dimensional belief so that the dimensionality of the state space will not increase with the number of classes.

Lastly, this paper has links to the economics literature on optimal learning. \cite{henry2019research} split the information acquisition and diagnostic decisions between two players and studied their strategic interactions.  \cite{che2019optimal} optimized the dynamic allocation of limited attention across two sources of information. \cite{ke2019optimal} considered choosing among two alternatives, with the option of purchasing information about each of them.  These studies all concern binary rather than multiple hidden classes.

Despite superficial similarities, the classical multi-class diagnosis and the multi-armed bandit problem are generally addressed in separate bodies of literature; see~\cite{Castanon1995} for a detailed comparison and discussion.

\subsection*{Scope and Contributions}
We consider a class of problems in which the conditional distributions of the observations, given the class membership, are related to each other through exponential tilting~\citep{anderson1979multivariate, qin1999empirical}. In this case, the unconditional distribution of the observation follows a semiparametric finite mixture model called the exponential tilting model (ETM), with each mixture component corresponding to a class. The ETM is a flexible model widely used in statistical practice. For example, when the conditional distribution associated with each class is normal, the unconditional distribution is a Gaussian mixture, which is a special case of ETM. The mixtures of other distributions from the same exponential family, e.g., gamma, binomial, and Poisson, are also special cases of ETM. 

In the ETM context, we show that the reachable belief states may be restricted to a series of low-dimensional, time-dependent manifolds with closed-form expressions, embedded in the high-dimensional probability simplex. The sparsity of the reachable space is driven by the low dimensionality of the sufficient statistic in ETM, which is intuitive, as well as by structural interrelations among the observation distributions, which is not always intuitive. We show that some high-dimensional problems have, in fact, a lower intrinsic dimension and then develop a matrix factorization approach to uncover the low dimensionality hidden in high-dimensional belief states. 

When the observations are univariate, we find that the intrinsic dimension of the belief states is $1$ or $2$ in many applications. This allows us to use a low-dimensional diagnostic statistic to \emph{reconstruct} the high-dimensional belief state and reformulate the dynamic program in a lower dimension, i.e., $1$ or $2$. This belief-reconstruction approach makes it possible to compute the optimal policy for a large number of classes. To the best of our knowledge, existing POMDP models in the operations research literature often use the belief state as the state variable~\citep{Krishnamurthy2016}, and few works exploit the properties of observation distributions to derive a sparse representation of the belief states. 

We characterize the structure of the optimal policy in the diagnostic-statistic space, which not only aids the computation but also provides further insights into the decision process. In numerical studies and a medical application, the optimal policy significantly outperforms MSPRT, an asymptotically optimal policy, in the regime of quick diagnosis, especially when some classes are difficult to distinguish. 

When the observations are multivariate, the intrinsic dimension is generally high, although special cases with lower dimensions do exist. We propose a low-rank matrix approximation scheme to find sparse representations of the high-dimensional belief space so that the belief states can be projected onto a series of low-dimensional manifolds. This approximate belief-reconstruction scheme works well when the reachable beliefs are distributed near the low-dimensional manifolds. We illustrate this approach with a maintenance application involving multiple sensors. 




\section{Sequential Multi-class Diagnosis}\label{sec: preliminaries} 
We first describe the problem of interest, which is a multi-class generalization of the sequential hypothesis testing problem~\citep{Wald1945}. The decision maker (DM) is presented with a system at time $0$, in which the class (or type) of the system is hidden and must be identified quickly before a deadline~$T$. The existing literature mostly considers the infinite-horizon case ($T=\infty$). This paper focuses on the finite horizon ($T<\infty$) in which the optimal policy is non-stationary. Time is partitioned into a sequence of epochs, indexed by $t=0, 1, \ldots, T$; one decision is to be made at each epoch. The hidden class, denoted by~$s$, belongs to the set $ \mathcal N \triangleq \{0,1,\ldots,N\}$. The DM is endowed with a prior~$\boldsymbol\theta=(\theta_0, \theta_1, \ldots, \theta_N)$, in which~$0\leq\theta_i \leq 1$ denotes the probability that class $i$ is true. A common prior is the uninformative prior with $\theta_i=1/(N+1)$ for all $i$. 

Although the true class cannot be directly observed, the DM can infer it from the (imperfect) observations. Let~$Y_t$ denote the observation obtained at time~$t$, discrete or continuous,  taking values in a set $\mathcal{Y}$ in the real line (the extension to the multivariate case will be explored in~\S\ref{sec: Extension to Multivariate Observations}). The $Y_t$'s are independent and identically distributed given the class membership. Let $f_i(y)$ denote the density corresponding to class $i \in \mathcal N$. We will use the notation of continuous distributions, but the results also apply to discrete distributions. $Y_t$ could also represent the change from the last observation, in which case the model can be used to differentiate among different linear growth rates.

Let $ \mathscr{F}_t=\sigma\{ Y_1, \ldots, Y_t \}$ be the sigma-algebra generated by $\{ Y_1, \ldots, Y_t \}$, the observation sequence collected up to time $t$. The DM chooses an action among the following alternatives: diagnose the system as class $i \in  \mathcal N $ and terminate the decision process; or wait until the next period, collect a new observation, and decide again. Denoting $ \mathscr{F}=\{ \mathscr{F}_t, t=0,1,\ldots\}$, the DM follows a sequential policy, $\delta=(\tau, d)$, in which $\tau$ is the stopping time with respect to $\mathscr{F}$, and a diagnostic decision rule~$d$ is an $\mathscr{F}_\tau$-measurable decision function taking values in the set~$\mathcal N$. The decision process is terminated at time~$\tau \leq T$ when the DM stops observing, and, if~$d=i$, diagnoses the system to be in class~$i$. Let~$\Delta$ denote the set of admissible policies in which the stopping and diagnostic decisions are based on the information available at time~$\tau$. If the hidden class is $i$,  a termination cost $a_{ij} \ge 0$ will be incurred if the diagnosis is~$j$. If the DM waits for more information, a delay cost $c_{i} \ge 0$ per period will be incurred. Here, we allow the delay cost to depend on the underlying class, which relaxes an assumption of MSPRT. 

The objective is to find a control policy that minimizes the expected total cost over the finite decision horizon, given the prior~$\boldsymbol\theta$. Specifically,  the minimum cost can be written as 
$
R^\star_{\boldsymbol\theta}= \inf_{\delta=(\tau \leqslant T, d)\in\Delta} \mathbb{E}_\delta   \{  \sum ^ N _{i=0} \tau  c_i  \mathbb{I}_{\{s=i\}}  +    \sum ^ N _{i=0} \sum ^ N _{j=0}  a_{ij}  \mathbb{I}_{\{s=i, d=j\}}  | \boldsymbol\theta \},
$
where $\mathbb{E}_\delta$ represents the expectation with respect to policy $\delta$. A popular cost structure is zero-one:  the termination costs are $a_{ij}=0$ for $j=i$ and $a_{ij}=1$ for $j\neq i \in\mathcal{N}$, and the delay cost is a constant $c_i=c$ for all $i\in\mathcal{N}$. Unless stated otherwise, all numerical examples in this paper use the zero-one cost structure. 

This problem can be formulated as a partially observable Markov decision process (POMDP) with static hidden states. The state in a POMDP is usually the posterior state probability vector $\Pi_t=(\pi_{0t}, \pi_{1t}, \ldots, \pi_{Nt})$, also known as the belief state. Here, $\pi_{jt}$ is the posterior probability for the hidden state (class)~$j$ given the information available at time~$t$. Its dependence on $t$ is suppressed at times in this paper for brevity. The belief state lies in an $N$-dimensional belief space 
$
S^{N} \triangleq \{\Pi=(\pi_0, \pi_1, \ldots, \pi_N) \in [0,1]^{N+1} \mid \pi_0+ \pi_1+ \cdots+ \pi_N=1\}.
$
It is well known that the belief state encapsulates all the historical information relevant for decision making~\citep{Bersekas1976}. By Bayes' rule, $\Pi_t=\mathcal{B}(\Pi_{t-1}, y_t)$, where $\mathcal{B}: S^{N} \times \mathcal{Y} \to S^{N}$ is the Bayesian operator defined as
$$
\mathcal{B}(\Pi, y)  \triangleq \frac{\Pi {G}(y)}{\Pi F(y) }. 
$$
${G}(y)\triangleq \text{diag}(f_0(y), f_1(y), \ldots, f_N(y))$ is a diagonal matrix; and $F(y)\triangleq(f_0(y), f_1(y), \ldots, f_N(y))^\mathsf{T}$ is a column vector. Note that $\Pi_t$ depends on the prior as~$\Pi_0= \boldsymbol\theta$. 

Let $V_t(\Pi)$ denote the minimum expected cost from period $t$ onward when the current belief state is $\Pi=(\pi_0, \ldots, \pi_N)$. Then, $\{V_t(\Pi),t=0,\ldots, T\}$ satisfy the following optimality equations: 
\begin{align}
\label{eq: finite-beliefstate-OE}
V_t(\Pi)&=\min\big\{V^0_t(\Pi), V^1_t(\Pi), \ldots, V^N_t(\Pi), V^w_t(\Pi)  \big\}, t=0, 1, \ldots, T-1, 
\end{align}
with $V^j_t(\Pi)=\sum^N_{i=0} \pi_i a_{ij}$, 
$$
V^w_t(\Pi)=\sum^N_{i=0} \pi_i c_{i}+\int_{\mathcal{Y}} V_{t+1} \big(   \mathcal{B}(\Pi, y)    \big)  \Pi F(y)dy,
$$ and $V_T(\Pi)=\min_j \{ \sum^N_{i=0} \pi_i a_{ij}  \}$. Here, $V^j_t(\Pi)$ is the expected cost of diagnosing the system in class~$j$ at time~$t$, while $V^w_t(\Pi)$ is the expected cost when the decision is to wait at $t$, collect a new observation~$y \in \mathcal{Y}$ at the next period, and follow the optimal policy thereafter. The optimal decision rule chooses the action that minimizes the right-hand side of~\eqref{eq: finite-beliefstate-OE}. Since the decision process lasts until a finite horizon, the optimal policy is generally non-stationary.

Let $\Gamma_{jt} \triangleq \{\Pi \in S^N: V_t(\Pi) =V^j_t(\Pi)  \}$ denote the optimal stopping region for class $j$. It is optimal to diagnose class~$j$ as soon as the belief state enters this region. If the belief state remains outside all stopping regions, it is optimal to wait. The standard approach to finding the optimal policy is to compute the optimality equations using value iteration~\citep{Bersekas1976}. Unfortunately, the computation suffers from the curse of dimensionality in the state space, since the $N$-dimensional probability simplex $S^N$ grows exponentially as the number of classes increases. Accordingly, the multi-class optimal policy has been generally considered computationally intractable.

\section{Belief Reconstruction}
\label{sec: belief reconstruction}
In this section, we show that the optimal policy can be computationally tractable when the conditional distributions $f_i, i\in\mathcal{N}$ are related through exponential tilting. We first describe the exponential tilting model, and then present our main result: a solution method that reconstructs the high-dimensional beliefs using a lower-dimensional sufficient statistic. 

\begin{figure}
\centering \includegraphics[width=6.5in]{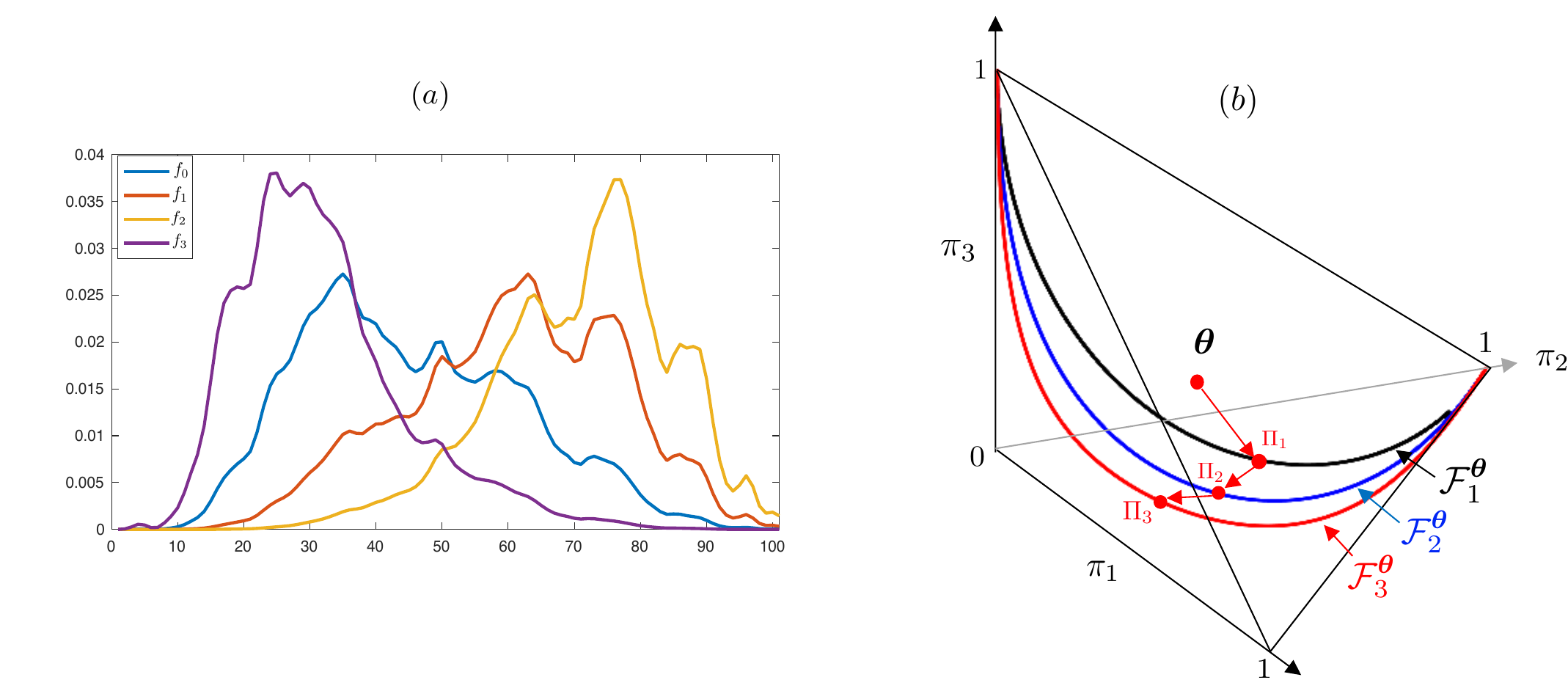} 
\caption{(a) ETM distributions with one tilting parameter ($p=1$); (b) the corresponding belief tracks ($r=1$)}
\label{Fig: ETM_Example1} 
\end{figure}

The exponential tilting model (ETM) is a semiparametric model used for finite mixture modelling~\citep{anderson1979multivariate}. For a mixture of $(N+1)$ component densities $f_0, \ldots, f_N$, the ETM assumes the following relation among the components: 
\begin{align}
\label{eq: ETM}
f_i(y) =f_0(y) \exp\{ \beta_{i0}+ \boldsymbol\beta^\mathsf{T}_i \boldsymbol h(y)  \} , \ \ \ i=1,\ldots, N. 
\end{align}
No parametric assumption is placed on the individual component (an example of ETM components is given in Figure~\ref{Fig: ETM_Example1}--a). Here, $\boldsymbol\beta_i=(\beta_{i1}, \ldots, \beta_{ip})^\mathsf{T}$ is a vector of tilting parameters, and $\boldsymbol h(y)=( h_1(y), \ldots, h_p(y)  )^\mathsf{T}$ is a vector of data transformations, referred to as the natural sufficient statistic. Some commonly used transformations include $y, y^2$, $\log(y)$ and $\log(1-y)$. The term~$\beta_{i0}=-\log \int  \exp(  \boldsymbol\beta^\mathsf{T}_i \boldsymbol h(y) ) f_0(y)dy$ is a normalizing constant. The baseline distribution $f_0$ can be estimated non-parametrically.  

The ETM has been widely used to model the unconditional distribution of data, in areas ranging from cancer diagnosis to meteorology~\citep{de2014spectral, li2017semiparametric} due to its flexibility and estimation efficiency~\citep{efron1996using}. There are several reasons for its broad application. First, distributions in the exponential families such as normal, binomial, and Poisson can serve as the components of the ETM (Table~\ref{list_distributions-ETM_EF}).\endnote{Note that the ETM condition is preserved under truncation of the distribution; this is a desirable property as most real data  vary only within a limited range.} Second, ETM is capable of representing relations beyond the exponential tilting. For example, if~$h(y)=\log y$, we obtain 
$f_i(y) =f_0(y) \exp\{ \beta_{i0}+\beta_i \log y  \}=f_0(y) y^{\beta_i} \exp(\beta_{i0})$. Third, ETM is closely related to logistic regression~\citep{qin1999empirical}. Consider an example with two classes: healthy ($s=0$), and unhealthy ($s=1$). The ETM assumption implies that, given an observation $y$, the probability of being unhealthy satisfies  
$
\text{logit} \{\Pr(s=1|y)\}= \beta^*_{10}+ \boldsymbol\beta^\mathsf{T}_1 \boldsymbol h(y), 
$
where $\text{logit}\{x\}=\log\{x/(1-x)\}$ is the link function for logistic regression and $\beta^*_{10}=\beta_{10}+\text{logit}(\theta_1)$.

\begin{table}\footnotesize
\centering
\smallskip
\caption{Commonly used univariate exponential families} 
\label{list_distributions-ETM_EF}
\begin{threeparttable}
\begin{tabular}{c c c c c }
\toprule
$f_i$&$\boldsymbol\beta^\mathsf{T}_i$&$\boldsymbol h^\mathsf{T}(y)$&$\beta_{i0}$&$p$ \\
\toprule
\textbf{Binomial ($n, p_i$)}& $ \log \frac{p_i}{1-p_i}-\log \frac{p_0}{1-p_0}$ & $ y$&$n\log(1-p_i)-n\log(1-p_0)$&$1$\\
\cmidrule(r){1-5} 
\textbf{Poisson ($\lambda_i$)}& $ \log\lambda_i-\log\lambda_0$&$ y$&$\lambda_0-\lambda_i $& $1$\\
\cmidrule(r){1-5} 
\textbf{Normal ($\mu_i, \sigma^2_i$)}& $ (\frac{\mu_i}{\sigma^2_i}-\frac{\mu_0}{\sigma^2_0}, \frac{1}{2\sigma^2_0}-\frac{1}{2\sigma^2_i})$&$(y, y^2)$&  $\frac{\mu^2_0}{2\sigma^2_0}-\frac{\mu^2_i}{2\sigma^2_i}+\log\sigma_0-\log\sigma_i$   & $2$\\
\cmidrule(r){1-5} 
\textbf{Gamma ($\alpha_i, \beta_i$)}& $ ( \alpha_i-\alpha_0, \beta_0-\beta_i)$&$ (\log y, y)$&$\alpha_i \log\beta_i-\alpha_0 \log\beta_0 +\log\frac{\Gamma(\alpha_0)}{\Gamma(\alpha_i)}$&$2$\\
\cmidrule(r){1-5} 
\textbf{Beta ($\alpha_i, \beta_i$)}& $ (\alpha_i-\alpha_0, \beta_i-\beta_0)$&$ (\log y, \log (1-y))$&$\log B(\alpha_i, \beta_i)-\log B(\alpha_0, \beta_0) $&$2$\\
\bottomrule
\end{tabular}
\end{threeparttable}
 \end{table}

Before presenting the solution method, we introduce some notations. Let $\boldsymbol e_j, j=1,\ldots, N$ denote an~$N$-dimensional unit row vector with $1$ at the $j$th component, and let $\boldsymbol e_0$ be the $N$-dimensional null row vector. Let $\Pi_t (y_1, \ldots, y_t, \boldsymbol \theta) =(1-\sum^N_{i=1} \pi_{it}, \pi_{1t}, \ldots, \pi_{Nt}) \in S^N$ denote the belief state, given the observation sequence $y_1, \ldots, y_t$ and prior $\boldsymbol \theta$.

\begin{definition} 
\label{def: B matrix}
Define the diagnostic parameter matrix as $\mathbb{B} \triangleq (\boldsymbol\beta_1, \ldots, \boldsymbol\beta_N)^\mathsf{T}$. 
\end{definition}
Each row of~$\mathbb{B}$ represents a class, and each column represents a tilting parameter. Using rank decomposition, this matrix can be decomposed as~$\mathbb{B}=\mathbb{RP}$, where $\mathbb{R}$ is an $N\times r$ matrix of full column rank called the \emph{reconstruction matrix}, and $\mathbb{P}$ is an $r\times p$ matrix of full row rank called the \emph{projection matrix}. The motivation behind rank decomposition and the technical details will be explained in~\S\ref{subsec: Rank Decomposition}. The properties of the reachable belief space are described in the following theorem (all proofs are in the electronic companion).
\begin{theorem}
\label{theorem 1}
If the class-specific distributions $f_0, \ldots, f_N$ satisfy the exponential tilting condition~\eqref{eq: ETM}, then the dimensionality of the reachable belief space is equal to the rank of~$\mathbb{B}$, namely, $r=\text{rank} (\mathbb{B})$. For $t\geqslant 1$, the belief state $\Pi_t (y_1, \ldots, y_t, \boldsymbol \theta)$  can be reconstructed from an $r$-dimensional diagnostic statistic $\boldsymbol x_t \triangleq \mathbb{P}  \sum^t_{m=1} \boldsymbol h(y_m)/t$. Specifically, $\pi_{jt} = \mathcal{T}^t_j(\boldsymbol x_t, \boldsymbol \theta )$, where the transform $\mathcal{T}^t_j: \Omega_t \times S^N \to [0,1]$ is defined as
\begin{align}
\label{eq: T transform with x}
\mathcal{T}^t_j(\boldsymbol x, \boldsymbol \theta )   
  &\triangleq  \frac{\theta_j \exp\big\{   t (\beta_{j0}+  \boldsymbol{e}_j \mathbb{R}   \boldsymbol x) \big\} }
 {\theta_0+\sum^N_{i=1} \theta_i \exp\big\{  t(\beta_{i0}+   \boldsymbol{e}_i \mathbb{R}  \boldsymbol x)
  \big\}} , \ \  j =1,\ldots, N,
\end{align}
where $\beta_{i0}=-\log \int  \exp \{  \boldsymbol{e}_i \mathbb{R P}    \boldsymbol h(y) \} f_0(y)dy$. Further, $\pi_{0t} =\mathcal{T}^t_0 \triangleq 1-\sum^N_{j=1} \mathcal{T}^t_j$. 
 \end{theorem}

The dimension of the reachable belief space is $r$, which will be referred to as the \emph{intrinsic dimension}. It never exceeds the number of tilting parameters~$p$ or the number of classes $N$. Popular distributions such as normal, binomial, Poisson, and gamma have only one or two parameters, and the corresponding reachable belief space is often sparse. 

When $r=1$, the reachable belief space consists of a set of curves embedded in the belief space. An example is shown in Figure~\ref{Fig: ETM_Example1}--(b), where the distributions are  $f_i(y) =f_0(y) \exp\{ \beta_{i0}+\beta_i \log y  \}$, as shown in Figure~\ref{Fig: ETM_Example1}--(a). Starting from the prior belief $\boldsymbol\theta$, if the decision is to wait and observe $y_1$, then the updated belief~$\Pi_1$ will fall on the curve~$\mathcal F^{\boldsymbol \theta}_1$. After the next observation $y_2$, regardless of the location of $\Pi_1$, the updated belief $\Pi_2$ will fall somewhere on the curve $\mathcal F^{\boldsymbol \theta}_2$.  In each period~$t$, the belief state $\Pi_t$ is constrained to lie on a \emph{belief track}~$\mathcal F^{\boldsymbol \theta}_t$; the space outside the belief tracks is never visited. 

When $r=2$, $\mathcal F^{\boldsymbol \theta}_2, \mathcal F^{\boldsymbol \theta}_3, \ldots$ will appear in the form of surfaces, or meshes for discrete observations, embedded in the belief space, which we refer to as \emph{belief surfaces}. The belief state will jump from one surface to the next in sequence as more data are collected. $\mathcal F^{\boldsymbol \theta}_1$ is still one-dimensional. 

Theorem~\ref{theorem 1} suggests that, in each period $t$, a belief state in higher dimension can be mapped into a coordinate system of lower dimension, represented by $r$ coordinates $\boldsymbol x_t=(x_{1t}, \ldots, x_{rt})$, just as points on a sphere can be described in terms of latitude and longitude. 

The belief tracks and surfaces can be fully characterized in closed form at the beginning of the decision process, as they are the mappings of the reachable sets of $\boldsymbol x_t$ into the belief space.
\begin{definition} 
\label{def: Omega set and belief tracks}
Define~$\Omega_t \triangleq \{ \mathbb{P}  \sum^t_{m=1}\boldsymbol h(y_m)/t : y_m \in \mathcal{Y}  \} \subset {\rm I\!R}^r$ as the set of  diagnostic statistic $\boldsymbol x_t $ in period $t$, and the manifold $\mathcal{F}^{\boldsymbol \theta}_t \triangleq \{ (\mathcal{T}^t_0(\boldsymbol x_t, \boldsymbol \theta ), \ldots, \mathcal{T}^t_N(\boldsymbol x_t, \boldsymbol \theta )) \in S^N: \boldsymbol x_t \in \Omega_t \}$ as the set of belief states reachable from the prior $\boldsymbol \theta$ in $t$ periods. 
\end{definition}
The set $\Omega_t$ can be computed from the support~$\mathcal{Y}$ using the method described in the Appendix~\S\ref{Appendix sec: Characterizing the State Space through Minkowski Sums}. For each $\boldsymbol x_t \in \Omega_t$, the corresponding belief state $\Pi_t =(\mathcal{T}^t_0(\boldsymbol x_t, \boldsymbol \theta ), \ldots, \mathcal{T}^t_N(\boldsymbol x_t, \boldsymbol \theta ) ) \in \mathcal{F}^{\boldsymbol \theta}_t$ can be represented in closed form. That is, the manifold $\mathcal{F}^{\boldsymbol \theta}_t$ is a mapping of $\Omega_t$ to the belief space; thus, $\mathcal F^{\boldsymbol \theta}_t$ can be parameterized by $\boldsymbol x_t$  in closed form. Note that $\mathcal{F}^{\boldsymbol \theta}_t$ are time-dependent since each period is associated with a distinct reachable belief space, and prior-dependent since they are the belief sets reachable from the prior~$\boldsymbol \theta$. 

\subsection{Drivers of Sparsity in the Belief Space}
\label{subsec: Rank Decomposition}
Theorem~\ref{theorem 1} suggests that the sparsity of the reachable belief space has two drivers, which correspond to two levels of dimension reduction in the state space: 
\begin{enumerate}
\item The first driver, which is intuitive, is the existence of a low-dimensional natural sufficient statistic in the ETM so that the $N$-dimensional belief state can be represented by the $p$-dimensional sufficient statistic. Thus, the first level of dimension reduction reduces the dimension of the state space from $N$ to $\min\{N, p\}$. 
\item The second driver, which is less intuitive, is a certain \emph{interrelation} among the class-specific tilting parameter vectors~$\boldsymbol\beta_1, \ldots, \boldsymbol\beta_N$, which allows the dimension to be reduced from $\min\{N, p\}$ further down to $r$, i.e., the dimensionality of~$\boldsymbol x_t$. We call this the second level of dimension reduction. An example with univariate distribution is given below. Another (less intuitive) example with multivariate observations is given in \S\ref{sec: Extension to Multivariate Observations} (Example~\ref{ex: MV Example 1}). 
\end{enumerate} 

\begin{example}
\label{ex: 2nd level reduction-Normal}
Consider the situation in which $f_i$ is normal with mean $\mu_i$ and variance $\sigma^2_i$. Define $\zeta_i \triangleq (\sigma^2_0\mu_i-\sigma^2_i\mu_0)/ (\sigma^2_i-\sigma^2_0), i=1,\ldots, N$. (If $\sigma_i=\sigma_0$, then $\zeta_i=\infty$). When $\zeta_i =\zeta$ for all $i=1, \ldots, N$, which represents a specific interrelation among all classes, the rows of $\mathbb{B}$ are linearly dependent. The reconstruction and projection matrices are given by 
\begin{align}
\mathbb{R}= \big( \frac{\sigma^2_0\mu_1 -\sigma^2_1\mu_0}{\sigma^2_0\sigma^2_1}, \dots, \frac{\sigma^2_0\mu_N -\sigma^2_N\mu_0}{\sigma^2_0\sigma^2_N} \big) ^\mathsf{T}, \ \ \ \ \mathbb{P}=(1, 1/2\zeta). \nonumber
\end{align}
Although the average transformed observation $( \sum^t_{m=1} y_m /t,  \sum^t_{m=1} y^2_m /t )^\mathsf{T}$ is two-dimensional, it can be projected to one dimension by multiplying by $\mathbb{P}$ as follows:
$$
\boldsymbol x_t = \sum^t_{m=1} ( y_m+{y^2_m}/{2\zeta} ).
$$
Here, the diagnostic statistic $\boldsymbol x_t$ is a \emph{nonlinear} feature sufficient for diagnosis: all classes can be effectively discriminated by this scalar. Consequently, the reachable belief space is one-dimensional. This situation arises when either the means or the variances of all classes are identical, or when the mean and variance are both class-specific yet $(\sigma^2_0\mu_i-\sigma^2_i\mu_0)/ (\sigma^2_i-\sigma^2_0)$ remains constant. In this situation, the information differentiating any two classes can also differentiate any other pair of classes. 
\end{example}

The key to the second-level dimension reduction is rank decomposition $\mathbb{B}=\mathbb{RP}$, which can be implemented through singular value decomposition: 
\begin{align}
\mathbb{B}=U\Sigma V^\mathsf{T}
=
\begin{bmatrix}
U_r &U_{N-r} 
\end{bmatrix}
\begin{bmatrix}
&\Sigma_r & \boldsymbol 0 \\
&\boldsymbol 0 & \boldsymbol 0
\end{bmatrix}
\begin{bmatrix}
& V^\mathsf{T}_r \\
&V^\mathsf{T}_{p-r}
\end{bmatrix}
=U_r ( \Sigma_r V^\mathsf{T}_r)=\mathbb{RP}, \nonumber
\end{align}
where  $U$  is an $N \times N$ orthogonal matrix; $\Sigma$ is an $N \times p$ rectangular diagonal matrix with non-negative singular values on the diagonal; and $V$ is a $p\times p$ orthogonal matrix. The block matrix~$U_r$ represents the first $r$ columns of $U$; it has full column rank and can be selected as the $N\times r$ reconstruction matrix $\mathbb{R}$. The $r\times r$ block matrix $\Sigma_r$ is the diagonal matrix containing $r$ nonzero singular values, and the $r\times p$ matrix $V^\mathsf{T}_r$ consists of the first $r$ rows of $V^\mathsf{T}$. The product $\Sigma_r V^\mathsf{T}_r$ is an $r\times p$ matrix having full row rank, chosen to be the projection matrix $\mathbb{P}$. When~$\mathbb{B}$ has full rank, we may also choose $\mathbb{R}=\mathbb{B}$ and  $\mathbb{P}=\mathbb{I}$ (see Appendix~\ref{Appendix: An Example of Belief Reconstruction} for an example). 

\begin{figure}
\centering \includegraphics[width=5.9in]{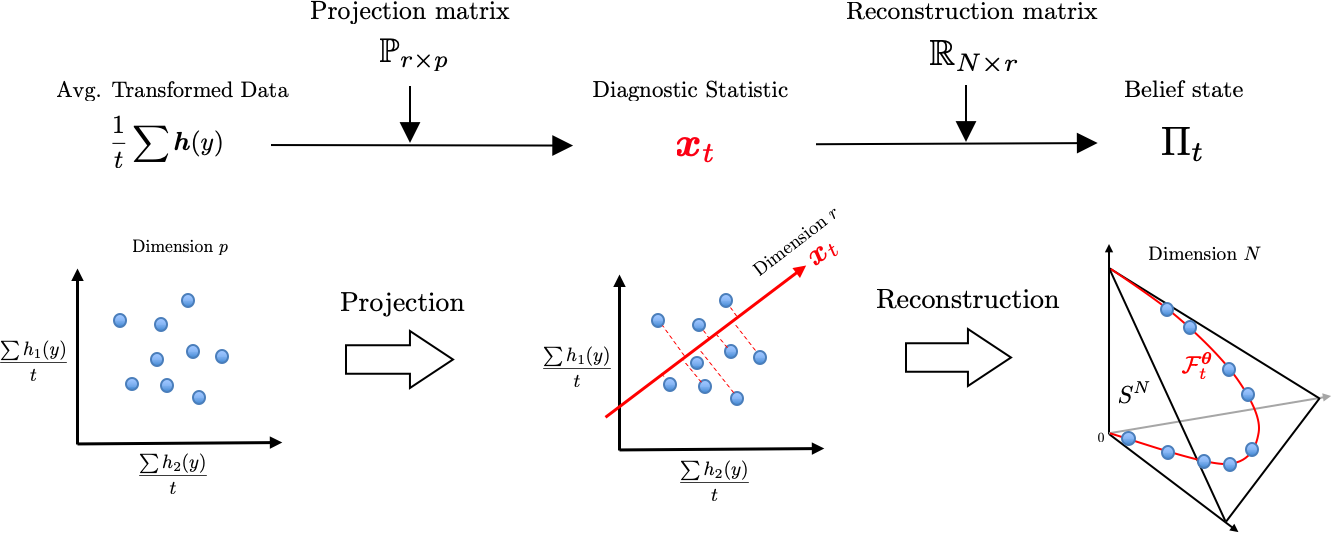} 
\caption{Schematic illustration of the data projection (i.e., the second-level dimension reduction) and belief reconstruction. Note that $\boldsymbol x_t$ is linear in $\boldsymbol h(y)$ but is often \emph{nonlinear} in $y$.}
\label{Fig: ProjectionAndReconstruction} 
\end{figure}

Figure~\ref{Fig: ProjectionAndReconstruction} summarizes the data projection and belief-reconstruction process. Samples of~$\sum^t_{m=1}\boldsymbol h(y_m)/t$, the $p$-dimensional average transformed observation at period $t$, are shown in the scatter plot at the bottom left. The average transformed observation is a sufficient statistic for all historical information and is the result of the first-level reduction. The projection matrix projects the sufficient statistic to an $r$-dimensional space, producing the diagnostic statistic $\boldsymbol x_t$, the result of the second-level reduction. No diagnostic information is lost because the projection matrix, which carries structural information about all classes, preserves the direction along which these classes can be most effectively discriminated. Then, the $N\times r$ reconstruction matrix $\mathbb{R}$ maps the $r$-dimensional statistic $\boldsymbol x_t$ to a space of $N$ dimension so that the posterior probabilities $\pi_1, \ldots, \pi_N$ can be reconstructed exactly following Theorem~\ref{theorem 1}. The reconstructed beliefs $\Pi_t$ lie on the $r$-dimensional manifold $\mathcal{F}^{\boldsymbol \theta}_t$, shown at the bottom right.  The observation is mapped to different manifolds in different time periods. 

It is worth noting that the diagnostic statistic $\boldsymbol x_t$ is often a nonlinear feature of $y$ because the transformation~$\boldsymbol h(y)$ is often nonlinear, e.g., $y^2$, $\log(y)$, or $\log(1-y)$. In the computer science literature, there is a stream of research that uses simulation to seek a sparse approximation of the reachable state space~\citep{roy2005finding}, in which the beliefs in different time periods are projected to the same manifold. In contrast, the belief reconstruction method is able to recover the belief exactly and does not require simulation. In addition, the beliefs in different periods are projected to different manifolds.  

\subsection{Reformulating the Optimality Equations}
The properties of the ETM allow the computation to focus on lower-dimensional manifolds. Based on Theorem~\ref{theorem 1},  we can reformulate the original optimality equations~\eqref{eq: finite-beliefstate-OE} using the diagnostic statistic as the state. Let~$J_t(\boldsymbol x_t; \boldsymbol \theta ) \triangleq V_t(\Pi_t)$ represent the minimum expected cost-to-go from period~$t$ onward,  given the prior $\boldsymbol \theta$ and the diagnostic statistic~$\boldsymbol x_t$. A reformulation appears below.
 \begin{proposition}
\label{prop: reconstructedValueFunction}
Under the exponential tilting model~\eqref{eq: ETM}, the optimality equations~\eqref{eq: finite-beliefstate-OE} can be reformulated as follows:
\begin{align}\label{Eq: theorem: 1D-Finite-OE-Main}
J_t(\boldsymbol x_t; \boldsymbol \theta ) & =\min\big\{J^0_t(\boldsymbol x_t; \boldsymbol \theta ), J^1_t(\boldsymbol x_t; \boldsymbol \theta ), \ldots, J^N_t(\boldsymbol x_t; \boldsymbol \theta ), J^w_t(\boldsymbol x_t; \boldsymbol \theta )    \big\},  \ \ \ \boldsymbol x_t \in \Omega_t, \  t=1, 2,  \ldots , T-1,\\
J_T(\boldsymbol x_T; \boldsymbol \theta ) & =\min\big\{J^0_T(\boldsymbol x_T; \boldsymbol \theta ), J^1_T(\boldsymbol x_T; \boldsymbol \theta ), \ldots, J^N_T(\boldsymbol x_T; \boldsymbol \theta )  \big\} \nonumber\\
\label{Eq: theorem: 1D-Finite-OE-Main-stopping}
J^j_t(\boldsymbol x_t; \boldsymbol \theta ) &=\sum^N_{i=0}  \mathcal{T}^t_{i}(\boldsymbol x_t, \boldsymbol \theta) a_{ij}, \ \ \ j \in \mathcal N, \\
\label{Eq: theorem: 1D-Finite-OE-Main-wait}
J^w_t(\boldsymbol x_t; \boldsymbol \theta ) &=\sum^N_{i=0} \mathcal{T}^t_i(\boldsymbol x_t, \boldsymbol \theta) c_{i}+\int_\mathcal{Y} J_{t+1} \Big( \frac{  t \boldsymbol  x_t+ \mathbb{P} \boldsymbol h(y) }{t+1} ; \boldsymbol \theta  \Big) \sum^N_{i=0}  \mathcal{T}^t_i(\boldsymbol x_t, \boldsymbol \theta) f_i(y) dy. 
\end{align}
\end{proposition}
The proof is straightforward and hence omitted. For $t=0$, the optimality equation is still given by~\eqref{eq: finite-beliefstate-OE}, except that 
$
V^w_0(\boldsymbol\theta)=\sum^N_{i=0} \theta_i c_{i}+\int_{\mathcal{Y}} J_1(\mathbb{P}  \boldsymbol h(y) ; \boldsymbol \theta )  \boldsymbol\theta F(y) dy. 
$

\begin{figure}
\centering \includegraphics[width=4.0in]{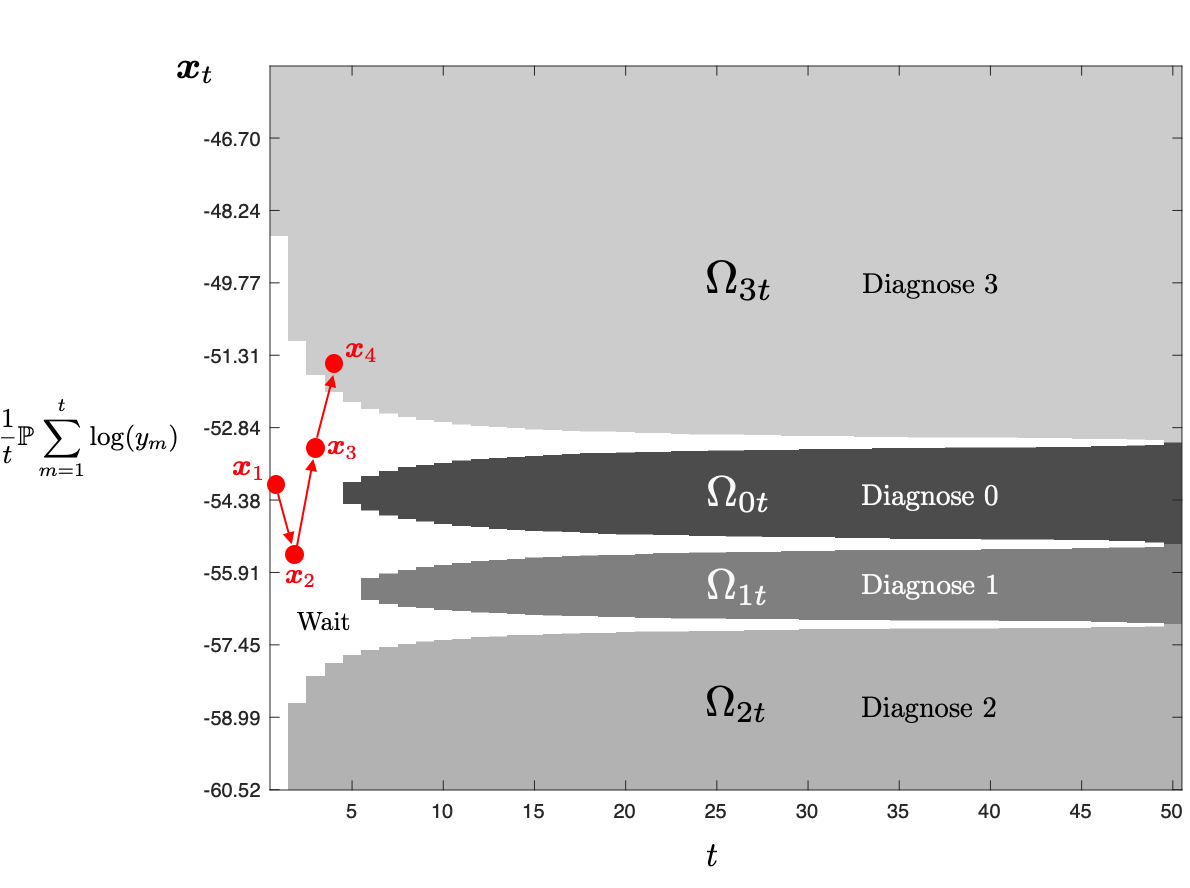} 
\caption{An illustration of the optimal policy implemented by belief reconstruction ($r=1$). Parameters: $N=3$; $T=50$; $c_i=0.03, \theta_i=0.25$, for all $i=0,\ldots, 3$; the $f_i$'s are as shown in Figure~\ref{Fig: ETM_Example1}--(a); and the termination costs are zero-one.} 
\label{Fig: 1D_chart_Example_20180702} 
\end{figure}

With this reformulation, the state space $\Omega_t$ has dimension $r\leqslant 2$ in many cases. Therefore, the optimal policy can be easily computed for a large number of classes, since the dimensionality of the state space is bounded by the number of tilting parameters. In the diagnostic-statistic space, the optimal stopping region for class $j$ is denoted by $\Omega_{jt\theta} \triangleq \{\boldsymbol x \in \Omega_t: J_t(\boldsymbol x; \boldsymbol \theta )  =J^j_t(\boldsymbol x; \boldsymbol \theta )  \}$, which we write as $\Omega_{jt}$ for brevity. It is optimal to stop and diagnose class $j$ once $\boldsymbol  x_t$ enters $\Omega_{jt} $ in period~$t$. If $\boldsymbol  x_t \notin \Omega_{jt}$ for all $j\in \mathcal N$, then it is optimal to wait. Figure~\ref{Fig: 1D_chart_Example_20180702} gives an example with four classes. Starting from the waiting region, it is optimal to stop and diagnose class $i$ as soon as the diagnostic statistic enters the stopping region~$\Omega_{it}$.\endnote{When $r=1$, $\Omega_{it}$ are stopping intervals rather than regions, but we shall use the term region  for the general value of $r$.}

In the classical POMDP formulation, the stopping regions $\Gamma_{jt}, j\in\mathcal{N}$ are prior-independent, but the belief state $\Pi_t$ depends on the prior. In contrast, the stopping regions $\Omega_{jt}, j\in\mathcal{N}$ in the diagnostic statistic space are prior-dependent, whereas the statistic $\boldsymbol  x_t$ does not depend on the prior. We discuss this point further in~\S\ref{sec: Scaffolding Algorithm}.

When $\Omega_t$ is discretized with 200 sample points, the optimal policy in Figure~\ref{Fig: 1D_chart_Example_20180702} takes only 6.89 sec to calculate on a computer with a 2.6GHz dual-core Intel Core i5 processor. Incorporating more classes will not inflate the dimensionality, but rather will add more stopping regions. Updating the diagnostic statistic~$\boldsymbol x_t$ amounts to taking a simple weighted average of the existing statistic and the projected new observation: 
$$
\boldsymbol  x_{t+1}=\frac{t\boldsymbol  x_t + \mathbb{P}\boldsymbol h( y)}{t+1}.
$$

\subsection{Toward Practical Implementation}
\subsubsection{Specification of the ETM.}
To implement the sequential diagnosis policy, it is necessary to specify the conditional densities $f_0, \ldots, f_N$, ideally based on historical data. The way to estimate the densities depends on whether the true class label can be observed in the historical data. If the class label is observable, i.e., it is known from which class the historical data is taken, then standard density estimation techniques can be used. If the class label is not observable, i.e., it is unknown from which class the data is taken, then a finite mixture model can be fit to the data to capture the underlying heterogeneity. In this case, the unconditional density of the historical data $y$ is modelled as
$
f(y)=\sum^N_{i=0} \alpha_i f_i(y), 
$
where $\alpha_0, \ldots, \alpha_N$ are the mixing probabilities that sum to 1, and~$f_i$ is the component density for the underlying class or subgroup~$i$. An efficient estimation of the mixture model often involves some assumptions about the densities. In fact, the ETM assumption in~\eqref{eq: ETM} is used to facilitate estimation by offering a good balance between model parsimony and flexibility. The ETM parameters can be estimated using the expectation-maximization (EM) algorithm. Of course, it may not be possible to know what the subgroups that have been found truly represent, but interpretations can be made using expert judgement as in interpretations of clustering analysis. 

\subsubsection{Characterization of $\Omega_t$.}
The reformulated optimality equations in Proposition~\ref{prop: reconstructedValueFunction} can be solved by backward induction. This requires the characterization of the state space~$\Omega_t = \{   \sum^t_{m=1} \mathbb{P}\boldsymbol h(y_m)/t : y_m \in \mathcal{Y}  \} $ for all $t=1,\ldots, T$ so that the range of the state variable $\boldsymbol x_t$ can be specified in each period, the discretization of $\Omega_t$ can then be performed when necessary. In Appendix~\ref{Appendix sec: Characterizing the State Space through Minkowski Sums}, we show that $\Omega_t$ is an iterated Minkowski sum of $\{\mathbb{P} \boldsymbol h(y), y\in\mathcal{Y}\}$ that can be characterized either analytically or using the state-of-the-art algorithms for Minkowski sums. 

\subsubsection{Bayesian Robustness.} 
\label{sec: Scaffolding Algorithm}

In Bayesian decision theory, the optimal action often depends on the prior. Since the prior is obtained from some elicitation process that may be imprecise, it is often valuable to perform a sensitivity analysis around the original prior, a process also known as Bayesian robustness analysis~\citep{berger2013statistical}. It is common to consider a set of priors in the neighbourhood of the original prior~$\boldsymbol\theta$, known as the $\epsilon$-contamination set: $\Gamma_{\boldsymbol\theta}=\{\boldsymbol\theta' \in S^N: \boldsymbol\theta' =(1-\epsilon) \boldsymbol\theta + \epsilon \boldsymbol q, \boldsymbol q\in \mathcal {Q} \}$, where $0<\epsilon<1$, and $\mathcal {Q}$ is the set of possible contaminations specified by the DM. 

Bayesian robustness analysis is convenient in the standard POMDP formulation because the stopping regions in the belief space $\Gamma_{jt}$ do not depend on the prior belief; only the posterior state~$\Pi_t$ does. The DM can easily recalculate the posterior belief under a new prior to check if the optimal action would change. However, the computation of $\Gamma_{jt}$ is intractable for large $N$. The belief reconstruction approach can efficiently scale to large $N$, but its stopping regions $\Omega_{jt}$ are prior-dependent. Although it is straightforward to resolve the Bellman equations for every single prior in~$\Gamma_{\boldsymbol\theta}$, the process can be time-consuming. 

In Appendix~\S\ref{ECsec: Bayesian Robustness Analysis}, we show that the results obtained from one prior in $\Gamma_{\boldsymbol\theta}$ may be reused by a set of other priors in~$\Gamma_{\boldsymbol\theta}$. We also propose a scaffolding algorithm that can efficiently outline substantial subsets of $\Gamma_{jt}$ in the belief space, where the robustness analysis is easier to perform.

\section{Structural Results}
\label{sec: Structure of the Optimal Policy}
Next, we characterize the qualitative structure of the optimal policy in the diagnostic-statistic space, which can lead to computational savings and provide operational insights. Specifically, understanding the structure allows us to identify certain subsets of the stopping regions without additional computation. Moreover, there may exist a critical stopping time (before the end of the horizon) at which the waiting region is empty and the decision process must terminate. Additional structural results are presented in Appendix~\S\ref{ECsec: Additional Structural Properties}. 

\subsection{Properties of the Stopping Regions}
\label{sec: Properties of the stopping regions}

\begin{figure}
\centering \includegraphics[width=6.8in]{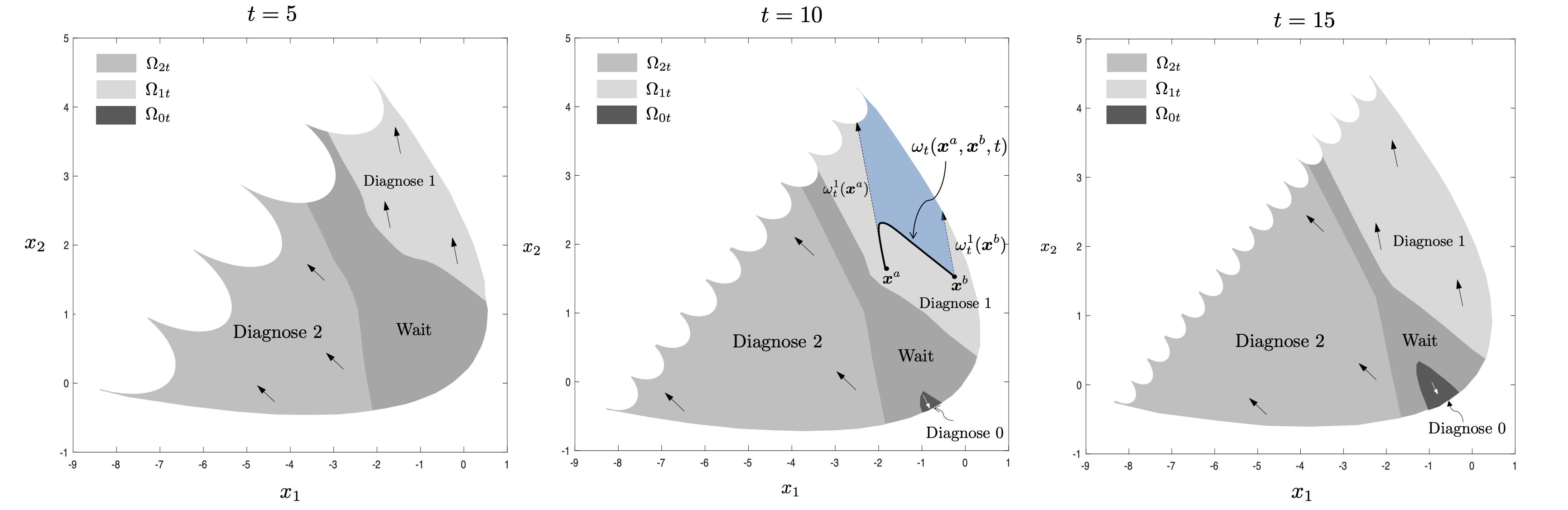} 
\caption{Illustrations of the optimal stopping regions and the $\omega$-convexity property for $r=2$ and $N=2$}
\label{Fig: 2Dchart} 
\end{figure}

Figure~\ref{Fig: 2Dchart} shows that the optimal stopping regions may be non-convex. But the following theorem suggests that they are still structured in the sense of a generalized convexity. In contrast to a standard convex set, which contains the straight line connecting any two points in the set, the stopping regions in Figure~\ref{Fig: 2Dchart} exhibit what we call $\omega$-convexity: a \emph{curve} of a particular shape that connects two points is contained in the same set. For expositional convenience, we denote $ \mathcal{T}^t (\boldsymbol x, \boldsymbol \theta) \triangleq (\mathcal{T}^t_0(\boldsymbol x, \boldsymbol \theta ), \ldots, \mathcal{T}^t_N(\boldsymbol x, \boldsymbol \theta ))$ as the belief state corresponding to the diagnostic statistic $\boldsymbol x$ in period $t$. 

\begin{theorem}[$\boldsymbol\omega$-convexity of the stopping regions]
\label{prop: w(x1,x2) belongs to Omegai}
\mbox{}
\begin{enumerate}
\item If $a_{ii} < a_{ij}$ for all $j \neq i$ and $\boldsymbol x^a \in \Omega_{it}$ for some $t$, then the set $\omega^i_{t} (\boldsymbol x^a) \triangleq  \big\{  \boldsymbol x \in \Omega_t \big| \boldsymbol e_j \mathbb{R} (\boldsymbol x-\boldsymbol x^a)=0, \forall j \neq i \in \mathcal{N} ,  \boldsymbol e_i \mathbb{R} (\boldsymbol x-\boldsymbol x^a)> 0  \big\}$ is a subset of $\Omega_{it}$. 
\item   If $\boldsymbol x^a, \boldsymbol x^b \in \Omega_{it}$ for some $t$, then the set $\omega_s (\boldsymbol x^a, \boldsymbol x^b, t) \triangleq \big\{\boldsymbol x \in \Omega_s \big|  \mathcal{T}^s(\boldsymbol x, \boldsymbol \theta )  =\rho \mathcal{T}^t (\boldsymbol x^a, \boldsymbol \theta) +  (1-\rho) \mathcal{T}^t (\boldsymbol x^b, \boldsymbol \theta) , \rho\in(0,1)  \big\}$ is a subset of $\Omega_{is}$ for all $s\geqslant t$. 
\end{enumerate}
\end{theorem}

Part 1 of the theorem can be interpreted as follows: the condition $a_{ii} < a_{ij}$ for all $j \neq i$ ensures that a correct diagnosis is less costly than an incorrect one. The $\omega$-convexity is driven by the convexity of the stopping regions in the belief space, as well as by the properties of the nonlinear manifolds $\mathcal{F}^{\boldsymbol \theta}_t$. Suppose we have identified a point inside a stopping region, say $\boldsymbol x^a\in\Omega_{1t}$, as shown in the middle panel of Figure~\ref{Fig: 2Dchart}. We can draw a straight line extending from $\boldsymbol x^a$ following the  direction of the arrow, which is determined by the first row of the reconstruction matrix~$\mathbb{R}$ (the stopping region for class $i$ corresponds to the $i$th row of $\mathbb{R}$). If this line belongs to the space $\Omega_t$, it will hit the boundary of $\Omega_t$ and produce a line segment, denoted by $\omega^1_{t} (\boldsymbol x^a)$, which the theorem suggests belongs to the same stopping region as $\boldsymbol x^a$. Note that the arrow direction is class-specific and is uniquely characterized by the corresponding row of the matrix~$\mathbb{R}$. Thus, two points in the same stopping region, $\boldsymbol x^a, \boldsymbol x^b$, can generate two parallel line segments,  $\omega^1_{t} (\boldsymbol x^a)$ and $\omega^1_{t} (\boldsymbol x^b)$, respectively, that belong to the same stopping region. 

Part 2 of the theorem characterizes a curve connecting $\boldsymbol x^a$ and $\boldsymbol x^b$ from the same stopping region, and its intersection with $\Omega_t$, if the intersection nonempty, denoted by $\omega_t (\boldsymbol x^a, \boldsymbol x^b, t)$,  will belong to the same stopping region as its endpoints $\boldsymbol x^a$ and $\boldsymbol x^b$. The shape of the curve depends on the endpoints $\boldsymbol x^a, \boldsymbol x^b$, as well as on the time $t$. Given two points $\boldsymbol x^a$, $\boldsymbol x^b$ from the same stopping region, the two parts of the theorem suggest that the region enclosed by the curve $\omega_t (\boldsymbol x^a, \boldsymbol x^b, t)$, the parallel lines $\omega^1_t(\boldsymbol x^a)$, $\omega^1_t(\boldsymbol x^b)$, and the boundary of $\Omega_t$ is a subset of the stopping region $\Omega_{1t}$, as illustrated in the middle panel of Figure~\ref{Fig: 2Dchart}. Additional interpretations of the $\omega$-convexity and its computational benefits are discussed in Appendix~\ref{sec: Omega convexity over time}. 

\begin{figure}
\centering \includegraphics[width=6.0in]{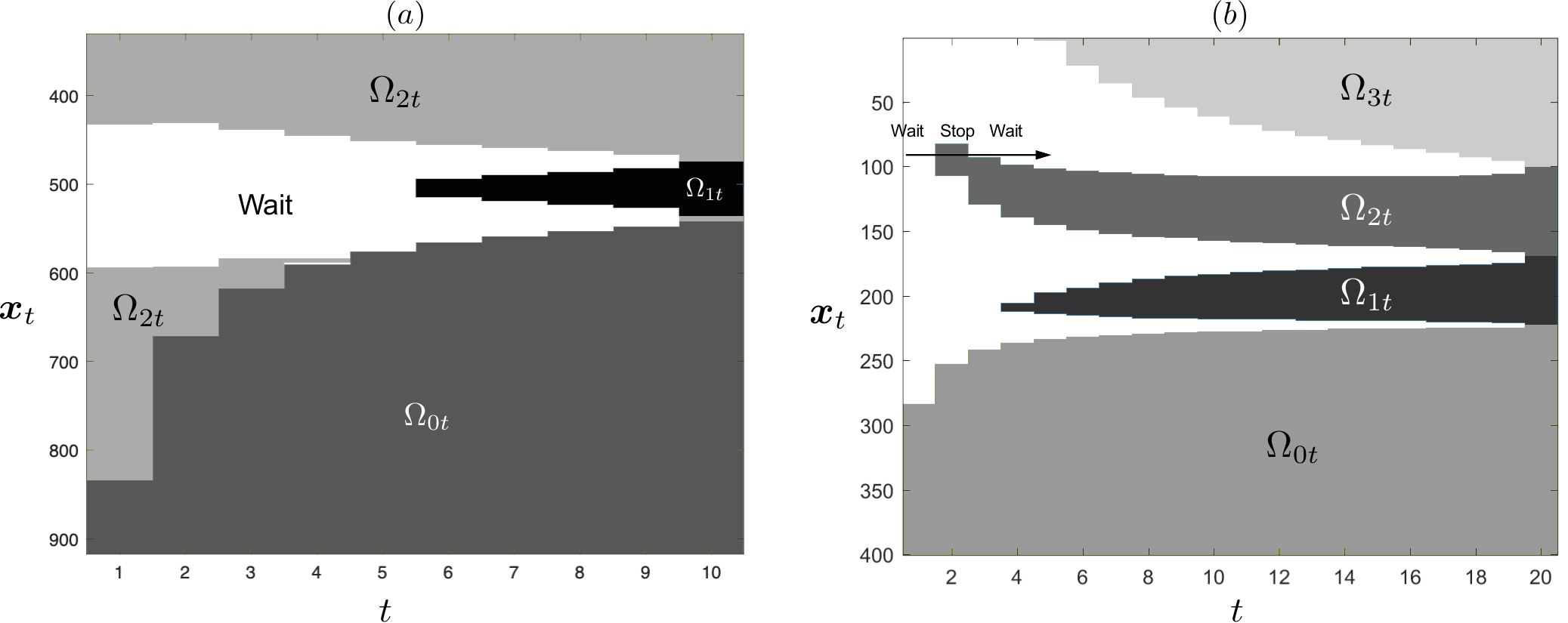} 
\caption{(a) The stopping region $\Omega_{2t}$ may have an exclave. Parameters: $N=2$; $c_i=0.03$; $\theta_i=1/3$; $\sigma^2_i=2.4$, for $i=0,1,2$; and $(\mu_0, \mu_1, \mu_2)=(-2, 0, 1.5)$. Termination costs are zero-one except $a_{12}=15$; $a_{21}=0.05$.  (b) When $\boldsymbol x_t$ remains fixed and $t$ increases, the optimal action may change from stop back to wait. Parameters: $N=3$; $c_i=0.075$, for $i=0,\ldots,3$; $\boldsymbol\theta=(1/3, 1/2, 1/15, 1/10)$; $(\mu_0, \mu_1, \mu_2, \mu_3)=(0,1.5,2,-1.5)$; and $(\sigma^2_0, \sigma^2_1, \sigma^2_2, \sigma^2_3)=(1,1.5, 5/3, 0.5)$. Termination cost: zero-one.}
\label{Fig: DisconnectedOmega} 
\end{figure}

Note that the arrows or curves may not intersect with $\Omega_t$. As a result, $\omega^i_{t} (\boldsymbol x^a)$, $\omega^i_{t} (\boldsymbol x^b)$, or $\omega_s (\boldsymbol x^a, \boldsymbol x^b, t)$ may be empty. In this case, the stopping region can be noncontiguous, since some points from the same stopping region may not be connected by a line or curve in that region. Figure~\ref{Fig: DisconnectedOmega}--(a) illustrates such a case, in which $\Omega_{2t}$ is noncontiguous with an exclave surrounded by other control regions. This example involves three normal distributions with common variance, and the means are ordered as $\mu_0<\mu_1<\mu_2$. Interestingly, exclave stopping regions can  appear only in $\Omega_{0t}$ and $\Omega_{2t}$, but not in $\Omega_{1t}$. A detailed analysis of this case can be found in Appendix~\ref{ECsec: connected middle stopping interval}. 



When the sample average $\boldsymbol x_t$ is fixed, the uncertainty of the belief declines as the sample size~$t$ increases. This might suggest that the optimal policy should be less inclined to wait. However, Figure~\ref{Fig: DisconnectedOmega}--(b) illustrates that more observations do not necessarily decrease the incentive to wait. For example, suppose we observe 85 three times in a row; Then, the optimal decision would change from wait to stop and diagnose class 2, and then back to wait. This is because the value of information is not necessarily concave in the sample size. In Appendix~\ref{sec: Counter-intuitive Properties of the Stopping Regions}, we explore the drivers of these counterintuitive properties in greater detail. 


\subsection{Critical Stopping Time}
\label{sec: Optimal Premature Termination}
Next, we uncover a phenomenon unique to sequential multi-class diagnosis. Consider the diagnosis over an infinite horizon, with observations taking real values. In two-class diagnosis, the waiting region in $\Omega_t$ is nonempty for each period, and thus there is a sample path along which it is optimal to keep waiting.\endnote{In two-class sequential diagnosis, a well-known structural result is that it is optimal to wait when the posterior probability remains between two thresholds~\citep{Wald1945}. This structure carries over to the diagnostic-statistic space if the observations take real values because each posterior probability in the waiting interval can be mapped to a diagnostic statistic that must lie in the waiting region of $\Omega_t$. } However, in multi-class diagnosis, the stopping time may be bounded: that is, there may exist a finite \emph{critical stopping time} at which all possible sample paths are cut off.

\begin{figure}
\centering \includegraphics[width=6.0in]{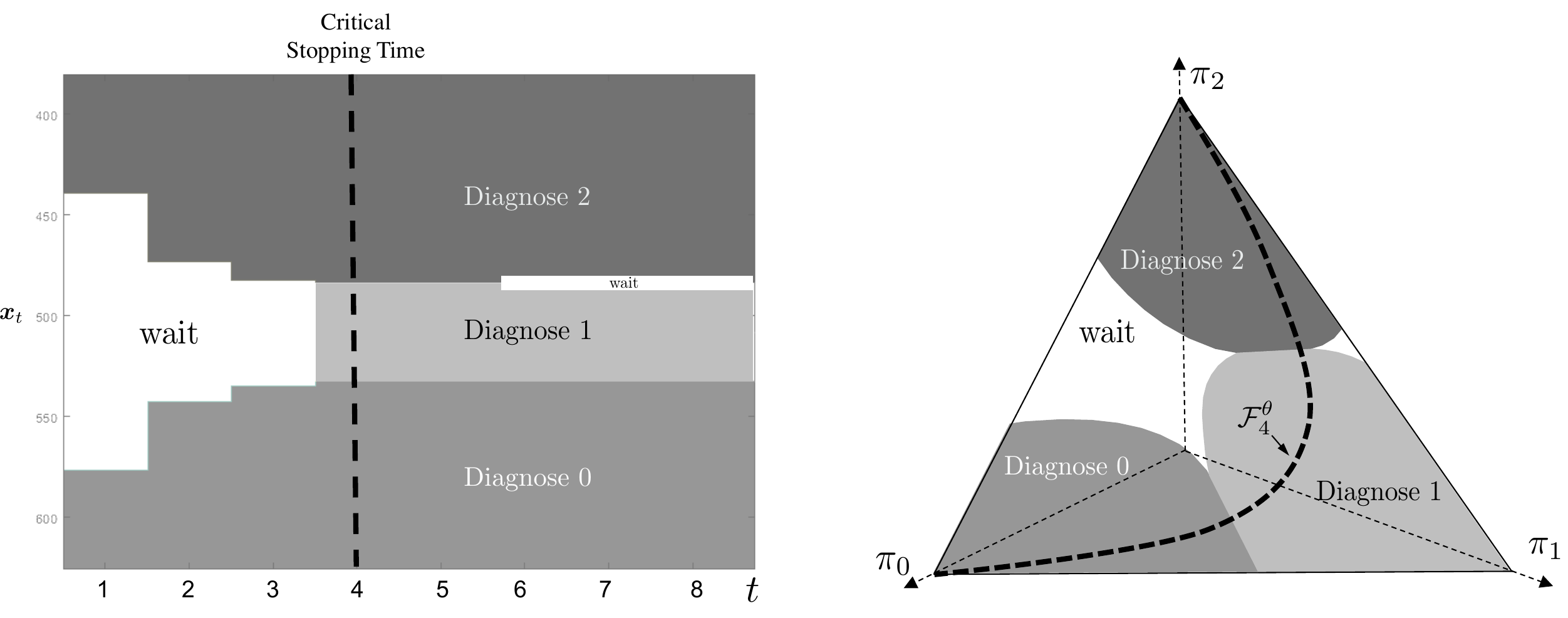} 
\caption{Illustration of the critical stopping time (left) and the corresponding belief track (right) }
\label{Fig: criticalHorizon} 
\end{figure}

In finite-horizon problems, the critical stopping time can be shorter than the horizon length. An example is given in the left-hand panel of Figure~\ref{Fig: criticalHorizon}, in which the decision process is allowed to run for $T=10$ periods, but it is never optimal to progress beyond the $4$th period, i.e., the critical stopping time. Note that the optimal waiting interval in that period is empty, so $x_t$ will hit the wall and must stop. This is because the belief track $\mathcal{F}^{\boldsymbol \theta}_4$ at that period is fully covered by the stopping regions, as shown in the right-hand panel, so the optimal action is to stop for all belief states on the belief track. 


 
We now consider a classical infinite-horizon problem with zero-one termination cost. Let $|A|$ denote the maximum number of duplicated elements in~$A$. For example, for $A=\{1,2,1,2,2\}$, we have $|A|=3$. Define the \emph{degree of confusion} as
$
n_s \triangleq\max_{ \boldsymbol  x\in\Omega_t, t=1,\ldots, T} | \beta_{i0}+  \boldsymbol{e}_i \mathbb{R}  \boldsymbol x , i\in\mathcal{N} |,
$
which is the maximum number of densities intersecting at the same point.\endnote{For example, consider three normal densities $f_0,f_1$, and $f_2$ with identical variance but distinct means $\mu_0< \mu_1<\mu_2$.  Two densities $f_i$ and $f_j$ intersect at $(\mu_i+\mu_j)/2$, but it is impossible for three densities to intersect. Accordingly, $n_s=2$ in this case. Obviously, the degree of confusion never exceeds the number of classes, namely, $n_s \leqslant N+1$. In addition, $n_s=1$ represents the trivial case where a single observation can reveal the true class; for this reason we  consider only cases with~$n_s \geqslant 2$. } The following theorem provides a sufficient condition for the existence of a finite critical stopping time: 

\begin{theorem}
\label{corollary: cutoff horizon in MSPRT}
Consider an infinite-horizon  ($T=\infty$) problem with uninformative prior~$\theta_i=\frac{1}{N+1}, i\in \mathcal N$, and zero-one cost structure. If $c>(n_s-1)/n_s$, then there exists a finite critical stopping time at which the decision process must terminate. When $c\geqslant N/(N+1)$, the critical stopping time is 0.
\end{theorem}

In two-class diagnosis ($N=1$, $n_s=2$), the number of classes is equal to the degree of confusion. According to Theorem~\ref{corollary: cutoff horizon in MSPRT}, it is optimal to stop at $t=0$ if $c>1/2$; otherwise, it is never optimal to force the decision process to stop at a predetermined time. In contrast, in multi-class diagnosis ($N\geqslant 2$), the number of classes may exceed the degree of confusion ($N+1>n_s$). If $(n_s-1)/n_s<c<N/(N+1)$, it can be optimal to initiate the decision process but  run only up to a finite critical stopping time.  


\section{Comparison with a Heuristic Policy}
\label{sec: Optimal vs. MSPRT}
We now compare the performance of the optimal policy with a state-of-the-art heuristic policy known as the M-ary sequential probability ratio test (MSPRT)~\citep{Baum1994}. We first conduct numerical studies to understand when and why the optimal policy can significantly outperform MSPRT, and then illustrate the benefits of the optimal solution in a healthcare application. 

\subsection{Limitations of MSPRT}
For MSPRT, a separate threshold is set on each posterior class probability. The decision process stops whenever at least one posterior class probability exceeds the corresponding threshold, and the diagnosis is given to the class with the largest posterior probability. Specifically, the thresholds are parameterized by $A=\{A_0, A_1, \ldots, A_N\}$, and the stopping time $\tau_A$ is the first $t\geqslant 1$ such that $\pi_{jt}>1/(1+A_j)$ for at least one $j\in\mathcal{N}$.  The diagnosis is given to $d=\argmax_j \pi_{j, \tau_A}$. MSPRT is asymptotically optimal when the delay cost approaches zero, in which case all thresholds simultaneously approach one. 

Despite its mathematical tractability, MSPRT has several limitations in practice: First, the thresholds of MSPRT are optimized based on an open-loop approximation to a closed-loop control problem. Given the threshold parameters $A$, the asymptotic expressions of the stopping time and error probabilities are derived using  nonlinear renewal theory~\citep{woodroofe1982nonlinear}, which provides an explicit approximation to the cost function. It is then  straightforward to optimize $A$ by applying the first-order condition. However, the open-loop approximation may have poor response to stochastic variations. 

Second, as an asymptotically optimal approach, the performance of MSPRT may deteriorate in the non-asymptotic regimes. For example, medical diagnoses often need to be made quickly and hence may be outside the asymptotic regime. This is especially the case when the observations are noisy, since more observations are needed to reach the asymptotic regime.  \cite{Baum1994} showed (in their Figure 2) that the threshold structure of MSPRT provides a good approximation to the  optimal stopping regions when these regions are small and close to the vertices of the belief space. However, in quick diagnosis with noisy observations, the stopping regions can be large and extend to the center of the belief space.   

Third, MSPRT is developed under the assumption that the cost structure is zero-one, so both the delay and misdiagnosis costs are class-invariant; However, the costs in practice can be class-specific because some errors or delays can be worse than others. For example, false negatives often have more serious implications than false positives in medical diagnosis, in which case, the threshold structure of MSPRT can be an oversimplification. 

\subsection{Numerical Studies}
Consider three classes represented by normal distributions with equal variance~$\sigma^2$ and different means,~$\mu_1<\mu_0<\mu_2$. In MSPRT, the threshold for a particular class is determined only by that class and the nearest class (in terms of the Kullback-Leibler divergence), thereby decoupling the joint $N$-class diagnosis problem into $N$ standalone two-class problems. For example, the threshold for class~1 depends only on itself and class 0, the nearest class, ignoring class~2. Similarly, if class~0 and~2 are closer, i.e.,~$|\mu_2-\mu_0|<|\mu_0-\mu_1|$, then the thresholds for class~0 and class~2 are identical; both ignore class~1. Such ignorance introduces considerable bias when some classes appear similar. 

Table~\ref{MSPRTvsOptimal2-NewCloud102_m1_07_fullPicture-Brief version} compares the expected total cost of MSPRT and the optimal policy when~$\mu_2$ is near~$\mu_0$ (without loss of generality). Here, the prior is~$(\theta_0, \theta_1, \theta_2)=(1/3,1/3,1/3)$, the cost structure is zero-one. We observe that MSPRT can incur up to~$66\%$ higher costs than the optimal policy.

Specifically, MSPRT performs poorly because in the regime favoring class~0 or~2, i.e., $f_0(y)\approx f_2(y) > f_1(y)$,  ignoring class~2 may lead to significant overestimation of the posterior probability of class~1, since $\pi_1=(1+\theta_0 f_0(y)/\theta_1 f_1(y)+\theta_2 f_2(y)/ \theta_1 f_1(y))^{-1}\ll (1+\theta_0  f_0(y)/ \theta_1 f_1(y))^{-1}$. Furthermore, in the regime that favors class~1, namely, $ f_1(y)> f_0(y)\approx f_2(y)$, we have $\pi_0\approx \pi_2=(1+\theta_0 f_0(y)/\theta_2 f_2(y)+\theta_1 f_1(y)/ \theta_2 f_2(y))^{-1}\ll (1+\theta_0  f_0(y)/ \theta_2 f_2(y))^{-1}$, so ignoring class~1 leads to the overestimation of the probabilities of class~0 and~2.  

\begin{table}\footnotesize
\centering
\caption{Cost of the MSPRT vs. the optimal policy}
\label{MSPRTvsOptimal2-NewCloud102_m1_07_fullPicture-Brief version}
\begin{tabular}{cccccccc}
  \toprule
\multicolumn{3}{c}{$(\mu_0, \mu_1, \sigma^2, T)=( 0, -0.7,  1, 80)$}& $\mu_2=0.01$ & $\mu_2=0.02$ & $\mu_2=0.03$ & $\mu_2=0.04$ & $\mu_2=0.05$  \\
  \midrule
    \multirow{3}{*}{$c=0.001$} & \multicolumn{2}{c}{MSPRT} & 0.5758&	0.5737&	0.5495&	0.4051&	0.3267 \\
       & \multicolumn{2}{c}{Optimal}      &0.3515	&0.3451	&0.3378	&0.3292	&0.3196	 \\
       & \multicolumn{2}{c}{MSPRT Increased cost}     &{63.78\%}	&{66.20\%}	&{62.69\%}	&{23.06\%}	&{2.21\%}	\\
    \midrule
        \multirow{3}{*}{$c=0.005$} & \multicolumn{2}{c}{MSPRT}      &0.5789	&0.5785	&0.5768	&0.5758	&0.5744		  \\
       & \multicolumn{2}{c}{Optimal}      	&0.4147	&0.4100	&0.4053	&0.4004	&0.3955	\\
       & \multicolumn{2}{c}{MSPRT Increased cost}        &{39.59\%}	&{41.07\%}	&{42.31\%}	&{43.79\%}	&{45.22\%}	\\
  \midrule
        \multirow{3}{*}{$c=0.010$} & \multicolumn{2}{c}{MSPRT}      &0.5842	&0.5823	&0.5811	&0.5793	&0.5773	  \\
       & \multicolumn{2}{c}{Optimal}      	&0.4674	&0.4633	&0.4593	&0.4552	&0.4511	\\
       & \multicolumn{2}{c}{MSPRT Increased cost}        &{25.67\%}	&{26.50\%}	&{27.25\%}	&{27.96\%}	&{29.13\%}	\\
  \bottomrule
\end{tabular}
\end{table}


\begin{figure}
\centering \includegraphics[width=6.0in]{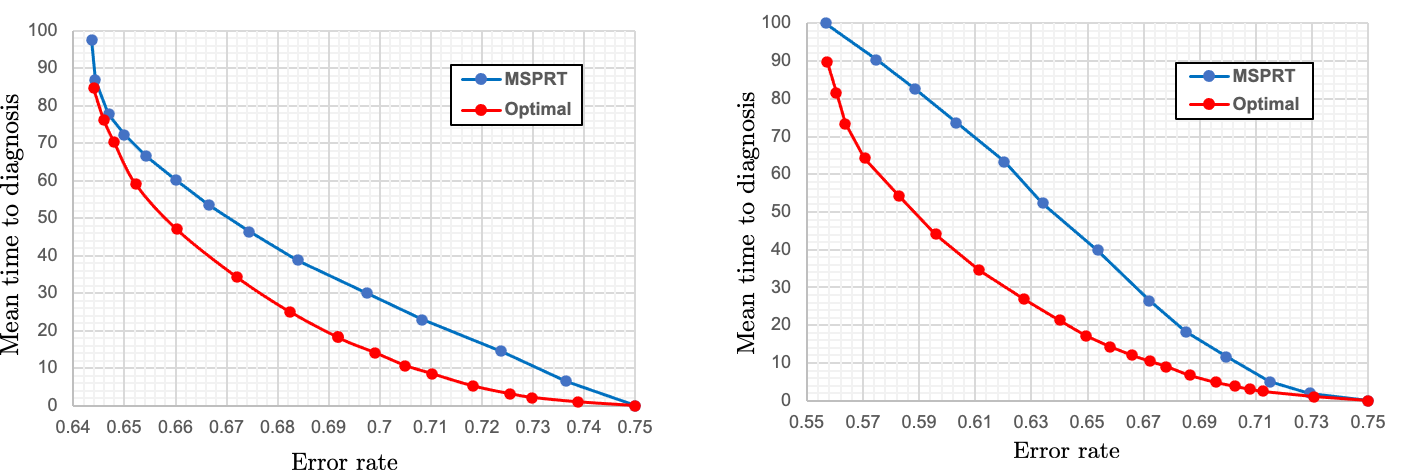} 
\caption{The optimal policy reaches a diagnosis with the same accuracy more quickly than MSPRT. Left: $(\mu_0,\mu_1,\mu_2,\mu_3)=(0,-0.1,0.01,-0.01)$; right: $(\mu_0,\mu_1,\mu_2,\mu_3)=(0,-0.1,0.01,0.1)$}
\label{Fig: StasOptimalMSPRT} 
\end{figure}

We also compare the two policies in terms of the mean stopping time and error rate. For a given delay cost $c$, we compute the optimal and MSPRT policies and use simulation to estimate the corresponding expected stopping times and error rates. By increasing~$c$, we can  produce a sequence of decreasing stopping times and a corresponding sequence of increasing error rates. Plotting them together produces a diagnosis frontier showing the expected time taken to reach a specific error rate under each policy, as shown in Figure~\ref{Fig: StasOptimalMSPRT}.\endnote{The error rates in Figure~\ref{Fig: StasOptimalMSPRT} are intrinsically high due to the difficult nature of early diagnosis in low signal-to-noise-ratio regimes.}  Here, the goal is to diagnose four classes, each having a normal distribution with unit variance, within $T=100$ periods under the uninformative prior.  We observe that the optimal policy can reduce the diagnosis time by over half in shorter-time regimes (the left-hand panel) and even in the longer-time regimes (the right-hand panel), suggesting that it is valuable in medical and safety-critical applications, where a slight improvement in diagnostic speed could have life-or-death implications.


We have seen that the optimal policy performs considerably better than MSPRT when some classes are difficult to distinguish. Such scenarios, having low signal-to-noise ratios, are ubiquitous in practice. For example, subnoise fault signals in jet engines are hard to differentiate from  normal signals. Next, we consider a medical application that illustrates the advantage of the optimal policy.  

\subsection{A Medical Application}
\label{sec: PD example}

Parkinson's disease (PD) is a chronic degenerative disorder of the central nervous system affecting 6.2 million people globally in 2015. It is a complex disease with known heterogeneity in its symptoms, progression rate, and response to treatment. To date, PD management programs are mostly one-size-fits-all, and no widely accepted scheme is available to classify the patients into finer subtypes. However, it is widely recognized that the early identification of PD subtypes is important as the right care for each patient could be initiated sooner.   

In personalized medicine, primary care physicians can assign a new patient to the standard management protocol, follow up with the patient periodically up for a prescribed time, and then decide when to stop monitoring and triage the patient to a more specialized protocol to receive targeted management.  

The diagnosis of PD is  based mainly on symptoms, and different subtypes exhibit only subtle differences in the initial symptoms. \cite{zhang2019data} identified three subtypes of PD with distinct severity and prognosis by analyzing data from the Parkinson progression marker initiative. The demographic and clinical characteristics of each subtype are shown in Table~\ref{Personalized Med: Class Parameter Table}. Here, we choose the observation variable to be the Montreal cognitive assessment (MoCA) score; this test is widely used to assess the cognitive impairment of PD patients.  

For each PD subtype, the MoCA follows a normal distribution whose parameters are summarized in Table~\ref{Personalized Med: Class Parameter Table}. Since differences in the MoCA are small across the subtypes, the diagnosis is intrinsically difficult, so multiple follow-up assessments are often necessary.\endnote{The change in MoCA between two assessments could also be used as the observation variable. However, the statistical distribution of this change is not well documented in the medical literature. } In this example, the horizon length is $T=20$, and the termination costs are zero-one. The optimal policy minimizes the expected number of assessments required to reach a target accuracy rate. Given a new patient's age and gender, we can use the demographic information in Table~\ref{Personalized Med: Class Parameter Table} and Bayes' rule to calculate the subtype probability vector~$\boldsymbol\theta$ as the prior (see Appendix~\ref{EC: Details of the medical application}). The corresponding optimal policy can then be shared by other patients of the same age and gender.

\begin{table}\footnotesize
\centering
\smallskip
\caption{Group characteristics of the three subtypes of Parkinson's  disease identified by \cite{zhang2019data}} 
\label{Personalized Med: Class Parameter Table}
\begin{threeparttable}
\begin{tabular}{c c c c}
\toprule
 & \textbf{Subtype 1} & \textbf{Subtype 2} & \textbf{Subtype 3} \\
\toprule
Prevalence (\%) & $43.13$  & $22.96$ & $33.91$\\
\cmidrule(r){1-4} 
Male/Female (\%) & $63.6/36.3$  & $58.8/41.2$ & $68.9/31.1$ \\
\cmidrule(r){1-4} 
 Age of diagnosis (years)\tnote{*}  & $62.66 (9.55)$  & $64.61 (9.20)$ & $69.16 (8.82)$  \\
\cmidrule(r){1-4} 
MoCA \tnote{*} & $27.98 (1.86)$  & $27.09 (2.40)$ & $24.62 (4.06)$\\
\bottomrule
\end{tabular}
\begin{tablenotes}\footnotesize
\item[*] Normal distribution: mean (standard deviation)
\end{tablenotes}
\end{threeparttable}
 \end{table}


\begin{figure}
\centering \includegraphics[width=6.5in]{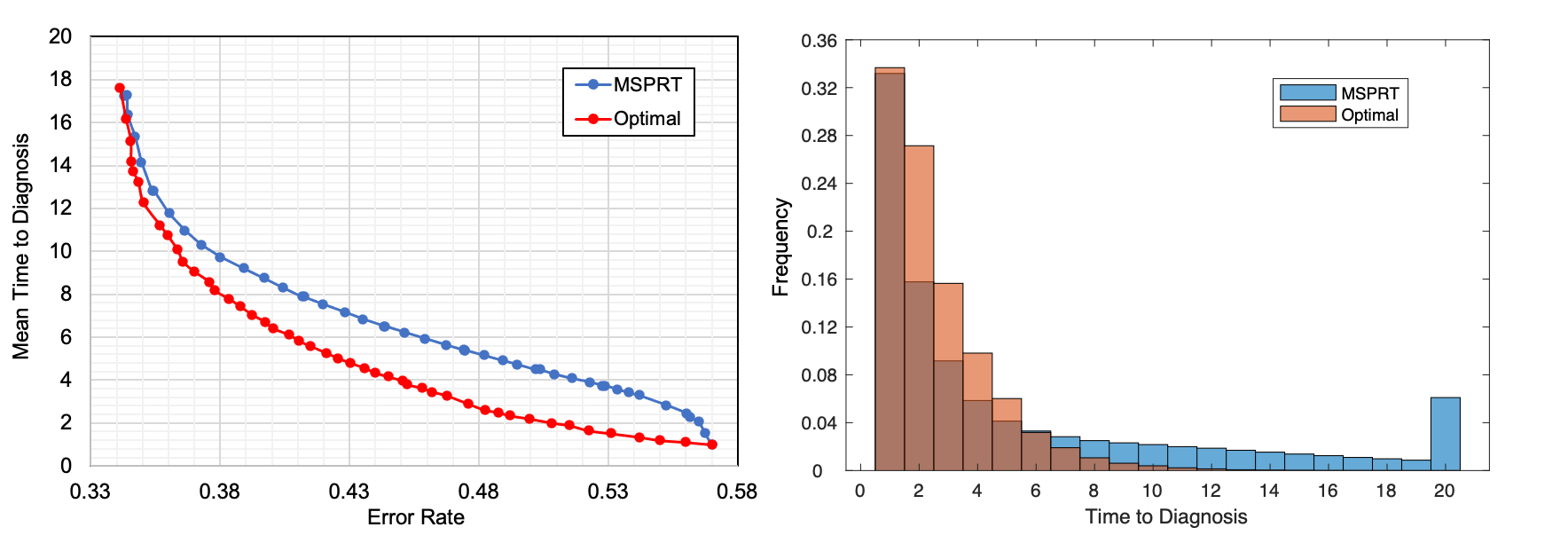} 
\caption{Comparison of the MSPRT and the optimal policy in the rapid triage of a 65-year-old male PD patient into 3 subtypes. Left: Time-error frontiers; right: histograms of the time to diagnosis with an error rate of 0.48}
\label{Fig: PD_OptimalvsMSPRT} 
\end{figure}

In the left-hand panel of Figure~\ref{Fig: PD_OptimalvsMSPRT}, we compare the optimal policy with the MSPRT in terms of the error rate and the mean time to diagnosis. We see that, when the diagnosis needs to be made quickly, the optimal policy reaches the target accuracy rate significantly faster than the MSPRT. For example, the optimal policy, on average, needs only  $2.6$ periods to achieve an accuracy rate that would have taken the MSPRT 5.2 periods. The diagnosis is time-sensitive because even a slightly earlier initiation of the effective treatment can slow, or even halt, the disease progression~\citep{murman2012early}. The right-hand panel of Figure~\ref{Fig: PD_OptimalvsMSPRT} shows the histograms of the time to diagnosis under different policies, both of which have the same error rate (48\%). We observe that, with the optimal policy, the time to diagnosis lies mostly in the lower range, whereas, with MSPRT, the histogram has a longer tail. Although the error rate of $48\%$ may seem high, it is due to the intrinsic difficulty of diagnosis with a small sample size and noisy observations. The role of primary care doctors is to triage the patients into different disease management programs, in which a more accurate diagnosis can be made after more extensive tests have run. 

Nevertheless, numerical evidence suggests that MSPRT is a good alternative to the optimal policy when the error rate is required to be small (e.g., $\leqslant 5\%$) and when it is possible to collect a large number of samples. 




\section{Extension to Multivariate Observations} 
\label{sec: Extension to Multivariate Observations}
So far, we have considered the case of univariate observations, but in some applications, the DM may have access to multivariate observations. For example, a jet engine is often equipped with multiple sensors that generate different streams of monitoring data (e.g., vibration, temperature, and speed data), which can be jointly analyzed for fault diagnosis. We now extend the idea of belief reconstruction to make effective approximations for the multivariate case and illustrate the proposed method in a maintenance application. 

\begin{table}\footnotesize
\centering
\smallskip
\caption{Examples of multivariate exponential families ($d \geq 2$)}
 \label{list_distributions-MV}
\begin{threeparttable}
\resizebox{\columnwidth}{!}{%
\begin{tabular}{c c c }
\toprule
 &{Multivariate Normal} &{Multinomial} \\
\cmidrule(r){1-3} 
Parameters & $ (\boldsymbol\mu_i, \boldsymbol\Sigma_i)$ & $ (n, p_{i1}, \ldots, p_{id}, 1-\sum^d_{j=1} p_{ij})$ \\
\cmidrule(r){1-3} 
$\boldsymbol\beta^\mathsf{T}_i$ & $(\boldsymbol\mu^\mathsf{T}_i  \boldsymbol\Sigma^{-1}_i - \boldsymbol\mu^\mathsf{T}_0  \boldsymbol\Sigma^{-1}_0 , \frac{1}{2} \text{vec} (\Sigma^{-1}_0)^\mathsf{T} - \frac{1}{2}\text{vec} (\Sigma^{-1}_i) ^\mathsf{T} ) $ & $ \big(\log \frac{p_{i1}}{1-\sum^d_{j=1} p_{ij}} -\log \frac{p_{01}}{1-\sum^d_{j=1} p_{0j}} , \ldots, \log \frac{p_{id}}{1-\sum^d_{j=1} p_{ij}}   - \log \frac{p_{0d}}{1-\sum^d_{j=1} p_{0j}} \big)$ \\
\cmidrule(r){1-3} 
$\boldsymbol h^\mathsf{T} (y)$ & $(y^\mathsf{T}, \text{vec}(yy^\mathsf{T})^\mathsf{T}  ) $ & $(y_1, \ldots, y_d) $ \\ 
\cmidrule(r){1-3} 
$\beta_{i0}$ & $ \frac{1}{2} (\boldsymbol\mu^\mathsf{T}_0 \Sigma^{-1}_0 \boldsymbol\mu_0 -\boldsymbol\mu^\mathsf{T}_i \Sigma^{-1}_i \boldsymbol\mu_i) +\frac{1}{2}\log |\Sigma_0| -\frac{1}{2} \log |\Sigma_i| $ & $n\log(1-\sum^d_{j=1}p_{ij}) -n \log(1-\sum^d_{j=1}p_{0j}) $ \\
\cmidrule(r){1-3} 
$p$  & $d(d+3)/2$ & $d$ \\
\bottomrule
\end{tabular}
}
\end{threeparttable}
 \end{table}

Table~\ref{list_distributions-MV} lists the parameters of the multivariate normal and multinomial distributions in the ETM form, where $d$ denotes the dimension of the observations. It can be seen that the number of tilting parameters~$p$ can be large, in which case, the computation may become intractable for large~$N$. In addition, the cost-to-go function involves an integration over the $d$-dimensional observation space.  To overcome these computational challenges, we will develop an approximation scheme based on the second-level dimension reduction, which involves exploiting the interrelation among $f_i$'s. In fact, some high-dimensional problems may actually lie in a lower dimensional subspace due to certain interclass relations. We provide a numerical example below. 


\subsection{Motivating Examples}

\begin{figure}
\centering \includegraphics[width=5.5in]{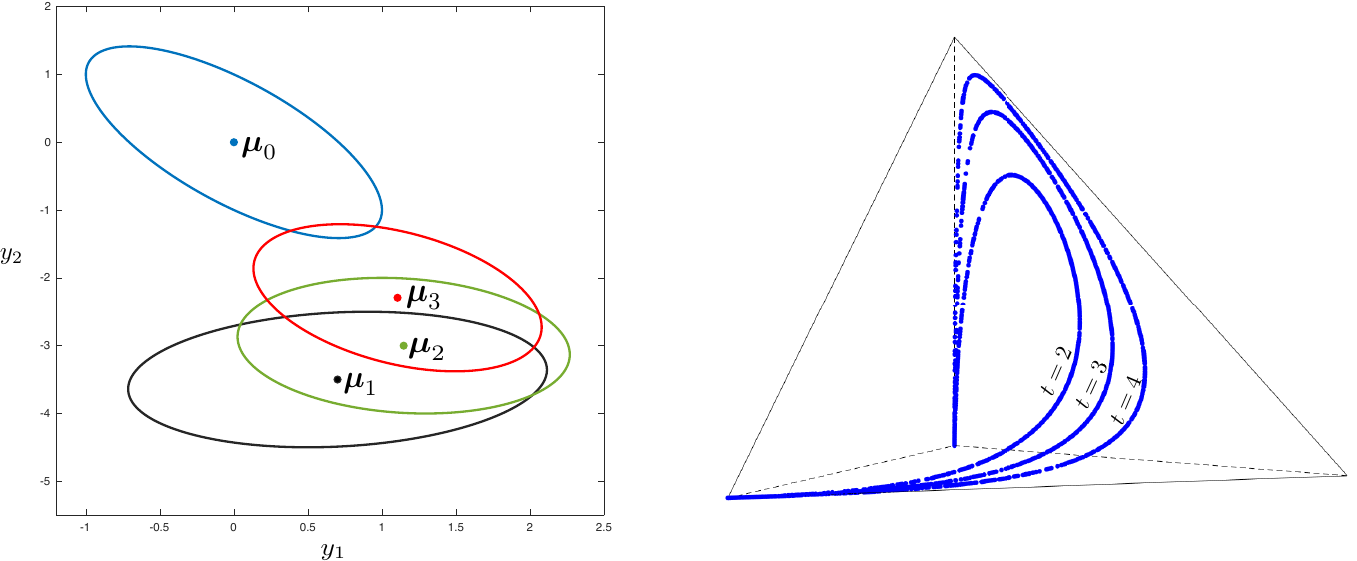} 
\caption{The four bivariate normal classes in the left-hand panel have intrinsic dimension of 1.} 
\label{Fig: SurprisingExample} 
\end{figure}

\begin{example}
\label{ex: MV Example 1}
Consider four classes with bivariate normal observations $f_i(y_1, y_2)=N(\boldsymbol \mu_i, \boldsymbol\Sigma_i), i=0,\ldots, 3$, where 
 $\boldsymbol \mu_0=(0, 0)^\mathsf{T}, \boldsymbol \mu_1=(0.7,-3.5)^\mathsf{T}$, 
\begin{align}
\boldsymbol\Sigma_0=
\begin{bmatrix}
&1&-1\\
&-1 & 2\\
\end{bmatrix}, 
\boldsymbol\Sigma_1=
\begin{bmatrix}
&2&0.2\\
&0.2 & 1\\
\end{bmatrix}, 
 \nonumber
\end{align}
and $(\boldsymbol \mu_2, \boldsymbol\Sigma_2), (\boldsymbol \mu_3, \boldsymbol\Sigma_3)$ are given by 
\begin{align}
\label{conditions for rank-one MVs-1}
\boldsymbol\Sigma_i  &= \big(\rho_i(\boldsymbol\Sigma^{-1}_1-\boldsymbol\Sigma^{-1}_0) + \boldsymbol\Sigma^{-1}_0 \big)^{-1}, \\
\label{conditions for rank-one MVs-2}
\boldsymbol \mu_i   &= \big( \rho_i (\boldsymbol \mu^\mathsf{T}_1 \boldsymbol\Sigma^{-1}_1 -\boldsymbol \mu^\mathsf{T}_0 \boldsymbol\Sigma^{-1}_0) \boldsymbol\Sigma_i + \boldsymbol \mu^\mathsf{T}_0 \boldsymbol\Sigma^{-1}_0 \boldsymbol\Sigma_i \big)^\mathsf{T}, 
\end{align}
with~$\rho_2=0.8$ and $\rho_3=0.5$. The equiprobability contours of these densities are shown in the left-hand panel of Figure~\ref{Fig: SurprisingExample}. Although $N=3$ and $p=5$, the intrinsic dimension is only $r=\text{rank}(\mathbb{B})=1$.\endnote{The conditions \eqref{conditions for rank-one MVs-1}--\eqref{conditions for rank-one MVs-2} imply that the rows of $\mathbb{B}$ are linearly dependent, i.e., $(\boldsymbol\mu^\mathsf{T}_i  \boldsymbol\Sigma^{-1}_i - \boldsymbol\mu^\mathsf{T}_0  \boldsymbol\Sigma^{-1}_0 , \frac{1}{2} \text{vec} (\Sigma^{-1}_0)^\mathsf{T} - \frac{1}{2}\text{vec} (\Sigma^{-1}_i) ^\mathsf{T} ) = \rho_i (\boldsymbol\mu^\mathsf{T}_1  \boldsymbol\Sigma^{-1}_1 - \boldsymbol\mu^\mathsf{T}_0  \boldsymbol\Sigma^{-1}_0 , \frac{1}{2} \text{vec} (\Sigma^{-1}_0)^\mathsf{T} - \frac{1}{2}\text{vec} (\Sigma^{-1}_1) ^\mathsf{T} ) $. Hence, the rank of $\mathbb{B}$ is 1. } We can see that the belief states are restricted on the belief curves shown in the right-hand panel of the figure. In this case, the four classes can be discriminated by a scalar 
$
{\boldsymbol x}_t = \sum^t_{m=1} \big( -0.9820  y_{1m} + 5.0081  y_{2m}  -1.0241  y^2_{1m} -1.515 y_{1m} y_{2m} +0.0140 y^2_{2m} \big)/t, 
$
without relinquishing the diagnostic information.  
\end{example}
Equations~\eqref{conditions for rank-one MVs-1} and~\eqref{conditions for rank-one MVs-2} characterize a specific relation among~$f_0,\ldots, f_N$, which allows the second-level dimension reduction. From an inspection of the equiprobability contours in Figure~\ref{Fig: SurprisingExample}, it is perhaps not easy to detect that the intrinsic dimension is 1. After all, the covariance matrices are not identical, and the centroids are not restricted to lie on a line. Nonetheless, the rank decomposition of~$\mathbb{B}$ can reveal the low dimensionality hidden in high-dimensional beliefs. 

Incidentally, we observe that~${\boldsymbol x}_t$ is a \emph{nonlinear} transformation of the original observations. Accordingly, the second-level dimension reduction is different from the classical linear dimension reduction methods,  such as principal component analysis or reduced-rank linear discriminant analysis. 

Of course, in practice, it is rare for the diagnostic parameter matrix $\mathbb{B}$ to be rank-deficient. However, in some situations, a full-rank matrix may be adequately approximated by a low-rank matrix. Consider a small perturbation to the distribution parameters originally satisfying~\eqref{conditions for rank-one MVs-1}--\eqref{conditions for rank-one MVs-2}. The new matrix is unlikely to be singular, but it may not be too far from a matrix having rank 1. We provide a numerical example below. 

\begin{example}
\label{ex: MV Example 2}
Consider four classes with bivariate normal observations $f_i(y_1, y_2)=N(\boldsymbol \mu_i, \boldsymbol\Sigma_i), i=0,\ldots, 3$, where $\boldsymbol \mu_0=(0,0), \boldsymbol \mu_1=(1,1), \boldsymbol \mu_2=(1.4,1.6),\boldsymbol \mu_3=(-1.6,-1.4)$, and 
\begin{align}
\boldsymbol\Sigma_0=
\begin{bmatrix}
&2.0&1.5\\
&1.5 & 2.5\\
\end{bmatrix}, 
\boldsymbol\Sigma_1=
\begin{bmatrix}
&2.2&1.5\\
&1.5 & 2.2\\
\end{bmatrix}, 
\boldsymbol\Sigma_2=
\begin{bmatrix}
&2.1&1.5\\
&1.5 & 2.5\\
\end{bmatrix}, 
\boldsymbol\Sigma_3=
\begin{bmatrix}
&2.0&1.5\\
&1.5 & 2.6\\
\end{bmatrix}. 
 \nonumber
\end{align}
The corresponding matrix~$\mathbb{B}$ has rank 3, and its three singular values in order are $0.9325, 0.3133$, and $0.0681$. Note that the first singular value is much larger than the others, suggesting that $\mathbb{B}$ is similar to a matrix having rank 1. The simulated sample belief states in a given period are concentrated near a curve, as shown in Figure~\ref{Fig: ApproximateBeliefReconstruction}, which implies that the four classes may be approximately discriminated by a scalar. 
\end{example}

When the high-dimensional beliefs are concentrated near some low-dimensional manifolds, we can project the beliefs to these manifolds so that the dynamic programming can be performed in a lower dimension. Next, we describe the details of the projection scheme. 


\subsection{Approximate Belief Reconstruction} 
\label{sec: Approximate Belief Reconstruction}
\begin{figure}
\centering \includegraphics[width=2.5in]{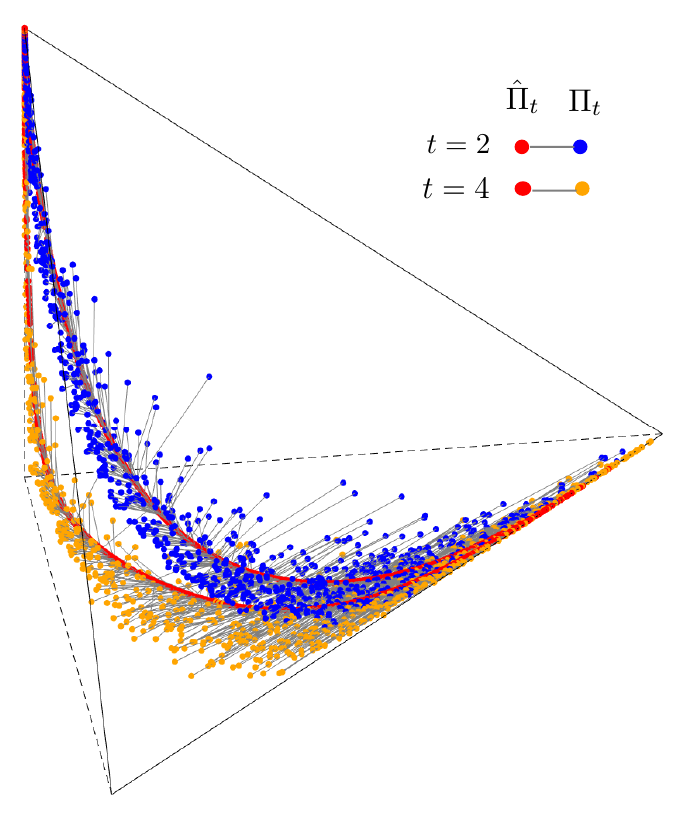} 
\caption{Illustration of the approximate belief reconstruction for Example~\ref{ex: MV Example 2}. Beliefs on three-dimensional manifolds ($r=3$) can be projected to a series of time-specific belief curves ($k=1$) in a nonlinear fashion. }
\label{Fig: ApproximateBeliefReconstruction} 
\end{figure}

We seek sparse representations of the high-dimensional reachable belief space by projecting it to a series of distinct low-dimensional subspaces (one for each period, as illustrated in Figure~\ref{Fig: ApproximateBeliefReconstruction}). Although not all belief spaces have good sparse representations, the proposed scheme can reveal meaningful low-dimensional representations when they exist.
 
The approximation scheme is based on low-rank matrix approximation. When the intrinsic dimension~$r=\text{rank}(\mathbb{B})$ is large, we can find a matrix~${\mathbb{B}}_k$ of lower rank, say $k<r$, to approximate $\mathbb{B}$ by minimizing some approximation error. Here, choosing an appropriate error measure is important. We select a mathematically convenient criterion called the Frobenius norm, defined as~$|| \mathbb{B}- {\mathbb{B}}_k ||_F $, where $|| D ||_F=\sqrt{\sum^N_{i=1}\sum^p_{j=1} D^2_{ij} }$ measures the total element-wise squared error. Under the Frobenius norm measure, the resulting low-rank matrix has an elegant form; we will show later that it also has a geometric interpretation in some cases. Given the singular value decomposition $\mathbb{B}=U\Sigma V^\mathsf{T}$, the Eckart-Young-Mirsky theorem suggests that the rank-$k$ matrix minimizing the Frobenius error is 
$$
{\mathbb{B}}_k = U\Sigma_k V^\mathsf{T}, 
$$
where $\Sigma_k$ is derived from $\Sigma$ by setting the $(r-k)$ smallest singular values on the diagonal to zero. The reduced rank $k$ is selected based on the computational resources available; laptop computers today can easily handle $k\leq 3$. The low-rank matrix can be decomposed as~${\mathbb{B}}_k =\mathbb{R}_k \mathbb{P}_k $, where the approximate reconstruction matrix $\mathbb{R}_k$ is obtained by selecting the~$k$ leading columns of~$U$, corresponding to the~$k$ leading singular values, and the approximate projection matrix $\mathbb{P}_k$ is constructed using the~$k$ leading rows of~$\Sigma_k V^\mathsf{T}$.

In exact belief reconstruction, we use the original matrix $\mathbb{B}$ to reconstruct the belief $\Pi_t$, which lies on an $r$-dimensional manifold. In approximate belief reconstruction, we use the low-rank matrix~${\mathbb{B}}_k$ to find a belief $\hat\Pi_t$, which lies on a $k$-dimensional manifold as an approximation to~$\Pi_t$ (see Figure~\ref{Fig: ApproximateBeliefReconstruction}). Equivalently, in the sufficient-statistic space, this corresponds to compressing the  $r$-dimensional state variable $\bar{\boldsymbol x}_t  \triangleq \sum^t_{m=1}\boldsymbol h(y_m)/t$ into the approximate diagnostic statistic $\hat{\boldsymbol x}_t \triangleq \mathbb{P}_k \sum^t_{m=1}\boldsymbol h(y_m)/t$ in $k$ dimensions. The approximate belief is parameterized by~$\hat{\boldsymbol x}_t$ as follows:
\begin{align}
\hat\pi_{jt}  = \hat{\mathcal{T}}^t_{j}(\hat{\boldsymbol x}_t, \boldsymbol \theta) =\frac{\theta_j \exp\big\{   t (\hat \beta_{j0}+  \boldsymbol{e}_j \mathbb{R}_k \hat{\boldsymbol x}_t )  \big\} }
 {\sum^N_{i=1} \theta_i \exp\big\{  t (\hat\beta_{i0}+   \boldsymbol{e}_i \mathbb{R}_k  \hat{\boldsymbol x}_t )
  \big\} +\theta_0} , \ \  j =1,\ldots,  N, \nonumber
\end{align}
where $\hat\beta_{i0}=-\log \int  \exp \{  \boldsymbol{e}_i \mathbb{R}_k \mathbb{P}_k    \boldsymbol h(y) \} f_0(y)dy$. 

Note that the stopping functions $J^j_t(\bar{\boldsymbol x}_t; \boldsymbol \theta )$ are in closed form, with $\mathbb{R}=\mathbb{B}$, and hence require no approximation. Only the waiting function~$J^w_t(\bar{\boldsymbol x}_t; \boldsymbol \theta )$ needs to be approximated by
\begin{align}
\label{eq: 824-bvsffw}
\hat J^w_t(\bar{\boldsymbol x}_t; \boldsymbol \theta ) \triangleq & \sum^N_{i=0} \mathcal{T}^t_i(\bar{\boldsymbol x}_t, \boldsymbol \theta) c_{i}+
\int Q_{t+1} \Big(  \frac{ t \mathbb{P}_k \bar{\boldsymbol x}_t+  \boldsymbol x}{t+1} ; \boldsymbol \theta  \Big) \sum^N_{i=0}  {\mathcal{T}}^t_i( \bar{\boldsymbol x}_t, \boldsymbol \theta) \boldsymbol g^k_i( \boldsymbol x  ) d\boldsymbol x, 
\end{align}
where $\boldsymbol g^k_i$ denotes the $k$-dimensional density of $\mathbb{P}_k \boldsymbol h(y)$ when $y$ has a density $f_i$, and $Q_t, t=1,\ldots, T$ are $k$-dimensional functions that can be computed using the following recursive equations:\endnote{The integrals in~$\hat J^w_t, Q^w_t$ and~$\hat\beta_{i0}$ can be computed by changing the variable of integration from~$y$ to its compressed version~$\hat y=\mathbb{P}_k \boldsymbol h(y)$ and estimating the corresponding densities~$\hat f_i (\hat y), i=0,\ldots, N$ by simulation. }  
\begin{align}
Q_t(\hat{\boldsymbol x}_t; \boldsymbol \theta ) & =\min\big\{Q^0_t(\hat{\boldsymbol x}_t; \boldsymbol \theta ), Q^1_t(\hat{\boldsymbol x}_t; \boldsymbol \theta ), \ldots, Q^N_t(\hat{\boldsymbol x}_t; \boldsymbol \theta ), Q^w_t(\hat{\boldsymbol x}_t; \boldsymbol \theta )    \big\},  \hat{\boldsymbol x}_t \in \hat\Omega_t, \  t=1, 2,  \ldots , T-1, \nonumber\\
Q_T(\hat{\boldsymbol x}_T; \boldsymbol \theta ) & =\min\big\{Q^0_T(\hat{\boldsymbol x}_T; \boldsymbol \theta ), Q^1_T(\hat{\boldsymbol x}_T; \boldsymbol \theta ), \ldots, Q^N_T(\hat{\boldsymbol x}_T; \boldsymbol \theta )  \big\}, \nonumber\\
Q^j_t(\hat{\boldsymbol x}_t; \boldsymbol \theta ) &=\sum^N_{i=0}  \hat{\mathcal{T}}^t_{i}(\hat{\boldsymbol x}_t, \boldsymbol \theta) a_{ij}, \ \ \ j \in \mathcal N, \nonumber\\
Q^w_t(\hat{\boldsymbol x}_t; \boldsymbol \theta ) &=\sum^N_{i=0} \hat{\mathcal{T}}^t_i(\hat{\boldsymbol x}_t, \boldsymbol \theta) c_{i}+\int Q_{t+1} \Big( \frac{  t \hat{\boldsymbol  x}_t+  \boldsymbol x }{t+1} ; \boldsymbol \theta  \Big) \sum^N_{i=0}  \hat{\mathcal{T}}^t_i( \hat{\boldsymbol x}_t, \boldsymbol \theta)  \boldsymbol g^k_i( \boldsymbol x  ) d\boldsymbol x . \nonumber
\end{align}
Note that the integrals in the above are over $k$-dimensional space rather than the $d$-dimensional observation space. The approximate belief-reconstruction policy $\pi_{abr}$ works by replacing the waiting function $J^w_t$ with its approximation $\hat J^w_t$. Specifically, upon receiving the multivariate observations $\bar{\boldsymbol x}_t$,  the DM chooses the decision that attains the minimum value after comparing the corresponding exact stopping functions~$J^0_t,\ldots, J^N_t$ given in~\eqref{Eq: theorem: 1D-Finite-OE-Main-stopping}, as well as the approximate waiting function $\hat J^w_t$ given in~\eqref{eq: 824-bvsffw}. 

\subsection*{Quantifying the Performance Loss}
Next, we consider quantifying the performance loss incurred by the approximate policy $\pi_{abr}$ relative to the optimal policy $\pi^*$. Let $J^{\pi_{abr}}(\boldsymbol\theta)$ and $J^*(\boldsymbol\theta)$ denote the expected total cost of following~$\pi_{abr}$ and~$\pi^*$, respectively, given the prior~$\boldsymbol\theta$. Clearly, $J^{\pi_{abr}}(\boldsymbol\theta) \geqslant J^*(\boldsymbol\theta)$. Here, $J^{\pi_{abr}}(\boldsymbol\theta)$ can be estimated using Monte Carlo simulation as follows: we take $T$ iid samples from the density~$f_i$, which is chosen with probability~$\theta_i$, to generate a sample path and then apply the policy $\pi_{abr}$ and calculate the corresponding total cost. Averaging the costs over all the sample paths yields an estimate of $J^{\pi_{abr}}$. Since the computation of $J^*(\boldsymbol\theta)$ is often expensive for multivariate observations, we will construct a lower bound~$J^{lb}(\boldsymbol\theta)$ and use it to bound the performance loss as $J^{\pi_{abr}}(\boldsymbol\theta)-J^*(\boldsymbol\theta) \leqslant J^{\pi_{abr}}(\boldsymbol\theta)-J^{lb}(\boldsymbol\theta)$, or, equivalently, to bound the relative performance loss as $(J^{\pi_{abr}}-J^* )/J^* \leqslant (J^{\pi_{abr}}-J^{lb})/J^{lb}$. 

The lower bound is given in the following proposition, in which we use~$\boldsymbol g$ to denote the density of $\boldsymbol h(y)$ when $y$ has a density $f_0$, and use $\boldsymbol g^{*n}$ to denote the $n$th convolution power of $\boldsymbol g$. 

\begin{proposition} [Upper bound on performance loss]
\label{prop: Bound on performance loss}
$
0 \leqslant J^{\pi_{abr}}(\boldsymbol\theta)-J^*(\boldsymbol\theta) \leqslant J^{\pi_{abr}}(\boldsymbol\theta)-J^{lb}(\boldsymbol\theta), 
$
where $J^{lb}(\boldsymbol\theta) =\min\{\sum^N_{i=0} \theta_i a_{i0}, \ldots, \sum^N_{i=0} \theta_i a_{iN}, \sum^N_{i=0} \theta_i c_{i} +\mathcal{L}^T_0(\boldsymbol \theta )\}$ and 
$$
\mathcal{L}^T_0(\boldsymbol \theta ) = \int  \min_{j\in \mathcal N} \Big\{   \theta_0 a_{0j} +  \sum^N_{i=1} \theta_i \exp\big\{T \beta_{i0}+ \boldsymbol{e}_i \mathbb{B}   \boldsymbol h     \big \} a_{ij}  \Big\}   \boldsymbol g^{*T}(\boldsymbol h)  d \boldsymbol h.  
$$
\end{proposition}

Here, $\boldsymbol g^{*T}$ represents the density of $\sum^T_{t=1} \boldsymbol h(y_t)$, the sum of $T$ transformed iid observations, in which each $y_t$ is drawn from~$f_0$. The term $\mathcal{L}^T_0(\boldsymbol \theta)$ in the lower bound can be interpreted as the expected termination cost when the diagnosis is delayed to the last period~$T$. It can be estimated using Monte Carlo simulation: in each simulation run, we draw $T$ samples $y_1,\ldots, y_T$ from $f_0$ to generate a sample of $\boldsymbol h = \sum^T_{t=1} \boldsymbol h(y_t)$; then, we minimize the term in the curly bracket. Averaging the minimum values across all the simulation runs produces an estimate of $\mathcal{L}^T_0(\boldsymbol \theta)$. 

The lower bound $J^{lb}(\boldsymbol\theta)$ is the expected cost of a clairvoyant policy in which, after the first decision period, the DM is given access to all future observations for free. Since it can only help to know more information, $J^{lb}(\boldsymbol\theta)$ is a lower bound of $J^*(\boldsymbol\theta)$. A rigorous proof is given in Appendix~\ref{EC proof: bound on performance loss}. Note that  $J^{lb}(\boldsymbol\theta)$ depends on all the model inputs, including the cost parameters, the observation densities, the prior, and the horizon length. Numerical evidence in the upcoming example indicates that the bound in Proposition~\ref{prop: Bound on performance loss} is tight in problems with a short horizon and noisy observations. 



Next, we compare $\pi_{abr}$ and $\pi^*$ in terms of the mean delay time and the error rate. The total cost can be decomposed into the delay cost and the termination cost. For example, we can write $J^{\pi_{abr}}(\boldsymbol\theta)=J^{\pi_{abr}}_d(\boldsymbol\theta)+J^{\pi_{abr}}_s(\boldsymbol\theta)$, where $J^{\pi_{abr}}_d$ is the expected delay cost of $\pi_{abr}$, and $J^{\pi_{abr}}_s$ is the expected termination cost.  Similarly, the optimal cost can be decomposed as $J^*(\boldsymbol\theta)=J^*_d(\boldsymbol\theta)+J^*_s(\boldsymbol\theta)$, with $J^*_d$ and $J^*_s$ representing the optimal delay and termination cost, respectively. Under the zero-one cost structure, the expected delay cost represents the (scaled) mean delay time, and the expected termination cost represents the misdiagnosis probability or the error rate. The following proposition suggests that, compared with the optimal policy, $\pi_{abr}$ sacrifices accuracy for speed.

\begin{proposition}
\label{prop: Approx low-rank approximation characterize a subset of the waiting region}
It is optimal to wait if $\hat J^w_t =\min\{J^0_t,\ldots, J^N_t, \hat J^w_t \}$. Further, we have $J^{\pi_{abr}}_d (\boldsymbol\theta) \leqslant  J^*_d(\boldsymbol\theta)$ and $J^{\pi_{abr}}_s (\boldsymbol\theta) \geqslant  J^*_s(\boldsymbol\theta)$ for all $\boldsymbol\theta\in S^N$.
\end{proposition}
This proposition suggests that, if~$\pi_{abr}$ prescribes waiting, then waiting is indeed optimal. In other words, the stopping regions of $\pi^*$ are subsets of the stopping regions of~$\pi_{abr}$. Therefore, for any sample path, $\pi_{abr}$ stops no later than the optimal policy, so its delay cost is lower than optimal, i.e., $J^{\pi_{abr}}_d (\boldsymbol\theta) \leqslant  J^*_d(\boldsymbol\theta)$. Then, the expected termination cost of $\pi_{abr}$ must be higher than optimal: $J^{\pi_{abr}}_s (\boldsymbol\theta) \geqslant  J^*_s(\boldsymbol\theta)$ because $\pi_{abr}$ cannot outperform the optimal policy in terms of the total cost.

\subsection*{Numerical Example}
Consider Example~\ref{ex: MV Example 2} again, in which the matrix $\mathbb{B}$ has rank $r=3$ with a rank decomposition~$\mathbb{B}=U(\Sigma V^\mathsf{T})=\mathbb{RP}$, where 
\begin{align}
\mathbb{R}=
\begin{bmatrix}
-&0.3955   &-0.3482    &0.8499 \\
-&0.5589   &-0.6430    &-0.5236\\
&0.7288   &-0.6821    &0.0597 \\
\end{bmatrix}, 
\mathbb{P}=
\begin{bmatrix}
&-0.8208   &-0.4405   &-0.0229   &-0.0074   &-0.0074    &0.0345 \\
&0.1465   &-0.2717   &-0.0442    &0.0214    &0.0214   &-0.0043\\
&-0.0040    &0.0017   & 0.0064    &0.0251    &0.0251   &-0.0576 \\
\end{bmatrix}.
 \nonumber
\end{align}
As discussed earlier, a reasonable approximate rank is $k=1$. The approximate reconstruction matrix $\mathbb{R}_1$ is the first column of~$\mathbb{R}$, the column of $U$ that corresponds to the largest singular value, $0.9325$. The approximate projection matrix $\mathbb{P}_1$ is the first row of~$\mathbb{P}$, the first row of $\Sigma V^\mathsf{T}$ corresponding to the same singular value. Specifically, 
\begin{align}
\mathbb{R}_1=
\begin{bmatrix}
-&0.3955   \\
-&0.5589   \\
&0.7288   \\
\end{bmatrix}, 
\mathbb{P}_1=
\begin{bmatrix}
&-0.8208   &-0.4405   &-0.0229   &-0.0074   &-0.0074    &0.0345 \\
\end{bmatrix}. 
 \nonumber
\end{align}
Bivariate normal distributions have sufficient statistic~$\boldsymbol h (y)=(y_1, y_2, y^2_1, y_1y_2, y_2 y_1, y^2_2)^\mathsf{T} $. Hence, the approximate diagnostic statistic is
$
\hat{\boldsymbol x}_t \triangleq \mathbb{P}_1 \sum^t_{m=1}\boldsymbol h(y_m)/t= \sum^t_{m=1} ( -0.8208  y_{1m}-0.4405  y_{2m} -0.0229  y^2_{1m} -0.0148  y_{1m} y_{2m} +0.0345 y^2_{2m} )/t.  
$

\begin{figure}
\centering \includegraphics[width=5.5in]{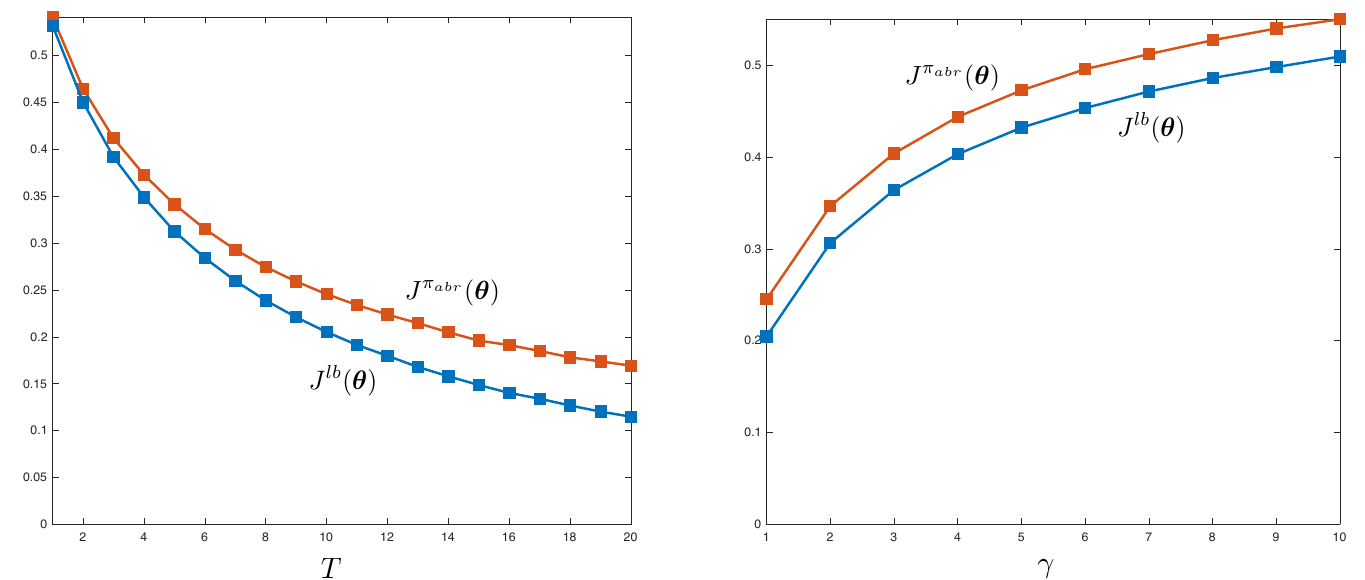} 
\caption{Total cost of the approximate policy $J^{\pi_{abr}}$ compared against a lower bound of the optimal cost $J^{lb}$}
\label{Fig: ABR_Bounds} 
\end{figure}

We now use Proposition~\ref{prop: Bound on performance loss} to evaluate the quality of approximation against a lower bound of the optimal cost. The parameters are chosen to be $\theta_i=1/4$, $c_i=0.001$, for all $i$, and the termination costs are zero-one. The estimations of $J^{\pi_{abr}}$ and $J^{lb}$ shown in Figure~\ref{Fig: ABR_Bounds} are based on $10^6$ simulation runs.\endnote{This number of runs is sufficient to guarantee that the $95\%$ confidence interval for the estimated cost has width less than 0.001.} We observe from the left-hand panel that the gap between $J^{\pi_{abr}}$ and $J^{lb}$ is small when the horizon is short. In the right-hand panel, we rescale the covariance as $\gamma \Sigma_i$ for all $i$ so that a larger~$\gamma$ represents noisier observations. The estimations are displayed for different values of $\gamma$, and the gap between $J^{\pi_{abr}}$ and $J^{lb}$ generally remains stable, but the relative difference $(J^{\pi_{abr}}-J^{lb})/J^{lb}$ decreases as the observations become noisier. 



\subsection*{A Geometric Interpretation}
The algebraic motivation behind the low-rank approximation is the minimization of the Frobenius approximation error. The following proposition suggests that it also has a geometric interpretation when the densities are multivariate normal with equal covariances. 

\begin{proposition}
\label{prop: LRA interpretation-MVN-Equal covariance}
When $f_i=N(\boldsymbol \mu_i, \boldsymbol\Sigma), i=0,\ldots, N$, we have $\mathbb{P} \boldsymbol h(y) = (v_1,\ldots, v_r)^\mathsf{T} y$, in which $v_1$ maximizes the following Rayleigh quotient:
\begin{align}
\label{eq: Rayleigh quotient}
\max_v \frac{v^\mathsf{T} \mathbb{S}^* v }{v^\mathsf{T} v},
\end{align}
where $\mathbb{S}^*=  \sum^N_{k=1} \boldsymbol s^\mathsf{T}_k \boldsymbol s_k$, and $\boldsymbol s_k=(\boldsymbol\mu_k- \boldsymbol\mu_0  )^\mathsf{T}  \boldsymbol\Sigma^{-1}$, for $k=1,\ldots, N$.  Similarly, $v_2$ is orthogonal to $v_1$ and maximizes $v^\mathsf{T}_2 \mathbb{S}^* v_2 /{v^\mathsf{T}_2 v_2}$, and so on. 
\end{proposition}

The interpretation is that the rank-one approximation finds the informative one-dimensional projection of the multivariate Gaussian that preserves the class separability as much as possible. Here, class separability is measured in terms of the variance of the projected classes, or~$v^\mathsf{T} \mathbb{S}^* v$. Intuitively, $\boldsymbol s_k$ can be viewed as a standardized~$\boldsymbol\mu_k$, and the projections of all $\boldsymbol s^\mathsf{T}_k$'s on the direction~$v$ have variance~$\sum^N_{k=1} (\boldsymbol s_k v)^\mathsf{T} \boldsymbol s_k v=v^\mathsf{T} \mathbb{S}^* v $. To make the variance larger, the projected class centroids need to be spread out further, or the  within-class variability in the projection needs to be smaller, since the standardization involves the inverse covariance matrix. In this sense, the maximization in~\eqref{eq: Rayleigh quotient} is equivalent to minimizing the overlap among the projected ellipsoids.

\subsection{A Maintenance Application}
\label{sec: CBM real example}
We apply the approximate belief-reconstruction method to online fault diagnosis for a smart triplex pump, which is equipped with multiple sensors that generate multivariate data streams. Triplex pumps are key pieces of equipment in offshore oil drilling platforms, and their unexpected downtime can be costly. Thus, quick identification of faults from the sensor data is essential. It is not always feasible to collect fault data from pumps in the field because such data are often scarce and operating a pump with faults may lead to catastrophic failure. A common solution is to develop a digital twin of the pump and use it to simulate the sensor data under various fault conditions.  We performed experiments on a digital twin developed by Simscape\texttrademark, which can simulate the detailed mechanical, hydraulic, and electrical behaviors of the pump under various conditions.   

\begin{figure}
\centering \includegraphics[width=6.0in]{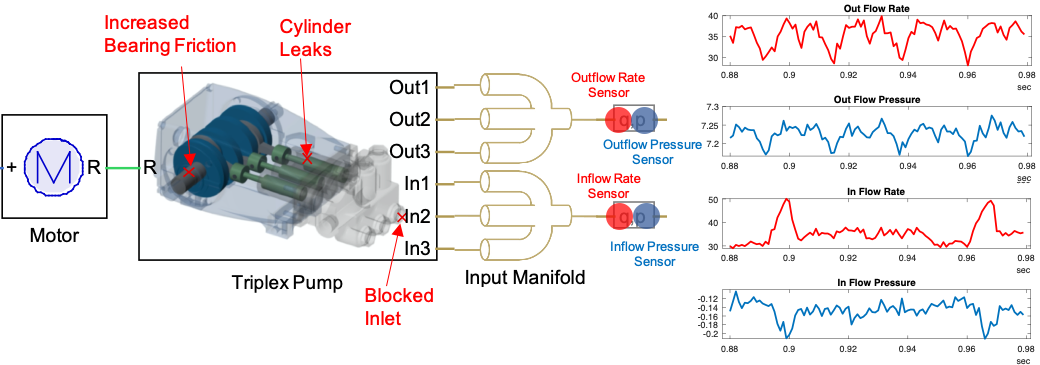} 
\caption{Digital twin of a hydraulic triplex pump with sensors (by Simscape\texttrademark). Signals shown correspond to a $0.1$ second record under the healthy state. }
\label{Fig: pdmRecipPump} 
\end{figure}

The digital twin can simulate three types of faults: blocked inlet, cylinder leaks, and increased bearing friction (see Figure~\ref{Fig: pdmRecipPump}). Since multiple faults may coexist, we consider a healthy state and $N=7$ fault states with different combinations of faults. Each of the fault states 1-3 includes only one type of fault, the fault states 4-6 include two types of faults, and the fault state 7 includes all three types of faults. The parameter settings (Table~\ref{Fault Parameter Table}) represent early-stage faults that are difficult to diagnose. The digital twin was operated for 90 sec under each health state. We recorded four signals every millisecond, including the inflow rate and pressure and the outflow rate and pressure. For each signal, we calculated the sample mean and the power in the low-frequency range (10-20 Hz) for every $0.2$ sec window. Thus, there are $d=8$ condition variables in total (Table~\ref{Condition variables Table}) that are updated every $0.2$ sec.

\begin{table}
\footnotesize
\centering
\smallskip
\caption{Fault parameters for the triplex pump, with fault conditions shown in bold} 
\label{Fault Parameter Table}
\begin{threeparttable}
\begin{tabular}{c c c c c c c c c  }
\toprule
Parameter & Healthy & Fault 1 & Fault 2 & Fault 3  & Fault 4 & Fault 5 & Fault 6  & Fault 7   \\
\cmidrule(r){1-9}  
Leak area & 0 & $\boldsymbol {1\times10^{-8}}$ & 0 & 0 &  $\boldsymbol {1\times10^{-8}}$ &  $\boldsymbol {1\times10^{-8}}$ & 0 &  $\boldsymbol {1\times10^{-8}}$  \\
\cmidrule(r){1-9}  
Blocking factor&  $80\%$   & $80\%$ & $\boldsymbol{79.5\%}$ &   $80\%$ &  $\boldsymbol{79.5\%}$ &  $80\%$ &  $\boldsymbol{79.5\%}$ &  $\boldsymbol{79.5\%}$ \\
\cmidrule(r){1-9} 
Bearing factor& 0 & 0  & 0& $\boldsymbol{5\times10^{-6}}$ &  0 &  $\boldsymbol{5\times10^{-6}}$ &  $\boldsymbol{5\times10^{-6}}$ &   $\boldsymbol{5\times10^{-6}}$    \\
\bottomrule 
\end{tabular}
\caption{\footnotesize Condition variables for the triplex pump}
\label{Condition variables Table}
\begin{tabular}{c c c c c c c c }
\toprule
 \multicolumn{4}{c}{Inflow Rate: Mean ($Y_1$), Low-freq. power ($Y_2$)} &    \multicolumn{4}{c}{Inflow Pressure: Mean ($Y_3$), Low-freq. power ($Y_4$)} \\
 \cmidrule(r){1-8} 
  \multicolumn{4}{c}{Outflow Rate: Mean ($Y_5$), Low-freq. power ($Y_6$)} &    \multicolumn{4}{c}{Outflow Pressure: Mean ($Y_7$), Low-freq. power ($Y_8$)} \\
\bottomrule 
\end{tabular}
\end{threeparttable}
 \end{table}

\begin{figure}
\centering \includegraphics[width=6.5in]{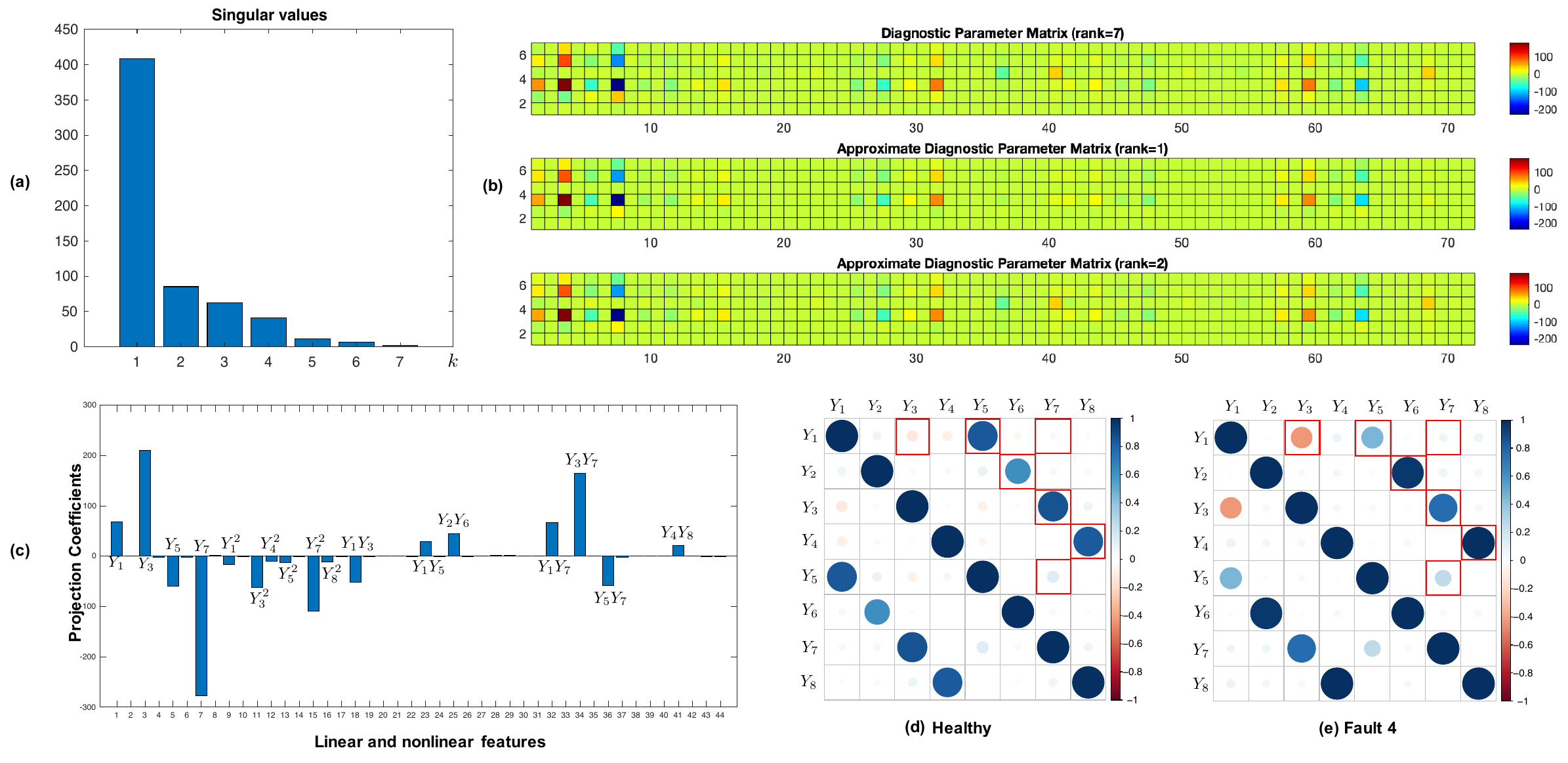} 
\caption{Triplex pump fault diagnosis: Illustrations of (a) singular values of the $\mathbb{B}$ matrix; (b) the $\mathbb{B}$ matrix (top) and its low-rank approximations ${\mathbb{B}}_1$ (middle), and ${\mathbb{B}}_2$ (bottom); (c) projection coefficients from  $\mathbb{P}_1$ for different data features (useful nonlinear features such as interactions are automatically extracted); correlation matrices for (d) the healthy state and (e) fault 4. }
\label{Fig: Pump_ApproximateBMatrices} 
\end{figure}

The initial 0.8-sec record was excluded from the analysis as the pump is in a transient phase.  The rest of the data are then normalized by Box-Cox transformation, with normality verified by Mardia's test ($p$-value: 0.19). We also standardized the data to ensure that all variables are on a similar scale before estimating the multivariate normal parameters. The diagnostic parameter matrix $\mathbb{B}$ has rank 7, suggesting that the optimal policy is computationally expensive. We used the singular value decomposition approach to find a low-rank approximation of $\mathbb{B}$. Figure~\ref{Fig: Pump_ApproximateBMatrices}--(a) shows that the first singular value of $\mathbb{B}$ is much larger than the others, suggesting that the behaviour of~$\mathbb{B}$ is similar to that of a rank-one matrix; hence, the reachable belief space is sparse. Indeed, Figure~\ref{Fig: Pump_ApproximateBMatrices}--(b) suggests that the rank-one approximation matrix $\mathbb{B}_1$ is already close to the original matrix, and the improvement from a rank-two approximation is small, not to mention the adding complexity of implementing the latter. Thus, it is reasonable to choose the rank-one approximation ($k=1$). 

The projection matrix (vector) $\mathbb{P}_1$ compresses a 72-dimensional feature vector\endnote{For a $d$-dimensional multivariate normal, $\boldsymbol h^\mathsf{T}(y)=(y^\mathsf{T}, \text{vec}(yy^\mathsf{T})^\mathsf{T}  ) $ contains $d(d+1)$ elements, but some are repeated because $yy^\mathsf{T}$ is symmetric. The effective number of variables is $d(d+3)/2$. } $\sum^t_{m=1}\boldsymbol h(y_m)/t$ into a scalar:
\begin{align}
\hat{x}_t = &\sum^t_{m=1} \big( 64.85  y_{1m}+210.42  y_{3m} -62.57  y_{5m} -277.57 y_{7m}  + \ldots -16.83  y^2_{1m} -62.72  y^2_{3m} + \ldots \nonumber\\
& \hspace{1cm} -51.11 y_{1m}y_{3m} + 29.19  y_{1m}y_{5m} + \ldots + 20.95 y_{4m}y_{8m} + \ldots \big)/t, \nonumber
\end{align}
which can be thought of as a health index. Note that $\hat{x}_t$ contains not only the linear features but also the nonlinear features of the condition variables (e.g., $Y^2_4$, $Y_1Y_3$). Figure~\ref{Fig: Pump_ApproximateBMatrices}--(c) plots the coefficients of these features obtained from the projection matrix $\mathbb{P}_1$. It is interesting to observe that over half the coefficients are near zero. Also, it is possible to give at least partial interpretations of the health index. For example, the coefficients for $Y_1$ and $Y_5$ are of similar magnitude but opposite signs, suggesting that the difference between the inflow and the outflow rates could be important diagnostic information, possibly a measure of leakage. Certain interaction terms such as $Y_1Y_3$ can distinguish between faults with distinct correlation structures. To illustrate this, we highlight these interaction terms using red boxes in the correlation matrix plots in (d) and (e). Different health states can exhibit distinct correlation patterns, especially over the interaction terms enclosed in the red boxes.  

The online decisions follow the procedures for $\pi_{abr}$ described in~\S\ref{sec: Approximate Belief Reconstruction}. The stopping time is based on the diagnostic statistic $\hat{x}_t$, but the diagnosis is based on the averaged original observations $\sum^t_{m=1}\boldsymbol h(y_m)/t$. As long as $\hat{x}_t$ remains inside the waiting region, by Proposition~\ref{prop: Approx low-rank approximation characterize a subset of the waiting region}, the waiting is optimal. When the termination costs are zero-one; $T=5$; and $\theta_i=1/8, c_i=0.001$, for all~$i$, the performance loss incurred by the approximation relative to the optimal policy is bounded by $(J^{\pi_{abr}}-J^* )/J^* \leqslant 16.47\%$. The approximate policy has a shorter delay time but a higher error rate than the optimal policy. 

\section{Summary and Discussion}
\label{sec: discussion}
Optimal sequential multi-class diagnosis is an important open problem with broad applications. The standard POMDP formulation uses the vector of posterior class probabilities as the state of a dynamic program and suffers from the curse of dimensionality in the state space. 

In practice, the class-specific observation densities are often estimated from the historical data. For unlabeled historical data, the densities can be estimated as components of a finite mixture model, which usually involves certain assumptions about the interrelation among the components, e.g., exponential tilting. It is thus natural to inherit the assumptions in the decision model. We found that the observation densities carry important opportunities for dimension reduction: when the densities are related through exponential tilting, the reachable belief space is often sparse. The high-dimensional vector of posterior probabilities can be reconstructed from a low-dimensional diagnostic statistic, making it possible to reformulate the dynamic program in lower dimension. Under this new framework, the optimal policies can be found for sequential diagnosis problems with large numbers of classes. 

The sparsity of the belief space is a consequence of the existence of the low-dimensional sufficient statistic of the ETM as well as, less intuitively, the interrelations among the tilting parameters. A key step in the proposed framework is rank decomposition, which reveals, exactly or approximately, the low dimensionality hidden in a higher dimension. Unlike the belief compression method for solving general POMDPs~\citep{roy2005finding}, the proposed framework can generate sparse representations that are time-dependent and requires no simulation. The dimension reduction is also different from the classical linear dimension reduction methods such as reduced-rank linear discriminant analysis, because the diagnostic statistic can contain nonlinear features of the data. 

Compared with the state-of-the-art heuristic policy, the optimal policy can significantly shorten the time to diagnosis, especially when the signal-to-noise ratio is low. Accordingly, it is highly valuable for time-sensitive applications such as medical diagnosis and the maintenance of safety-critical equipment. The low-dimensional belief tracks and surfaces offer unique opportunities to probe into the center of the belief space, a regime where the asymptotic analyses tend to perform poorly. 

The approximate belief reconstruction method can be generalized in multiple directions. For example, when the class membership changes in the course of diagnosis, our preliminary experiments show that the reachable belief space can still be sparse. Thus, it would be interesting to study how to modify the low-rank approximation so that it takes into account the class evolution. When the DM can choose among multiple sampling modes that differ in cost and information quality, the diagnostic matrix can be expanded by incorporating more columns associated with the sampling modes. In addition, when the observations do not follow the ETM, we may approximate them with an ETM. The projection scheme as well as the bound on performance loss can be improved. It is also important to understand how the performance loss depends on the Frobenius error or other appropriate error measures. 

\theendnotes

\ACKNOWLEDGMENT{The author thanks Xueze Song (University of Illinois at Urbana-Champaign) for the assistance of numerical studies, and thanks Opher Baron (University of Toronto), Yuri Levin, Mikhail Nediak, and Anton Ovchinnikov (Queen's University) for helpful suggestions. Constructive comments and suggestions from the area editor, associate editor, and three anonymous referees have resulted in significant improvements of the paper.  }

\section*{Biography}
\begin{description}
    \item Jue Wang is an assistant professor of management analytics in the Smith School of Business at Queen's University. He received his PhD from the University of Toronto.  His research interests are sequential decision making in maintenance and revenue management. 
\end{description}

\bibliographystyle{informs2014} 
\bibliography{Reference}

\newpage

\ECSwitch


\ECHead{Appendix}

 \section{Proof of Theorem~\ref{theorem 1}}
  \proof{Proof.}
For~$j=1,\ldots, N$, it follows from the definition of $\mathcal{T}^t_j$ and $\boldsymbol x_t$ that 
 \begin{align} 
\mathcal{T}^t_j(\boldsymbol x_t, \boldsymbol \theta )   
  &=\frac{\theta_j \exp\big\{   t(\beta_{j0}+  \boldsymbol{e}_j \mathbb{R}   \boldsymbol x_t) \big\} }
 {\sum^N_{i=1} \theta_i \exp\big\{  t(\beta_{i0}+   \boldsymbol{e}_i \mathbb{R}  \boldsymbol x_t)
  \big\} +\theta_0}  
  =\frac{\theta_j \exp\big\{   t\beta_{j0}+  \boldsymbol{e}_j \mathbb{RP}   \sum^t_{m=1}\boldsymbol h(y_m) \big\} }
 {\sum^N_{i=1} \theta_i \exp\big\{  t\beta_{i0}+   \boldsymbol{e}_i \mathbb{RP}   \sum^t_{m=1}\boldsymbol h(y_m)
  \big\} +\theta_0}.  \nonumber
   \end{align}
   Since $\boldsymbol{e}_j \mathbb{RP}  =\boldsymbol{e}_j \mathbb{B} =  \boldsymbol\beta^\mathsf{T}_j$, we have 
 \begin{align}   
 \mathcal{T}^t_j(\boldsymbol x_t, \boldsymbol \theta )  
&=\frac{ \frac{\theta_j }   {\theta_0 } \exp\big\{ t\beta_{j0}+  \boldsymbol\beta^\mathsf{T}_j  \sum^t_{m=1} \boldsymbol h(y_m)  \big\}  }
 {\sum^N_{i=1} \frac{\theta_i }   {\theta_0 } \exp\big\{t \beta_{i0}+ \boldsymbol\beta^\mathsf{T}_{i} \sum^t_{m=1} \boldsymbol h(y_m) \big\}   +1}.  \nonumber
    \end{align}
 Next, we make use of the properties of the exponential tilting model to rewrite the above equation as 
 \begin{align}  
   \mathcal{T}^t_j(\boldsymbol x_t, \boldsymbol \theta )  
   &=\frac{ \frac{\theta_j  \prod^t_{m=1} f_0  (y_m) \prod^t_{m=1} \exp\{\beta_{j0}+\boldsymbol\beta^\mathsf{T}_j \boldsymbol h(y_m) \} }{\theta_0  \prod^t_{m=1} f_0 (y_m)}} {\sum^N_{i=1} \frac{\theta_i  \prod^t_{m=1} f_0  (y_m) \prod^t_{m=1} \exp\{\beta_{i0}+ \boldsymbol\beta^\mathsf{T}_i \boldsymbol h(y_m) \} }{\theta_0  \prod^t_{m=1} f_0 (y_m)} +1} .  \nonumber
       \end{align}
  It then follows from the iid. assumption and Bayes' rule that
   \begin{align}   
     \mathcal{T}^t_j(\boldsymbol x_t, \boldsymbol \theta )  &=\frac{ \frac{\theta_j  \prod^t_{m=1} f_j (y_m) }{\theta_0  \prod^t_{m=1} f_0 (y_m)}} {\sum^N_{i=1} \frac{\theta_i  \prod^t_{m=1} f_i (y_m) }{\theta_0  \prod^t_{m=1} f_0 (y_m)} +1} =\frac{ \pi_{jt} / \pi_{0t}} {\sum^N_{i=1}  \pi_{it} / \pi_{0t} +1} \nonumber
  =\pi_{jt}.
 \end{align}
The last equality follows from $ \sum^N_{i=0}\pi_{it}=1$. Therefore, 
$$
\Big(1-\sum^N_{i=1} \mathcal{T}^t_{i} (\boldsymbol x_t , \boldsymbol \theta), \mathcal{T}^t_1 (\boldsymbol x_t , \boldsymbol \theta), \ldots, \mathcal{T}^t_N (\boldsymbol x_t , \boldsymbol \theta)\Big)=(\pi_{0t}, \ldots, \pi_{Nt})=\Pi_t (y_1, \ldots, y_t, \boldsymbol \theta).
$$ 
Since $\boldsymbol x_t$ lies in $r$ dimensions, the corresponding belief state must also lie in an $r$-dimensional manifold. \Halmos
 \endproof

 \section{Belief Reconstruction for Normal Distributions}
\label{Appendix: An Example of Belief Reconstruction}
Consider normal $f_i$ with mean $\mu_i$ and variance $\sigma^2_i$. Let $\zeta_i \triangleq (\sigma^2_0\mu_i-\sigma^2_i\mu_0)/ (\sigma^2_i-\sigma^2_0)$. If $\sigma_i=\sigma_0$, we let $\zeta_i=\infty$. We break the problem into two cases:
\begin{enumerate}
\item Case I: $\zeta_i$'s are non-identical ($r=2$). We can choose the reconstruction matrix as~$\mathbb{R}=\mathbb{B}$ and the projection matrix as the identity matrix, i.e., $\mathbb{P}=\mathbb{I}$. The  diagnostic statistic is~
$
\boldsymbol x_t = (x_1, x_2)^\mathsf{T}=  \big(  \frac{1}{t} \sum^t_{m=1}y_m,   \frac{1}{t} \sum^t_{m=1}y^2_m  \big)^\mathsf{T},
$
which is the natural sufficient statistic of normal distribution. The transformation $\mathcal{T}^t_j (\boldsymbol x_t; \boldsymbol \theta)$ can be specialized as
 \begin{align}
 \mathcal{T}^t_0(x_1,x_2; \boldsymbol \theta) &=\Big( \sum^N_{i=1}\frac{\theta_i}{\theta_0} \exp \Big\{ t \Big [
\frac{\sigma^2_0\mu_i -\sigma^2_i\mu_0}{\sigma^2_0\sigma^2_i} x_1 +  \frac{\sigma^2_i-\sigma^2_0}{2\sigma^2_0\sigma^2_i} x_2 
  -   \frac{\mu^2_i \sigma^2_0 - \mu^2_0 \sigma^2_i}{2 \sigma^2_0 \sigma^2_i} - \ln \Big(\frac{\sigma_i}{\sigma_0}\Big)        \Big]  
  \Big\} +1 \Big)^{-1}, \nonumber\\
\mathcal{T}^t_j(x_1,x_2; \boldsymbol \theta) &= \frac{\theta_j \exp\big\{ t\big [ \frac{\sigma^2_0\mu_j -\sigma^2_j\mu_0}{\sigma^2_0\sigma^2_j} x_1 +  \frac{\sigma^2_j-\sigma^2_0}{2\sigma^2_0\sigma^2_j} x_2 
  -  \frac{\mu^2_j \sigma^2_0 - \mu^2_0 \sigma^2_j}{2 \sigma^2_0 \sigma^2_j} -\ln \big(\frac{\sigma_j}{\sigma_0}\big)        \big]    \big\} }
 {\sum^N_{i=1} \theta_i \exp\big\{ t \big [ \frac{\sigma^2_0\mu_i -\sigma^2_i\mu_0}{\sigma^2_0\sigma^2_i} x_1 +  \frac{\sigma^2_i-\sigma^2_0}{2\sigma^2_0\sigma^2_i} x_2 
  -   \frac{\mu^2_i \sigma^2_0 - \mu^2_0 \sigma^2_i}{2 \sigma^2_0 \sigma^2_i} - \ln \big(\frac{\sigma_i}{\sigma_0}\big)        \big]    \big\} +\theta_0}, j=1,\ldots, N. \nonumber
\end{align}

\item Case II: $\zeta_i =\zeta$ for all $i=1, \ldots, N$ ($r=2$). The details are given in Example~\ref{ex: 2nd level reduction-Normal} in the main text. 

\end{enumerate}

\section{Characterizing the State Space through Minkowski Sums}
\label{Appendix sec: Characterizing the State Space through Minkowski Sums}
\begin{figure}
\centering \includegraphics[width=6.5in]{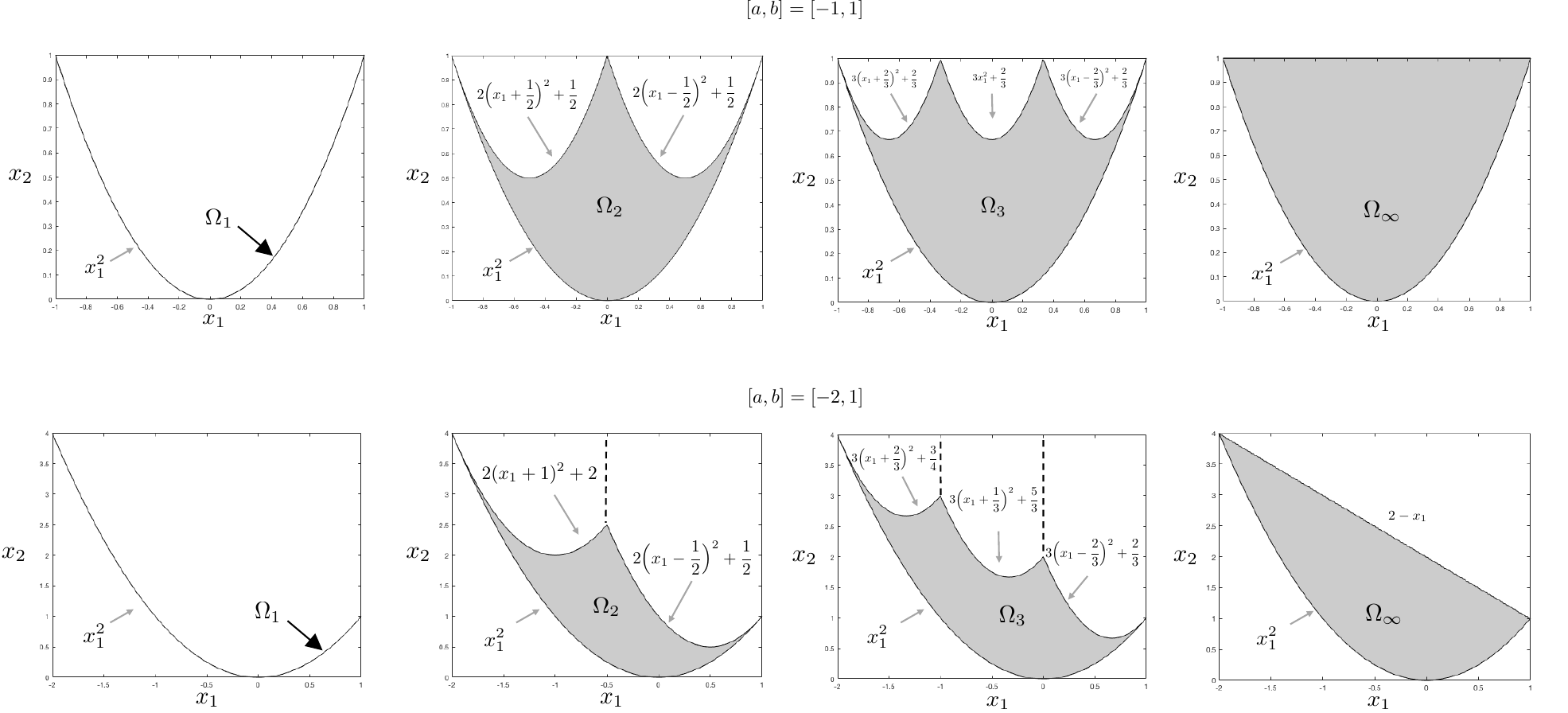} 
\caption{Examples of the diagnostic statistic set $\Omega_{t}$. Truncated normal distribution with symmetric support (upper panel) and asymmetric support (lower panel). }
\label{Fig: OmegaSetExamples} 
\end{figure}

 $\mathbb{P}\boldsymbol h(y)$ can be viewed as a parametric representation of a curve embedded in the $r$-dimensional space, where $y\in\mathcal{Y}$ is the parameter. For example, when $\mathbb{P}$ is the identity matrix and the observations follow a normal distribution with truncated support~$\mathcal{Y}=[a,b]$,  the curve is represented by $\mathbb{P}\boldsymbol h(y)=(y, y^2)^\mathsf{T}, y\in[a,b]$. An illustration of $\Omega_1$ is given in Figure~\ref{Fig: OmegaSetExamples}, where we use $x_1$ and $x_2$ to represent $y$ and~$y^2$, respectively. That is, the curve is represented by a parabola,~$x^2_1$, truncated at the interval $x_1\in[a,b]$. The upper panel shows a symmetric support with $[a,b]=[-1,1]$ and the lower panel shows an asymmetric support with $[a,b]=[-2,1]$.

The Minkowski sum of set $A$ and $B$ is defined as $A\oplus B= \{a+b, a\in A, b\in B\}$, formed by adding each element in $A$ to each element in $B$. In period $t=2$, the set is $\Omega_2 \triangleq \{    [  \boldsymbol h(   y_1)+ \boldsymbol h(y_2)] /2 : y_1, y_2  \in [a,b]  \} $, which is the Minkowski sum of~$\Omega_1$ with itself, scaled by~$1/2$. That is,~$\Omega_2 = (\Omega_1 \oplus \Omega_1)/2$, which denotes the set with all elements of $\Omega_1 \oplus \Omega_1$  divided by two.  Figure~\ref{Fig: OmegaSetExamples} shows that the set $\Omega_2$ is a region enclosed by three quadratic curves; two at the top and one at the bottom. In period $t=3$, the set $\Omega_3= (\Omega_1 \oplus \Omega_1 \oplus \Omega_1)/3$ is a bigger region enclosed by four quadratic curves;  three at the top and the same one at the bottom. In general, $\Omega_t$ has a scalloped shape and, as $t$ increases,  more scalloped edges appear at the top. When $t\to\infty$, the upper envelope approaches  a straight line and the set $\Omega_\infty$   remains bounded. Another observation from Figure~\ref{Fig: OmegaSetExamples}  is that $\Omega_t$ is not necessarily a subset of $\Omega_{t+1}$.  

We have seen that the curve, $\Omega_1$, is convex for normal distribution. Table~\ref{list_distributions-ETM_EF} suggests that it is also convex for the Gamma distribution, but concave for the Beta distribution. The following proposition gives the closed-form representations for the boundaries of $\Omega_t$ in these cases: 

\begin{proposition}
\label{prop: Omega set for normal densities}
Suppose $\mathbb{P} = \mathbb{I}$ and the curve $\Omega_1$ is represented parametrically by $(x_1, g(x_1)), x_1\in[a,b]$. If $g$ is convex, then
  $$\Omega_t =\left\{\left(x_{1}, x_{2}\right) \,\middle|\, x_{1} \in (a,b), g(x_1)  \le x_{2} \le
  \frac{1}{t}\left[ \lambda_1 g(a) +\lambda_2 g(b)+
  g\left( t x_1 - \lambda_1 a - \lambda_2 b  \right) \right] \right\}, $$
 where $\lambda_1 = \left\lfloor\frac{b-x_1}{b-a}t\right\rfloor$, $\lambda_2 = \left\lfloor\frac{x_1-a}{b-a}t\right\rfloor$, and $\lfloor x\rfloor$ represents the greatest integer less than or equal to~$x$. If $g$ is concave, then $\Omega_t $ is characterized by flipping the upper and lower bounds for $x_2$. 
\end{proposition}
\proof{Proof.}
When $\mathbb{P} = \mathbb{I}$, we have $\boldsymbol x_t  =(\sum_{m=1}^t y_m /t, \sum_{m=1}^t g(y_m) /t ) $. Let $x_1=\sum_{m=1}^t y_m/t$,  by the inequality of arithmetic and geometric means we have
    \begin{align}
      \sum_{m=1}^t g(y_m)                         & \ge  t g\left( \frac{ \sum^t_{m=1} y_m  }{t} \right)
                         = t g( x_1) \nonumber, 
    \end{align}
    for convex $g$,  which is the minimum value that is attained when $y_1 =  \cdots = y_t = x_1$. Therefore, the lower envelope of $\Omega_t$ is simply $x_2=g(x_1)$. 
    
Next,  we seek the upper envelope of $\Omega_t$. That is, the maximum of  $\sum_{m=1}^t g(y_m)$ given $\sum_{m=1}^t y_m/t=x_1$. By Jensen's inequality, in order to maximize $\sum_{m=1}^t g(y_m)$, we can have at most one observation that is not equal to $a$ or $b$. Otherwise, suppose $a<y_i<y_j<b$. Then, by decreasing $y_i$ by $\epsilon$ and increasing $y_j$ by $\epsilon$, we can have a larger $\sum_{m=1}^t g(y_m)$ while maintaining $\sum_{m=1}^t y_m$. There is a unique combination of the observations, $y_m, m=1,\ldots, t$, that attains the upper envelope, namely,
  $$x_1 = \frac{1}{t}\left(\underbrace{a + \cdots + a}_{\lambda_1\mbox{~items}}  + \underbrace{b + \cdots + b}_{\lambda_2\mbox{~items}} +  \left( t x_1 - \lambda_1 a - \lambda_2 b  \right) \right), $$
  where $\lambda_1 = \left\lfloor\frac{b-x_1}{b-a}t\right\rfloor$, $\lambda_2 = \left\lfloor\frac{x_1-a}{b-a}t\right\rfloor$. Thus, the maximum value of $x_2=\sum_{m=1}^t g(y_m)/t$ is given by
$\left( \lambda_1 g(a) +\lambda_2 g(b)+      g\left( t x_1 - \lambda_1 a - \lambda_2 b  \right) \right)/t $. Finally, it is easy to observe that, by choosing $\rho \in \left[0,1\right]$ and the observations from the set
\begin{align*}
  \mathcal{Y}_\rho = \Big( &\underbrace{\rho a + (1-\rho)x_1 , \cdots , \rho a + (1-\rho)x_1}_{\lambda_1\mbox{~items}}   , 
  \underbrace{\rho b + (1-\rho)x_1 ,  \cdots , \rho b + (1-\rho)x_1}_{\lambda_2\mbox{~items}} , \\
   &\mspace{2mu}\big(\rho\left( t x_1 - \lambda_1 a - \lambda_2 b  \right) +(1-\rho)x_1 \big) \Big),  
\end{align*}
every value in the interval $\left[ g( x_{1}), \frac{1}{t}\left( \lambda_1 g(a) +\lambda_2 g(b)+g\left( t x_1 - \lambda_1 a - \lambda_2 b  \right) \right) \right]$ can be reached. 

If $g$ is concave, a similar argument leads to the following result 
$$\Omega_t =\left\{\left(x_{1}, x_{2}\right) \,\middle|\, x_{1} \in (a,b),   \frac{1}{t}\left[ \lambda_1 g(a) +\lambda_2 g(b)+
  g\left( t x_1 - \lambda_1 a - \lambda_2 b  \right) \right] \le x_{2} \le  g(x_1) 
 \right\},
 $$
 in which the upper and lower bounds for $x_2$ are flipped compared to the case when $g$ is convex.~\Halmos
\endproof

When the projection matrix, $\mathbb{P}$, is not the identity matrix,  $\Omega_t$ may appear as a rotation of those in Figure~\ref{Fig: OmegaSetExamples}. An example is illustrated in Figure~\ref{Fig: 2Dchart} in the main text. When the observations are discrete, the boundaries of $\Omega_t$ may not have closed-form representations. They can, however, be efficiently computed using the state-of-the-art algorithms for Minkowski sums, such as Fukuda's algorithm~\citep{fukuda2004zonotope}.

 \section{Bayesian Robustness Analysis}
 \label{ECsec: Bayesian Robustness Analysis}
 In this section, we develop a ``scaffolding algorithm" for Bayesian robustness analysis. The computational procedure resembles a building construction process in which one sets up the scaffoldings before building more structures. 
 
 \subsection{Scaffoldings}
\label{subsec: Scaffoldings}   
We define a scaffolding as an $r$-dimensional subspace of the belief space that is \emph{closed} under Bayesian updating; that is, if the prior belongs to the scaffolding, then all subsequent posteriors stay in the scaffolding. 

\begin{figure}
\centering \includegraphics[width=6.5in]{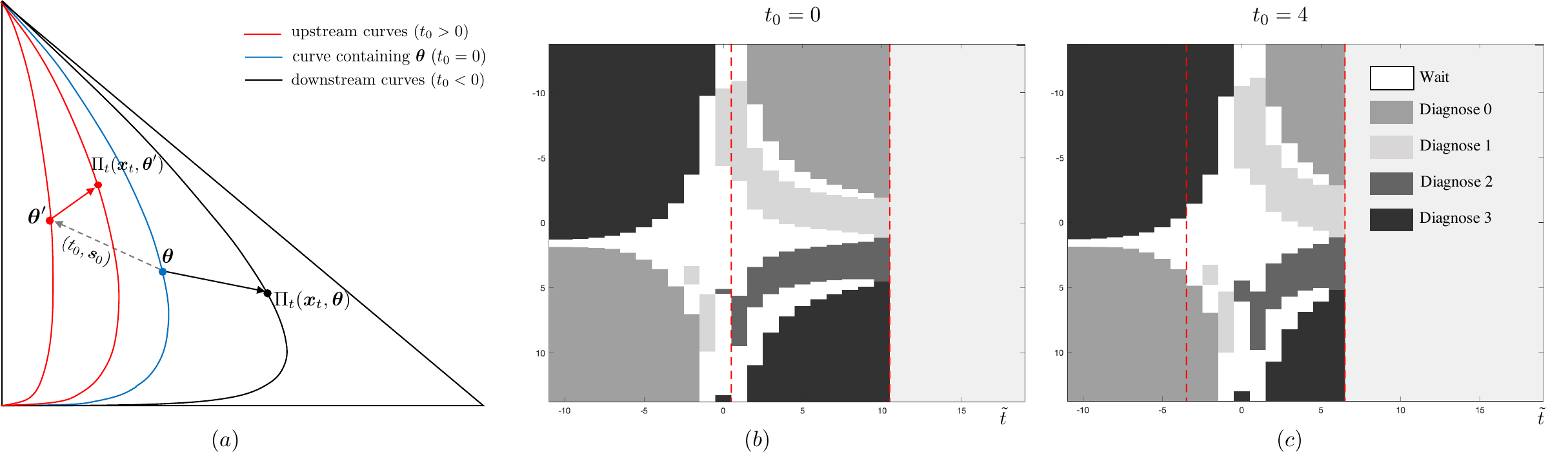} 
\caption{(a) Sample manifolds (belief curves) in the scaffolding $\Sigma(\boldsymbol\theta)$; (b, c) stopping regions in the scaffolding.  }
\label{Fig: UpDownStreamCurves} 
\end{figure}

\begin{definition}[Scaffolding]
For any $\boldsymbol\theta =(\theta'_0, \theta'_1,\ldots, \theta'_N ) \in S^N$, define a unique scaffolding 
\begin{align}
\Sigma(\boldsymbol\theta)=\Big \{ \boldsymbol\theta' \in S^N: \theta'_j   &=\frac{\theta_j \exp\big\{  - t_0 \beta_{j0} - \boldsymbol{e}_j \mathbb{R}   \boldsymbol s_0  \big\} }
 {\sum^N_{i=1} \theta_i \exp\big\{ - t_0 \beta_{i0} -  \boldsymbol{e}_i \mathbb{R}   \boldsymbol s_0   \big\} +\theta_0} ,   j =1,\ldots, N, t_0 \in \mathbb{N},  \boldsymbol s_0 \in {\rm I\!R}^r  \Big\}.  \nonumber
\end{align}
\end{definition}
The scaffolding $\Sigma(\boldsymbol\theta)$ depends on the original prior~$\boldsymbol\theta$ (referred to as \emph{seed prior} from now on) and consists of an infinite number of $r$-dimensional manifolds embedded in the belief space, indexed by $t_0=\ldots, -1, 0, 1, \ldots$, where one of them ($t_0=0$) contains $\boldsymbol\theta$. Any belief state in the scaffolding, say $\boldsymbol\theta'$, can be indexed by an integer $t_0 \in \mathbb{N}$ (the index of the manifold where $\boldsymbol\theta'$ resides) and a vector $\boldsymbol s_0 \in {\rm I\!R}^r $ (which specifies the exact location of $\boldsymbol\theta'$ on that manifold). When $t_0=0$ and $\boldsymbol s_0=\boldsymbol 0$, we have $\boldsymbol\theta'=\boldsymbol\theta$. If $t_0=0$ but $\boldsymbol s_0 \neq \boldsymbol 0$, then $\boldsymbol\theta'$ lies somewhere else on the same manifold that contains~$\boldsymbol\theta$. We call a manifold with $t_0<0$ a \emph{downstream} manifold, because if the prior is located on the manifold $t_0=0$, then all subsequent posteriors must lie on downstream manifolds. A manifold with $t_0>0$ is called an \emph{upstream} manifold. Sample manifolds in a scaffolding with $r=1$ are illustrated in Figure~\ref{Fig: UpDownStreamCurves}--(a). 

The set of belief states reachable from~$\boldsymbol\theta$ in $t$ periods, i.e., $\mathcal{F}^{\boldsymbol \theta}_t, t=1,2,\ldots$, is a proper subset of the scaffolding. More specifically, $\mathcal{F}^{\boldsymbol \theta}_t$ lies in the downstream manifold with index $t_0=-t<0$. Given the prior $\boldsymbol\theta$ and the cumulative sum of the first $t$ observations, $t  \boldsymbol x_t =  \sum^t_{m=1} \mathbb{P} \boldsymbol h(y_m)$, the posterior belief state $\Pi_t(\boldsymbol x_t, \boldsymbol \theta)=(\pi_{0t}, \ldots, \pi_{Nt})$ lies in $\mathcal{F}^{\boldsymbol \theta}_t$ and the posterior state probability is  
\begin{align}
\label{eq: 715-jhjkdkkjjsj}
\pi_{jt}    &=\frac{\theta_j \exp\big\{   t \beta_{j0}+  \boldsymbol{e}_j \mathbb{R}   t \boldsymbol x_t  \big\} }
 {\sum^N_{i=1} \theta_i \exp\big\{  t \beta_{i0}+  \boldsymbol{e}_i \mathbb{R}  t \boldsymbol x_t    \big\} +\theta_0} ,   j =1,\ldots, N, t\in \mathbb{N}^+, t  \boldsymbol x_t  \in  {\rm I\!R}^r. 
\end{align}

Consider a new prior in the same scaffolding, i.e., $\boldsymbol\theta' \in \Sigma(\boldsymbol\theta)$, with index $t_0$ and $\boldsymbol s_0$. The new prior~$\boldsymbol\theta'$ would revise the posterior belief to $\Pi_t(\boldsymbol x_t, \boldsymbol \theta')=(\pi'_{t0}, \ldots, \pi'_{tN})$, where
\begin{align}
\pi'_{jt}    &=\frac{\theta_j \exp\big\{  ( t -t_0) \beta_{j0}+  \boldsymbol{e}_j \mathbb{R}  ( t\boldsymbol x_t  -\boldsymbol s_0 ) \big\} }
 {\sum^N_{i=1} \theta_i \exp\big\{  (t-t_0) \beta_{i0}+  \boldsymbol{e}_i \mathbb{R}  ( t\boldsymbol x_t -\boldsymbol s_0  )   \big\} +\theta_0}, j =1,\ldots,  N . \nonumber
\end{align}
Note that the revised state probability has a similar structure as~\eqref{eq: 715-jhjkdkkjjsj}, suggesting that the revised posterior belief $\Pi_t(\boldsymbol x_t, \boldsymbol \theta')$ is also a member of~$\Sigma(\boldsymbol\theta)$. The difference is that the time index~$t$ and the cumulative sum $t\boldsymbol x_t$ are revised to  
\begin{align}
\tilde t \triangleq t-t_0 \text{ and }\ \boldsymbol s \triangleq t\boldsymbol x_t -\boldsymbol s_0, 
\end{align}
respectively. Thus, the revised belief state is located at $\boldsymbol s$ (referred to as the \emph{adjusted cumulative sum}) in the manifold $\tilde t$ (referred to as the \emph{adjusted time index}). If $t<t_0$, the revised belief falls on an upstream manifold, as shown in Figure~\ref{Fig: UpDownStreamCurves}--(a). Since the posterior has the same structure as the prior, the scaffolding can be also viewed as a class of conjugate priors with hyperparameters~$t_0, \boldsymbol{s}_0$.

Next, we will show that the DM may not need to solve the optimality equations for every new prior on the scaffolding. Since any posterior belief state in the scaffolding, say $\Pi_t \in \Sigma(\boldsymbol\theta)$, is indexed by $(\tilde t, \boldsymbol{s})$, we can rewrite the optimality equations~\eqref{eq: finite-beliefstate-OE} as follows
\begin{align}
 \tilde J_{t} (\tilde t, \boldsymbol{s}; \boldsymbol{\theta}) & = \min \{\tilde J_{t}^0(\tilde t, \boldsymbol{s}; \boldsymbol{\theta}), \cdots, \tilde J_{t}^N(\tilde t, \boldsymbol{s}; \boldsymbol{\theta}), \tilde J_{t}^w(\tilde t, \boldsymbol{s}; \boldsymbol{\theta})\}, \  t=1,\ldots, T-1, \tilde t \in \mathbb{N},  \nonumber \\
  \tilde J_{T}(\tilde t, \boldsymbol{s}; \boldsymbol{\theta}) & = \min \{ \tilde J_{T}^0(\tilde t, \boldsymbol{s}; \boldsymbol{\theta}), \cdots, \tilde J_{T}^N(\tilde t, \boldsymbol{s}; \boldsymbol{\theta})  \}, \tilde t \in \mathbb{N}, \nonumber\\
  \tilde J_{t} ^j(\tilde t, \boldsymbol{s}; \boldsymbol{\theta}) & = \sum_{i=0}^N \mathcal{\tilde  T}_i^{\tilde t}(\boldsymbol{s}, \boldsymbol{\theta})a_{ij},  \   j \in \mathcal N,  \  t=1,\ldots, T, \tilde t \in \mathbb{N},  \nonumber\\
  \tilde J_{t}^w(\tilde t, \boldsymbol{s}; \boldsymbol{\theta}) & = \sum_{i=0}^N \mathcal{\tilde T}_i^{\tilde t}(\boldsymbol{s}, \boldsymbol{\theta})c_{i} + \int_{\mathcal{Y}} \tilde J_{t+1}(\tilde t+1, \boldsymbol{s} + \mathbb{P}\boldsymbol{h}(y); \boldsymbol{\theta})\sum_{i=0}^N\mathcal{\tilde T}_i^{\tilde t}(\boldsymbol{s}, \boldsymbol{\theta})f_{i}(y)\mbox{d}y, \  t=1,\ldots, T-1, \tilde t \in \mathbb{N}, \nonumber
\end{align}
in which 
\begin{align}
\label{eq: T-tilde transform with cusum}
 \mathcal{\tilde T}^{\tilde t}_j(\boldsymbol s, \boldsymbol \theta )   
  &=\frac{\theta_j \exp\big\{   \tilde t \beta_{j0}+   \boldsymbol{e}_j \mathbb{R}   \boldsymbol s) \big\} }
 {\sum^N_{i=1} \theta_i \exp\big\{  \tilde t \beta_{i0}+   \boldsymbol{e}_i \mathbb{R}  \boldsymbol s)
  \big\} +\theta_0} , \ \  j =1,\ldots,  N, \tilde t \in \mathbb{N}. 
\end{align}
Here,~$\tilde J_{t} (\tilde t, \boldsymbol{s}; \boldsymbol{\theta})$ is the value function $V_t(\Pi)$ restricted on the scaffolding. It is a function of the number of decision periods $t$, and the pair $(\tilde t, \boldsymbol{s})$ which specifies the posterior belief given $\boldsymbol{\theta}$. The above equations  generalize those in Proposition~\ref{prop: reconstructedValueFunction}; the operator $\mathcal{T}^t_j(\boldsymbol x_t, \boldsymbol \theta )$ in Proposition~\ref{prop: reconstructedValueFunction} requires~$t$ to be a positive integer, but the operator $\mathcal{\tilde T}^{\tilde t}_j(\boldsymbol s, \boldsymbol \theta )$ in the above also allows for negative $\tilde t$. Another difference is that the state variable in the above equations is the adjusted cumulative sum $\boldsymbol{s} \in {\rm I\!R}^r$, whereas in Proposition~\ref{prop: reconstructedValueFunction} the state is the diagnostic statistic $\boldsymbol x_t \in \Omega_t$.\endnote{The numerical implementation of the optimality equations requires a finite truncation of the real space ${\rm I\!R}^r$. It is sufficient to consider the region in which the corresponding manifolds cover the $\epsilon$-contamination set. }  When $t_0=0$ and $\boldsymbol s_0=\boldsymbol 0$, the above optimality equations become equivalent to Proposition~\ref{prop: reconstructedValueFunction} where $\tilde t=t, \boldsymbol s=t \boldsymbol x_t$.

\subsubsection*{How to save computation.}
Consider a new prior on the scaffolding $\boldsymbol{\theta}' \in \Sigma(\boldsymbol\theta)$ with hyperparameters $(t_0, \boldsymbol{s}_0)$. Note that the belief updated by $(t, \boldsymbol{s})$ given the original prior~$\boldsymbol{\theta}$ is identical to the belief updated by~$(t-t_0, \boldsymbol{s}-\boldsymbol{s}_0)$ given the new prior $\boldsymbol{\theta}'$. Therefore,  
\begin{align}
\label{eq: theta and theta' identity}
\tilde J_{t}(t, \boldsymbol{s}; \boldsymbol{\theta}')=\tilde J_{t}(t-t_0, \boldsymbol{s}-\boldsymbol{s}_0; \boldsymbol{\theta}). 
\end{align}
This identity is the key to avoiding redundant computation. Suppose the DM has assumed the original prior $\boldsymbol \theta$ (i.e., $t_0=0, \boldsymbol s_0=\boldsymbol 0$) and computed the corresponding value functions $ \tilde J_{t} (\tilde t=t, \boldsymbol{s}; \boldsymbol{\theta})$, for $t=1,\ldots, T$ and $\boldsymbol{s} \in {\rm I\!R}^r$, by backward induction (here $\tilde t=t$ follows from $t_0=0$). If the new prior $\boldsymbol \theta'$ lies in the same manifold as $\boldsymbol \theta$ (i.e., the new prior has index $t_0=0$), then the DM does not need to repeat the computation because~$\tilde J_{t}(t, \boldsymbol{s}; \boldsymbol{\theta}')=\tilde J_{t}(t, \boldsymbol{s}-\boldsymbol{s}_0; \boldsymbol{\theta})$ by~\eqref{eq: theta and theta' identity}. The latter has already been computed under the original prior and, thus, can be reused by the new prior. When there are~$n$ sample priors in the manifold, reusing the value function is $n$ times faster than repeating the computation on every single prior.  

If the new prior $\boldsymbol \theta'$ lies in a different manifold (with index $t_0\neq 0$), then the DM needs to compute $\tilde J_{t}(t-t_0, \boldsymbol{s}; \boldsymbol{\theta}')$ for $t=1,\ldots, T$ using backward induction again. However, the resulting value function can be reused by any other prior in the same manifold as~$\boldsymbol \theta'$. Therefore, when the DM faces priors spreading over multiple manifolds, say $t_0\in \{-T_L, \ldots, T_U\}$, where $T_L$ and $T_U$ are arbitrary positive integers, it is sufficient to perform the backward induction only \emph{once} for each manifold; the results can be shared by all priors in that manifold following~\eqref{eq: theta and theta' identity}.

Additional computational saving is possible by exploiting the structural properties of the value functions.  In period $t$, $\tilde t$ takes on value from the set $\{ t-T_U, \ldots, t+T_L \}$. It is unnecessary to enumerate over all feasible combinations of~$t$ and~$\tilde t$ because the computational results for some combinations can be used to accelerate others, based on the following property
\begin{proposition}
\label{prop: monotone value function on the scaffolding}
If $\tilde J_{t} (\tilde t, \boldsymbol{s}; \boldsymbol{\theta})  = \tilde J_{t}^j(\tilde t, \boldsymbol{s}; \boldsymbol{\theta})$ for some $j\in\mathcal{N}$, then $\tilde J_{t'} (\tilde t, \boldsymbol{s}; \boldsymbol{\theta})  = \tilde J_{t'}^j(\tilde t, \boldsymbol{s}; \boldsymbol{\theta})$ for all $t'>t$. 
\end{proposition}


  \proof{Proof.}
We first use a relaxation argument to show that $\tilde J_{t} (\tilde t, \boldsymbol{s}; \boldsymbol{\theta})$ is nondecreasing in $t$ for fixed $\tilde t, \boldsymbol{s}$, and $\boldsymbol{\theta}$.  Note that $\tilde J_{t} (\tilde t, \boldsymbol{s}; \boldsymbol{\theta})$ is the expected total cost of following the optimal policy from period $t$ to the end of the horizon by starting from the belief state specified by $\tilde t, \boldsymbol{s}$, and $\boldsymbol{\theta}$. Consider a class of policies in which the decision process lasts no longer than $T'<T$ periods. The minimum expected total cost of this class of policies can be written as $\tilde J_{t+T-T'} (\tilde t, \boldsymbol{s}; \boldsymbol{\theta})$. Since this class of policies is a subset of the feasible policies, it cannot achieve a lower expected cost than the optimal policy, namely, $\tilde J_{t+T-T'} (\tilde t, \boldsymbol{s}; \boldsymbol{\theta}) \geqslant \tilde J_{t} (\tilde t, \boldsymbol{s}; \boldsymbol{\theta})$. Therefore, $\tilde J_{t} (\tilde t, \boldsymbol{s}; \boldsymbol{\theta})$ is nondecreasing in $t$. 

Next, note that $\tilde J_{t}^j(\tilde t, \boldsymbol{s}; \boldsymbol{\theta}), j\in\mathcal{N}$ is independent of $t$. Therefore, if $\tilde J_{t} (\tilde t, \boldsymbol{s}; \boldsymbol{\theta})  = \tilde J_{t}^j(\tilde t, \boldsymbol{s}; \boldsymbol{\theta})$, then we must have $\tilde J_{t'} (\tilde t, \boldsymbol{s}; \boldsymbol{\theta})  \geqslant \tilde J_{t'}^j(\tilde t, \boldsymbol{s}; \boldsymbol{\theta})$ for all $t'>t$ following the monotonicity result that we just proved. Finally, by the optimality equation we have $ \tilde J_{t} (\tilde t, \boldsymbol{s}; \boldsymbol{\theta}) \leqslant \tilde J_{t'}^j(\tilde t, \boldsymbol{s}; \boldsymbol{\theta})$. It follows immediately that $\tilde J_{t'} (\tilde t, \boldsymbol{s}; \boldsymbol{\theta})  = \tilde J_{t'}^j(\tilde t, \boldsymbol{s}; \boldsymbol{\theta})$ for all $t'>t$. 
 \Halmos
 \endproof

This proposition suggests that: if a state $(\tilde t, \boldsymbol{s})$ lies in the stopping region for state $j$ in period~$t$, then it must also lie in the stopping region for the same state in any future period $t'>t$. Note that in the stopping region the value function becomes $ \tilde J_{t'} (\tilde t, \boldsymbol{s}; \boldsymbol{\theta})=\tilde J_{t'} ^j(\tilde t, \boldsymbol{s}; \boldsymbol{\theta})$, which has a closed-form expression. Hence, there is no need to compute $\tilde J_{t'} (\tilde t, \boldsymbol{s})$ for all $t'>t$. 

\begin{example}
\label{ex: subsec: Scaffoldings}
 Figure~\ref{Fig: UpDownStreamCurves}--(b,c) displays sample stopping regions over the scaffolding. The parameters are $N=3$, $T=10$, and  $\boldsymbol{\theta}=(0.4854,    0.0243,    0.4854,    0.0049)$. The observation densities are $N(0,4), N(4,4), N(2,4)$, and $N(-1,4)$. The termination costs are zero-one and $c=0.027$. The vertical axis is the adjusted diagnostic statistic, i.e., $(t\boldsymbol x_t -\boldsymbol s_0)/(t-t_0) \in {\rm I\!R}$, except when $t=t_0$, in which case we use $t\boldsymbol x_t  - \boldsymbol s_0$. The range of~$t_0$ is truncated with $T_L=9,  T_U=11$ and the vertical axis is truncated at $[-20, 20]$ with 1000 discretization points. 

As mentioned earlier, the backward induction is required only once for each manifold. In Figure~\ref{Fig: UpDownStreamCurves}--(b) we choose the manifold $t_0=0$ in which $\tilde t=t=1,\ldots, 10$. The stopping regions are shown between the dash lines. Under the original prior~$\boldsymbol\theta$, the DM plots $(t, \boldsymbol x_t)$ in this region to find the optimal action (the first datapoint is plotted at $\tilde t=1$ and the last is plotted at $\tilde t=10$). This region can be shared by all priors lying in the same manifold as the original prior, making the implementation 1000 times faster than repeating the computation on every single prior in the manifold. Specifically, under a new prior with hyperparameters~$(t_0=0, \boldsymbol s_0)$, the DM will plot $(t,  \boldsymbol x_t-\boldsymbol s_0/t)$ in the same chart to find the optimal action. That is, revision of the prior only changes the vertical coordinate of the datapoint. A difference from the stopping regions computed by Proposition~\ref{prop: reconstructedValueFunction} is that the state space here is the real line ${\rm I\!R}$ instead of~$\Omega_t$. For implementation we truncate it at $[-20, 20]$.

If the new prior belongs to an upstream manifold, say $t_0=4$, then the DM will need to perform the backward induction again to compute Figure~\ref{Fig: UpDownStreamCurves}--(c).  In this case, $\tilde t=t-t_0=-3,\ldots, 6$. Again, all priors in this manifold can share the stopping regions shown between the dash lines (the DM will plot the first datapoint at $\tilde t=-3$, the last datapoint at $\tilde t=6$). However, Figure~\ref{Fig: UpDownStreamCurves}--(c) needs not to be computed from scratch given (b), because, by Proposition~\ref{prop: monotone value function on the scaffolding}, the stopping region in (b) is a subset of that in (c) where the value function is in closed-form. 
\end{example}

\begin{figure}
\centering \includegraphics[width=6.5in]{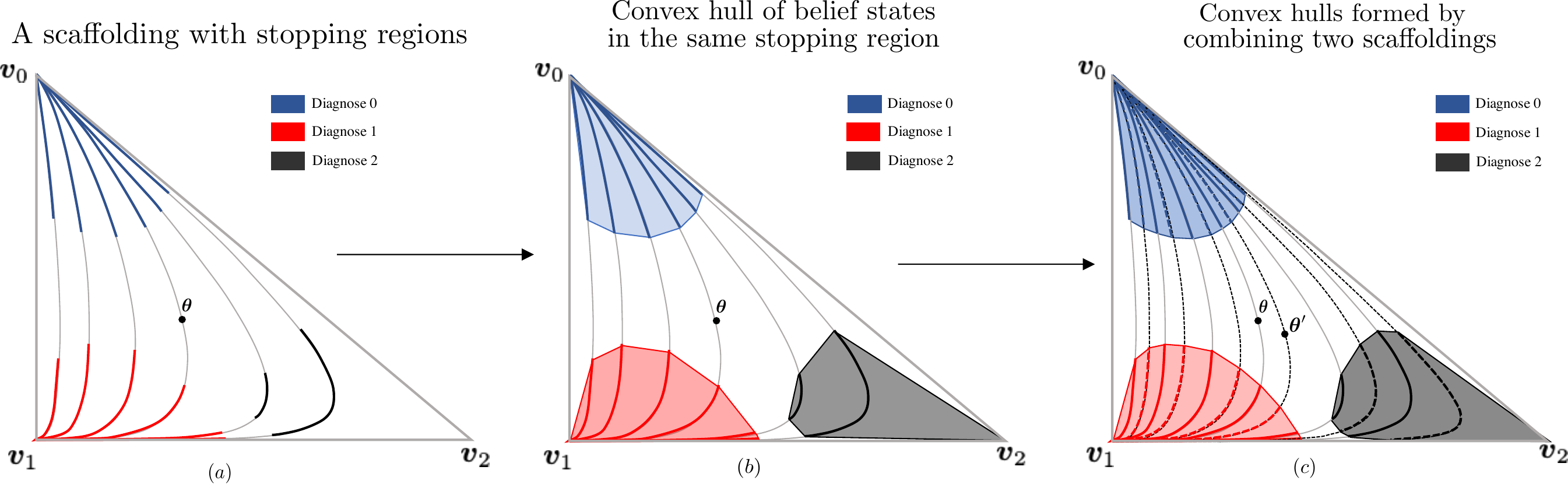} 
\caption{Constructing convex subsets of the stopping regions using scaffoldings }
\label{Fig: ConvexHull} 
\end{figure}

\subsection{Characterizing the subsets of the stopping regions}
\label{subsec: convex hull of scaffoldings}
Given the optimal stopping regions on a scaffolding, we can characterize a set of subsets of the stopping regions in the belief space, which can be used to approximate the true stopping regions. This process is illustrated in Figure~\ref{Fig: ConvexHull}. Since each state $(\tilde t, \boldsymbol{s})$ on the scaffolding corresponds to a belief state in~$S^N$, we can display the scaffolding in the belief space and color-code the stopping regions as Figure~\ref{Fig: ConvexHull}--(a). Let $\omega_i$ denote a set of belief states on the scaffolding corresponding to the same stopping action $i \in\mathcal{N}$, and let~$\boldsymbol v_i$ denote an~$(N+1)$-dimensional unit row vector with $1$ at the $(i+1)$th component, representing a vertex of the belief space. Then, the convex hull of $\omega_i$ and the vertex $\boldsymbol v_i$ must be a subset of the stopping region for class $i$, namely, 
$$
\text{Conv}(\omega_i, \boldsymbol v_i ) \subseteq  \Gamma_{it}.   
$$
This is because $\Gamma_{it}$ is convex and contains $\boldsymbol v_i$ (See the proof in \S\ref{EC:proof of Theorem 2}). The scaffolding in Figure~\ref{Fig: ConvexHull}--(a) produces three convex hulls in Figure~\ref{Fig: ConvexHull}--(b), which are subset approximations of the stopping regions in $S^2$. Higher-dimensional convex hulls can be efficiently computed using the quick hull algorithm \citep{barber1996quickhull}. 

Although the approximation produced by one scaffolding is relatively coarse, one can combine multiple scaffoldings to improve the approximation. Figure~\ref{Fig: ConvexHull}--(c) illustrates two interleaving scaffoldings,~$\Sigma(\boldsymbol\theta)$ and~$\Sigma(\boldsymbol\theta')$, parameterized by seed priors~$\boldsymbol\theta$ and~$\boldsymbol\theta'$, respectively. Combining them can produce larger and smoother convex hulls that better approximate $\Gamma_{it}$'s. The crux is selecting a set of ``orthogonal" seed priors satisfying two requirements: (1) the resulting scaffoldings should have minimal overlaps to minimize the redundancy; and (2) the number of scaffoldings should not grow exponentially with the number of classes. Similar goals also arise in the statistical design of experiments, where one aims to sample a high-dimensional space with a small number of samples. Inspired by this connection, we employ the {orthogonal array} (OA), a technique commonly used in experimental design, to select the seed priors. 

\subsection{Selecting seed priors with orthogonal array (OA)}
\label{sec: Seed prior}

\begin{figure}
\centering \includegraphics[width=6.5in]{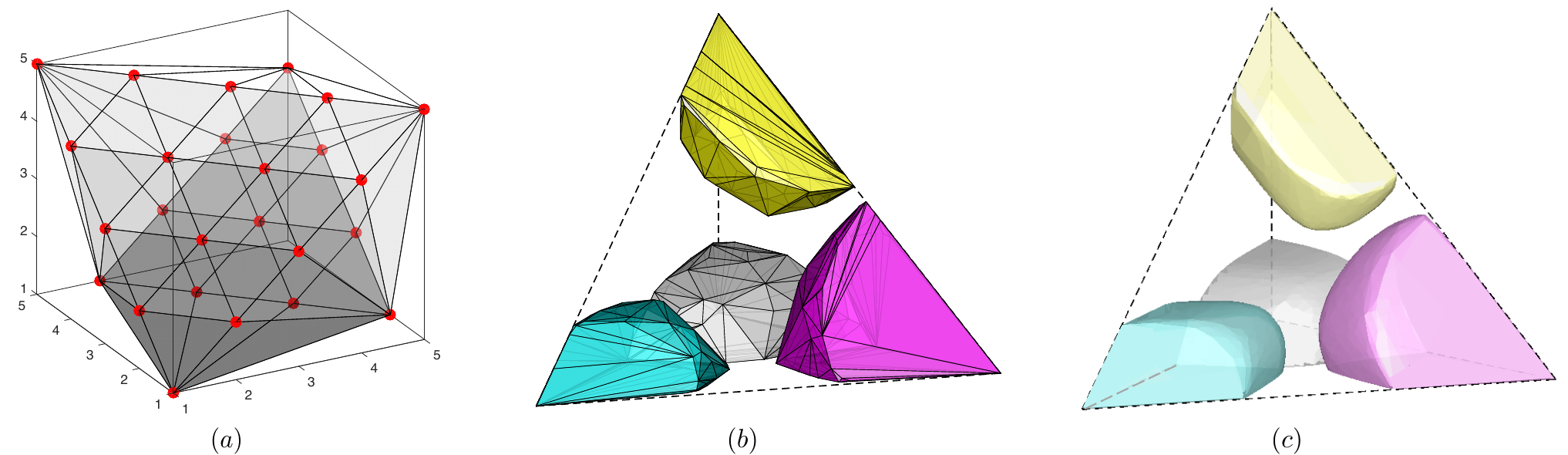} 
\caption{(a) Illustration of a Taguchi orthogonal array ($N=3, p=5$). Each red dot represents a seed prior. (b) Stopping regions constructed from 25 scaffoldings (comp. time: $0.56$ hrs), which approximate (c) the stopping regions constructed by full-space discretization (comp. time: $35. 88$ hrs).}
\label{Fig: Taguchi_cubes} 
\end{figure}

Consider an $N$-dimensional space in which each dimension corresponds to a factor that can be set to $p$ different levels. A full factorial design enumerating over all possible combinations of factor levels requires $n=p^N$ samples or ``runs". When $N$ is large, this design may require a prohibitively large number of runs. A fractional factorial design is a subset of a full factorial design and often requires considerably fewer runs. An orthogonal array (OA) is a fractional factorial design~\citep{hedayat2012orthogonal} denoted by $OA(n,N,p,d)$, where $d$ is called the strength. It has the following property: when we choose any $d$ factors, then each of the $p^d$ possible level combinations of these factors occur $\lambda$ times in the design. This implies that the total number of runs is $n=\lambda p^d$. It is common to set $d=2$, in which case the OA is also called Taguchi OA. An example with three factors, two levels (1 and 2), and $\lambda=1$ is given by 
\begin{align}
OA(4,3,2,2)=
\begin{bmatrix}
&1 \hspace{5mm} &1  \hspace{5mm} &1 \\
&1 \hspace{5mm}&2  \hspace{5mm}&2  \\
&2 \hspace{5mm}&1 \hspace{5mm}&2 \\
&2 \hspace{5mm}&2 \hspace{5mm}&1
\end{bmatrix}.
 \nonumber
\end{align}
Given any $d=2$ factors, there are $p^d=2^2=4$ possible combinations: $(1,1), (1,2), (2,1), (2,2)$. Considering any two ($d=2$) columns in $OA(4,3,2,2)$, we can see that each possible combination, e.g., $(2,1)$, appears only once ($\lambda=1$).  Although a full factorial design requires $p^n=2^3=8$ runs,  the OA design needs only 4 runs. When each factor can be set to more levels, such as $p=5$, the corresponding Taguchi $OA(25,3,5,2)$ is shown in Figure~\ref{Fig: Taguchi_cubes}--(a), which requires 25 instead of 125 runs. Next, we describe how to use this OA to select 25 priors. 

The scaffoldings are nonlinear manifolds with curvature. To apply the orthogonal arrays, we first use a transformation that ``straightens" the scaffoldings.  Note that a belief state $(\pi_0, \pi_1, \ldots, \pi_N)$ can be represented in terms of the log-likelihood ratio, 
\begin{equation*}
  \log\Big( \frac{\pi_i}{\pi_0} \Big) = \log \Big( \frac{\theta_i }{ \theta_0} \Big) +   t \beta_{i0}+  \boldsymbol{e}_j \mathbb{R}    \boldsymbol s, \ \ i=1,\ldots, N, 
\end{equation*}
as long as $\pi_0>0$. In this representation, a scaffolding with curvature is transformed into a hyperplane spanned by $(\beta_{10}, \ldots, \beta_{N0})$ and $( \boldsymbol{e}_1, \ldots, \boldsymbol{e}_N)\mathbb{R}$. The hyperplane can be represented by a vector consisting of the prior's log-likelihood ratios, $\log (\theta_i/\theta_0), i=1,\ldots, N$. By varying these ratios, one can translate the hyperplane across the log-likelihood ratio space. Unlike the scaffolding, which is a nonlinear manifold, the hyperplane in the log-likelihood ratio space is obviously linear. Since sampling a linear space is much easier, we will select the seed priors in the log-likelihood ratio space and then transform them back to the belief space. 

We use the orthogonal array to select the seed priors around the uninformative prior. Specifically, for $N=3$  the uninformative prior corresponds to the point $(0, 0, 0)$ in the  log-likelihood ratio space. We take samples from a hyper-cube centered at $(0, 0, 0)$ with length scale 7.5. For the orthogonal array we choose $p = 5$ and $d = 2$. The resulting $OA(25,3,5,2)$ in the log-likelihood ratio space is illustrated in Figure~\ref{Fig: Taguchi_cubes}--(a). Transforming them back to the belief space produces the 25 unique priors listed in Table~\ref{Seed priors Table}. Note that these priors are not necessarily orthogonal in the belief space. 

\begin{table}\footnotesize
  \centering
 \caption{Seed priors generated by Orthogonal Array}
 \label{Seed priors Table}
 \begin{tabular}{lcccc}
  \toprule
No.   & $\theta_0$ & $\theta_1$ & $\theta_2$  & $\theta_3$\\
  \midrule
1 &  0.934096470 & 0.021967843 & 0.021967843 & 0.021967843
\\
2 &  0.751751009 & 0.017679489 & 0.115284751 & 0.115284751
\\
3 &  0.330740576 & 0.007778273 & 0.330740576 & 0.330740576
\\
4 &  0.071097683 & 0.001672057 & 0.463615130 & 0.463615130
\\
5 &  0.011619033 & 0.000273253 & 0.494053857 & 0.494053857
\\
6 &  0.751751009 & 0.115284751 & 0.017679489 & 0.115284751
\\
7 &  0.433517880 & 0.066482120 & 0.066482120 & 0.433517880
\\
8 &  0.115284751 & 0.017679489 & 0.115284751 & 0.751751009
\\
9 &  0.019922201 & 0.003055169 & 0.129909072 & 0.847113558
\\
10 &  0.022884366 & 0.003509431 & 0.973068014 & 0.000538188
\\
11 &  0.330740576 & 0.330740576 & 0.007778273 & 0.330740576
\\
12 &  0.115284751 & 0.115284751 & 0.017679489 & 0.751751009
\\
13 &  0.021967843 & 0.021967843 & 0.021967843 & 0.934096470
\\
14 &  0.117036584 & 0.117036584 & 0.763174395 & 0.002752437
\\
15 &  0.022384166 & 0.022384166 & 0.951798946 & 0.003432723
\\
16 &  0.071097683 & 0.463615130 & 0.001672057 & 0.463615130
\\
17 &  0.019922201 & 0.129909072 & 0.003055169 & 0.847113558
\\
18 &  0.117036584 & 0.763174395 & 0.117036584 & 0.002752437
\\
19 &  0.070447374 & 0.459374585 & 0.459374585 & 0.010803455
\\
20 &  0.019591747 & 0.127754237 & 0.833062269 & 0.019591747
\\
21 &  0.011619033 & 0.494053857 & 0.000273253 & 0.494053857
\\
22 &  0.022884366 & 0.973068014 & 0.003509431 & 0.000538188
\\
23 &  0.022384166 & 0.951798946 & 0.022384166 & 0.003432723
\\
24 &  0.019591747 & 0.833062270 & 0.127754237 & 0.019591747
\\
25 &  0.010803455 & 0.459374585 & 0.459374585 & 0.070447374
\\

  \bottomrule
 \end{tabular}
\end{table}

Returning to Example~\ref{ex: subsec: Scaffoldings}, we use Taguchi $OA(25,3,5,2)$ to select 25 priors, as illustrated in Figure~\ref{Fig: Taguchi_cubes}--(a), and compute the scaffolding for each prior. Quick hull algorithm can compute the convex hulls of belief states within the same stopping region, and the resulting convex hulls are shown in Figure~\ref{Fig: Taguchi_cubes}--(b), which are subsets of the true stopping regions in the belief space. The entire process takes $0.56$ hrs on a laptop computer. For comparison, we also use the standard grid-based method to find the true stopping regions, as shown in Figure~\ref{Fig: Taguchi_cubes}--(c), which takes $35. 88$ hrs on the same computer. The convex hulls in Figure~\ref{Fig: Taguchi_cubes}--(b) account for $90.3\%$ volume of the stopping regions in (c) but needs only $1/60$ of the computation time. More precise approximations can be achieved by increasing the level (namely, $p$) of the Taguchi OA.

\subsection{Post-observation sensitivity analysis}
So far we have considered computing policies under different priors \emph{before} obtaining any data. Next, we consider the situation that the DM has obtained some data and needs to know how perturbations in the prior changes the optimal action. The theoretical foundation for this post-observation analysis is given below 
\begin{proposition}
\label{prop: convexity in prior}
Given the state $(\tilde t, \boldsymbol s)$ in period $t$, if it is optimal to stop and diagnose $i\in\mathcal{N}$ under the prior~$\boldsymbol \theta$ as well as under~$\boldsymbol \theta^n$, then it is also optimal to do the same under a prior $\boldsymbol \theta' \in \mathcal {D} (\boldsymbol \theta, \boldsymbol \theta^n) \triangleq \{ \boldsymbol \theta' \in S^N: \boldsymbol \theta' =\rho \boldsymbol \theta+(1-\rho) \boldsymbol \theta^n , \rho\in[0,1] \}$ in period $t$ and any period after~$t$.  
\end{proposition}

\proof{Proof.}
Given the observation sequence $\tilde y_t =\{y_1, \ldots, y_t \}$ and prior $\boldsymbol\theta$, the posterior belief state is given by  
\begin{align}
\label{eq: 2035-fuewhiuheiwu}
\Pi_t (\tilde y_t, \boldsymbol\theta)= \frac{\boldsymbol\theta \mathbb{G}(\tilde y_t)}  {\boldsymbol\theta \mathbb{F}(\tilde y_t) }, 
\end{align}
where
$$
\mathbb{G}(\tilde y_t)= \prod^{t}_{s=1} G(y_s), \ \ \mathbb{F}(\tilde y_t)= \prod^{t-1}_{s=1} G(y_s) F(y_t). 
$$
If the prior is revised to $\boldsymbol\theta^n$, the posterior becomes
\begin{align}
\label{eq: 2044-uu773yhyyswuu81}
\Pi_t (\tilde y_t, \boldsymbol\theta^n)= \frac{\boldsymbol\theta^n \mathbb{G}(\tilde y_t)}  {\boldsymbol\theta^n \mathbb{F}(\tilde y_t) }.
\end{align}
Now consider a new prior, $\boldsymbol \theta' =\rho \boldsymbol \theta+(1-\rho) \boldsymbol \theta^n$, which is a convex combination of $\boldsymbol \theta$ and $\boldsymbol \theta^n$. The corresponding posterior is 
\begin{align}
\Pi_t (\tilde y_t, \boldsymbol\theta')= \frac{\boldsymbol\theta' \mathbb{G}(\tilde y_t)}  {\boldsymbol\theta' \mathbb{F}(\tilde y_t) }
=& \frac{\rho \boldsymbol \theta \mathbb{G}(\tilde y_t) +(1-\rho) \boldsymbol \theta^n \mathbb{G}(\tilde y_t)}  {\boldsymbol\theta' \mathbb{F}(\tilde y_t) } \nonumber\\
=&  \frac{\rho \boldsymbol\theta \mathbb{F}(\tilde y_t) } {\boldsymbol\theta' \mathbb{F}(\tilde y_t) } \Pi_t (\tilde y_t, \boldsymbol\theta)  + \frac{(1-\rho) \boldsymbol\theta^n \mathbb{F}(\tilde y_t)}  {\boldsymbol\theta' \mathbb{F}(\tilde y_t) } \Pi_t (\tilde y_t, \boldsymbol\theta^n) 
\end{align}
where the last equality follows from \eqref{eq: 2035-fuewhiuheiwu} and \eqref{eq: 2044-uu773yhyyswuu81}. Next, note that 
\begin{align}
 \frac{\rho \boldsymbol\theta \mathbb{F}(\tilde y_t) } {\boldsymbol\theta' \mathbb{F}(\tilde y_t) }  + \frac{(1-\rho) \boldsymbol\theta^n \mathbb{F}(\tilde y_t)}  {\boldsymbol\theta' \mathbb{F}(\tilde y_t) } =1. 
\end{align}
Therefore, $\Pi_t (\tilde y_t, \boldsymbol\theta')$ is a convex combination of $\Pi_t (\tilde y_t, \boldsymbol\theta)$ and $\Pi_t (\tilde y_t, \boldsymbol\theta^n)$. If $\Pi_t (\tilde y_t, \boldsymbol\theta)\in \Gamma_{it}$ and $\Pi_t (\tilde y_t, \boldsymbol\theta^n)\in\Gamma_{it}$, it follows from the convexity of $\Gamma_{it}$ that $\Pi_t (\tilde y_t, \boldsymbol\theta')\in\Gamma_{it}$. Finally, using the same argument as that in the proof of Proposition~\ref{prop: monotone value function on the scaffolding}, one can easily show that $\Pi_t (\tilde y_t, \boldsymbol\theta')$ also falls in the stopping region $\Gamma_{it'}$ for $t'>t$. \Halmos
\endproof

This proposition suggests that, given the observations in hand, if two priors lead to the same optimal stopping action, then any convex combination of these two priors also leads to the same action. It holds because (i) stopping regions in the belief space are convex, and (ii) the convexity of posterior translates to the prior. Therefore, there is no need to recompute the policy for every prior on the line segment connecting $\boldsymbol \theta$ and $\boldsymbol \theta'$. If multiple priors lead to the same stopping action, then the convex hull of these priors preserves the optimality of that action. Recall that it is optimal to diagnose class $i$ at the vertex $\boldsymbol v_i$ of the belief space. Thus, by letting $\boldsymbol \theta^n=\boldsymbol v_i$ we have 
\begin{corollary}
Given the state $(\tilde t, \boldsymbol s)$ in period $t$, if it is optimal to stop and diagnose $i\in\mathcal{N}$ under the prior~$\boldsymbol \theta$, then it is optimal to do the same under a new prior $\boldsymbol \theta'\in \mathcal {D}_i (\boldsymbol \theta) \triangleq \{ (\theta_0, \ldots, \theta_{i-1}, r\theta_i, \theta_{i+1}, \ldots, \theta_N )/(1+(r-1)\theta_i)  \in S^N, r>1 \}$ in period $t$ and any period after~$t$.  
\end{corollary}
The proof is straightforward and thus omitted. This corollary suggests that, if it is optimal to stop and diagnose class $i$ under the prior $\boldsymbol \theta$, then it is also optimal to do so under a new prior  $\boldsymbol \theta'$ in which the probability of class $i$ is higher but the other class probabilities shrink proportionally.

\section{Proof of Theorem~\ref{prop: w(x1,x2) belongs to Omegai}}
\label{EC:proof of Theorem 2}
  \proof{Proof.}
It is well known that the value functions $V_t (\Pi)$ in a partially observable Markov decision process is concave~\citep{Krishnamurthy2016}. It then follows that $V^w_t (\Pi)$ is also concave, because the value function is positive homogeneous with degree one and concavity is preserved under summation and point-wise minimization~\citep{Krishnamurthy2016}. If two belief states lie on the same stopping region, namely, $\Pi_1, \Pi_2 \in\Gamma_{it}$, we must have 
$
V^i_t(\Pi_1) \leqslant V^w_t(\Pi_1) , V^i_t(\Pi_2) \leqslant V^w_t(\Pi_2).
$
Note that $V^i_t(\Pi)$ is linear in $\Pi$. Therefore, for any $\rho\in [0,1]$, the following inequality holds
$$
V^i_t(\rho\Pi_1+(1-\rho)\Pi_2)= \rho V^i_t(\Pi_1) +(1-\rho) V^i_t(\Pi_2) \leqslant \rho V^w_t(\Pi_1) +(1-\rho) V^w_t(\Pi_2).
$$
The concavity of $V^w_t$ implies that $\rho V^w_t(\Pi_1) +(1-\rho) V^w_t(\Pi_2) \leqslant V^w_t(\rho\Pi_1+(1-\rho)\Pi_2)$. Therefore, it is never optimal to wait in the state $\rho\Pi_1+(1-\rho)\Pi_2$. In addition, the linearity of the stopping value functions imply that $V^i_t(\rho\Pi_1+(1-\rho)\Pi_2) \leqslant V^j_t(\rho\Pi_1+(1-\rho)\Pi_2)$ for $j\neq i\in\mathcal{N}$. Thus,  $\rho\Pi_1+(1-\rho)\Pi_2 \in \Gamma_{it}$ and, hence, the stopping region $\Gamma_{it}$ is convex. 

 Next, we show that $V_t(\Pi) \leqslant V_{t+1}(\Pi)$ for all $\Pi \in S^N$ and $t$ by induction. Starting from $t=T-1$, 
 \begin{align}
 V_{T-1}(\Pi) 
 = & \min_{j\in\mathcal{N}} \Big\{ \sum^N_{i=0} \pi_i a_{ij} , \sum^N_{i=0}\pi_ic_i+ \int_{\mathcal{Y}}  \min_{j \in\mathcal{N}} \Big(\sum^N_{i=0} \pi_i a_{ij} f_i(y)  \Big) dy  \Big\} \nonumber\\
 \leqslant  &\min_{j\in\mathcal{N}} \Big\{ \sum^N_{i=0} \pi_i a_{ij} , \sum^N_{i=0}\pi_ic_i+ \min_{j \in\mathcal{N} }\sum^N_{i=0} \pi_i a_{ij}    \Big\}  \nonumber\\
 =&\min_{j\in\mathcal{N}} \Big\{ \sum^N_{i=0} \pi_i a_{ij} \Big\}= V_T (\Pi ), \nonumber
 \end{align}
 where the inequality follows from the concavity of the $\min$ operator and Jensen's inequality. Suppose $V_t(\Pi) \leqslant V_{t+1}(\Pi)$, it is straightforward to show that $V^w_{t-1}(\Pi) \leqslant V^w_{t}(\Pi)$. Then, $V_{t-1}(\Pi) \leqslant V_{t}(\Pi)$ follows immediately.  Hence, $V_t(\Pi) \leqslant V_{t+1}(\Pi)$ for all $\Pi \in S^N$ and $t$. An immediate consequence of this property is that the stopping region $\Gamma_{it}$ is expanding as $t$ increases. Let $i$ denote the state with the minimum expected termination cost for belief state $\Pi$ and suppose $\Pi\in\Gamma_{i(t-1)}$. Since $V^w_{t-1}(\Pi) \leqslant V^w_{t}(\Pi)$, then $V^i_{t-1}(\Pi) \leqslant V^w_{t-1}(\Pi)$ immediately implies $V^i_{t-1}(\Pi) \leqslant V^w_{t}(\Pi)$. Note that the stopping value function is time-independent, namely, $V^i_{t-1}(\Pi) =V^i_{t}(\Pi) $, hence we have $V^i_{t}(\Pi) <  V^w_{t}(\Pi)$. That is,  $\Pi \in \Gamma_{it}$. Thus, we have shown that $\Pi\in\Gamma_{i(t-1)}$ implies $\Pi\in\Gamma_{it}$, or $\Gamma_{i(t-1)}\subseteq \Gamma_{it}$.

Let $\boldsymbol v_i $ be the belief state with $\pi_i=1$ and $\pi_j=0$ for $j\neq i$. We will show that, when $a_{ii} < a_{ij}$ for all $j$, we have $\boldsymbol v_i \in \Gamma_{it}$ for all $t$. It is easy to see that $V^i_t(\boldsymbol v_i) =a_{ii} < a_{ij} = V^j_t(\boldsymbol v_i)$. Further, $V^w_t(\boldsymbol v_i) =c_{i}+V_{t+1} (  \boldsymbol v_i  ) \geqslant c_{i}+V_{t} (  \boldsymbol v_i  ) =c_{i}+V^i_{t} (  \boldsymbol v_i  ) \geqslant V^i_{t} (  \boldsymbol v_i  ) $, where we have used the fact that $ V_{t+1}(\Pi)  \geqslant V_t(\Pi) $ and $c_i \geqslant 0$. It is now clear that $V_t(\boldsymbol v_i) =V^i_t(\boldsymbol v_i) $ and $\boldsymbol v_i \in \Gamma_{it}$ for all $t$. 

If $\boldsymbol x^a \in \Omega_{it}$, then obviously $\mathcal{T}^t(\boldsymbol x^a, \boldsymbol \theta ) \in \Gamma_{it} $. Since we have shown that the vertex $\boldsymbol v_i$ also lies inside $\Gamma_{it} $ and $\Gamma_{it} $ is convex, the line segment connecting $\mathcal{T}^t(\boldsymbol x^a, \boldsymbol \theta )$ and $\boldsymbol v_i$ must also be  inside $\Gamma_{it} $. It can be easily verified with some algebra that the set  $\omega^i_{t} (\boldsymbol x^a)$ represents the intersection of this line segment with the reachable belief set in period $t$. Thus, it must be a subset of the stopping region $\Gamma_{it}$, thereby proving part one of the theorem. 

If $\boldsymbol x^a, \boldsymbol x^b \in \Omega_{it}$ for some $t$, then for any $\rho\in(0,1)$ the belief state  $\rho \mathcal{T}^t (\boldsymbol x^a, \boldsymbol \theta) +  (1-\rho) \mathcal{T}^t (\boldsymbol x^b, \boldsymbol \theta) $, which lies on the line segment connecting $\mathcal{T}^t (\boldsymbol x^a, \boldsymbol \theta)$ and $\mathcal{T}^t (\boldsymbol x^b, \boldsymbol \theta)$, must also belong to $\Gamma_{it}$, as $\Gamma_{it}$ is convex. Further, since $\Gamma_{i(t-1)}\subseteq \Gamma_{it}$, this line segment must be within all future stopping regions. For a future period $s \geqslant t$, if a diagnostic statistic belongs to the set $\omega_s (\boldsymbol x^a, \boldsymbol x^b, t)$, then it lies on this line segment that belongs to $\Gamma_{is}$. Since the optimal action is to stop and diagnose state $i$, this statistic must sit inside the stopping region $\Omega_{it}$. Thus, we have proved part two of the theorem.~\Halmos
  \endproof

\section{Additional Structural Results}
\label{ECsec: Additional Structural Properties}

\subsection{Properties of the reconstructed value functions}
\label{sec: properties of the reconstructed value functions}

In the standard formulation, the value function $V_t (\Pi)$ is concave in $\Pi$ \citep{Krishnamurthy2016}. However, the reconstructed value functions $J_t(\boldsymbol x_t; \boldsymbol \theta )$ are no longer concave in $\boldsymbol x_t$. The stopping value function is neither convex nor concave, and $J_t(\boldsymbol x_t; \boldsymbol \theta )$ can be a multimodal function with complex structure. Below we provide an equivalent reformulation in which the stopping value functions are log-convex, and is more amenable to structural analysis. Define 
$
\alpha_t( \boldsymbol x_t ;\boldsymbol\theta )  \triangleq  \theta_0+ \sum^N_{i=1}  \theta_i  \exp\big\{t(\beta_{i0}+ \boldsymbol{e}_i \mathbb{R}  \boldsymbol x_t )  \big \}
$ 
as the normalization term in $\mathcal{T}^t_j(\boldsymbol x_t, \boldsymbol \theta ) $, and let
$
 \mathcal{V}^j_t   (   \boldsymbol x_t; \boldsymbol\theta   )\triangleq \alpha_t( \boldsymbol x_t ;\boldsymbol\theta )  J^j_t(\boldsymbol x_t; \boldsymbol \theta ), j\in \mathcal N,  \ 
 \mathcal{V}^w_t   ( \boldsymbol x_t; \boldsymbol\theta  ) \triangleq \alpha_t(\boldsymbol x_t ;\boldsymbol\theta) J^w_t(\boldsymbol x_t; \boldsymbol \theta )$, and
$ \mathcal{V}_{t}   \big(\boldsymbol x_t; \boldsymbol\theta   \big) \triangleq  \alpha_t( \boldsymbol x_t; \boldsymbol\theta) J_t(\boldsymbol x_t; \boldsymbol \theta ) 
$
denote the \emph{scaled value functions} formed by multiplying $J^j_t, J^w_t$ and $J_t$ with a common normalization term, $\alpha_t( \boldsymbol x_t ;\boldsymbol\theta ) $.

 \begin{proposition} [Reformulation II]
\label{prop: unnormalizedValueFunction_convex}
The scaled value functions satisfy the following optimality equations: 
\begin{align}
\mathcal{V}_t(\boldsymbol x_t; \boldsymbol \theta ) & =\min\big\{\mathcal{V}^0_t(\boldsymbol x_t; \boldsymbol \theta ), \mathcal{V}^1_t(\boldsymbol x_t; \boldsymbol \theta ), \ldots, \mathcal{V}^N_t(\boldsymbol x_t; \boldsymbol \theta ), \mathcal{V}^w_t(\boldsymbol x_t; \boldsymbol \theta )    \big\},  \ \ \ t=1, 2,  \ldots , T-1,\nonumber\\
\mathcal{V}_T(\boldsymbol x_T; \boldsymbol \theta ) & =\min\big\{\mathcal{V}^0_T(\boldsymbol x_T; \boldsymbol \theta ), \mathcal{V}^1_T(\boldsymbol x_T; \boldsymbol \theta ), \ldots, \mathcal{V}^N_T(\boldsymbol x_T; \boldsymbol \theta )  \big\}, \nonumber\\
\mathcal{V}^j_t(\boldsymbol x_t; \boldsymbol \theta ) &=\theta_0 a_{0j}+ \sum^N_{i=1} \theta_i \exp\big\{t (\beta_{i0}+ \boldsymbol{e}_i \mathbb{R}  \boldsymbol x_t) \big \}   a_{ij}, \ \ \ j\in \mathcal N, \nonumber\\
\mathcal{V}^w_t(\boldsymbol x_t; \boldsymbol \theta ) &=\theta_0 c_0+ \sum^N_{i=1} \theta_i \exp\{t(\beta_{i0}+  \boldsymbol{e}_i \mathbb{R}  \boldsymbol x_t) \} c_i+\int_\mathcal{Y} \mathcal{V}_{t+1} \Big( \frac{  t \boldsymbol  x_t+ \mathbb{P} \boldsymbol h(y_{t+1}) }{t+1} ; \boldsymbol \theta  \Big)  f_0(y_{t+1}) d y_{t+1} . \nonumber
\end{align}
where $\mathcal{V}^j_t(\boldsymbol x_t; \boldsymbol \theta )$ is log-convex in $\boldsymbol x_t$, for $j\in \mathcal N$.
\end{proposition}

\proof{Proof.}
Multiplying $\alpha_t( \boldsymbol x_t ;\boldsymbol\theta )$ on both sides of~\eqref{Eq: theorem: 1D-Finite-OE-Main-stopping} yields
\begin{align}
\label{eq: 542-jiojeiowjkk}
\mathcal{V}^j_t(\boldsymbol x_t; \boldsymbol \theta )= \alpha_t(  \boldsymbol x_t; \boldsymbol\theta)  J^j_t(\boldsymbol x_t; \boldsymbol \theta )
 =&\theta_0 a_{0j}+ \sum^N_{i=1} \theta_i \exp\big\{t (\beta_{i0}+ \boldsymbol{e}_i \mathbb{R}  \boldsymbol x_t) \big \}   a_{ij}, \ \  \ j\in \mathcal N.  
 \end{align}
Note that $\exp\big\{t (\beta_{i0}+ \boldsymbol{e}_i \mathbb{R}  \boldsymbol x_t) \big \} $ is log-linear and, hence, log-convex in $\boldsymbol x_t$ for all $i$. Further, log-convexity is closed under summation (see \S~3.5.2 of~\cite{boyd2004convex}), therefore $\mathcal{V}^j_t(\boldsymbol x_t; \boldsymbol \theta )$ is log-convex.

Multiplying $\alpha_t( \boldsymbol x_t ;\boldsymbol\theta )$ on both sides of \eqref{Eq: theorem: 1D-Finite-OE-Main-wait} yields
 \begin{align}
&\mathcal{V}^w_t(\boldsymbol x_t; \boldsymbol \theta )=\alpha_t( \boldsymbol x_t ;\boldsymbol\theta ) J^w_t(\boldsymbol x_t; \boldsymbol \theta )  \nonumber\\
=& \sum^N_{i=0} \alpha_t( \boldsymbol x_t ;\boldsymbol\theta ) \mathcal{T}^t_i(\boldsymbol x_t, \boldsymbol \theta) c_{i}+\int_\mathcal{Y} J_{t+1} \Big( \frac{  t \boldsymbol  x_t+ \mathbb{P} \boldsymbol h(y_{t+1}) }{t+1} ; \boldsymbol \theta  \Big) \sum^N_{i=0} \alpha_t( \boldsymbol x_t ;\boldsymbol\theta )  \mathcal{T}^t_i(\boldsymbol x_t, \boldsymbol \theta) f_i(y_{t+1}) d y_{t+1} \nonumber\\
 =&\theta_0 c_0+ \sum^N_{i=1} \theta_i \exp\{t(\beta_{i0}+  \boldsymbol{e}_i \mathbb{R}  \boldsymbol x_t) \} c_i+\int_\mathcal{Y} J_{t+1} \Big( \frac{  t \boldsymbol  x_t+ \mathbb{P} \boldsymbol h(y_{t+1}) }{t+1} ; \boldsymbol \theta  \Big) \alpha_{t+1}\Big( \frac{ t\boldsymbol x_t +\mathbb{P} \boldsymbol h(y_{t+1})}{t+1}; \boldsymbol\theta\Big) f_0(y_{t+1}) d y_{t+1} , \nonumber
 \end{align}
where the last equality holds because
 \begin{align}
 \sum^N_{i=0} \alpha_t( \boldsymbol x_t ;\boldsymbol\theta )  \mathcal{T}^t_i(\boldsymbol x_t, \boldsymbol \theta) f_i(y_{t+1})  
 =& \Big[  \theta_0+\sum^N_{i=1} \theta_i\exp\Big\{  t (\beta_{i0}+  \boldsymbol{e}_i \mathbb{R}  \boldsymbol x_t) + \beta_{i0}+ \boldsymbol\beta^{\mathsf{T}}_i \boldsymbol h(y_{t+1}) \Big\} \Big] f_0(y_{t+1}) \nonumber\\
 =& \Big[  \theta_0+\sum^N_{i=1} \theta_i\exp\Big\{  t (\beta_{i0}+  \boldsymbol{e}_i \mathbb{R}  \boldsymbol x_t)  + \beta_{i0}+ \boldsymbol{e}_i \mathbb{RP}  \boldsymbol h(y_{t+1}) \Big\} \Big] f_0(y_{t+1})  \nonumber\\
 =& \Big[  \theta_0+\sum^N_{i=1} \theta_i\exp\Big\{  (t+1) \Big(\beta_{i0}+  \boldsymbol{e}_i \mathbb{R}  \frac{ t \boldsymbol x_t + \mathbb{P}  \boldsymbol h(y_{t+1}) }{t+1} \Big)  \Big\} \Big] f_0(y_{t+1})  \nonumber\\
 =& \alpha_{t+1}\Big( \frac{ t\boldsymbol x_t +\mathbb{P} \boldsymbol h(y_{t+1})}{t+1}; \boldsymbol\theta\Big) f_0(y_{t+1}).  \nonumber
 \end{align}
The expressions for $\mathcal{V}_t(\boldsymbol x_t; \boldsymbol \theta ) $ follow directly from multiplying both sides of~\eqref{Eq: theorem: 1D-Finite-OE-Main} by $\alpha_t( \boldsymbol x_t ;\boldsymbol\theta )$.  \Halmos
  \endproof

Since the value functions of all actions are multiplied by a common scalar, the optimal policy for Reformulation II is also optimal for the formulation in Proposition~\ref{prop: reconstructedValueFunction}. Yet the scaling simplifies the structural analysis of the optimal policy. 

Next, we set out to find bounds for the value functions. We first introduce some notations. Let~$\boldsymbol g$ denote the density of $\boldsymbol h(y)$ when $y$ has a density $f_0$, and let $*$ represent convolution. The $n$th convolution power (namely, $n$-fold iteration of the convolution) of $\boldsymbol g$ is denoted by
$
\boldsymbol g^{*n}   \triangleq    \underbrace{\boldsymbol g * \cdots *\boldsymbol g}_\text{$n$ iterations},  
$
which is the density of the sum $\sum^n_{t=1} \boldsymbol h(y_t)$. When $n=0$, we define $\boldsymbol g^{*0} =\delta_0$ as the Dirac delta distribution. 
\begin{definition}
\begin{align}
\label{def: lowerbound of V}
\mathcal{L}^\tau_t(\boldsymbol x; \boldsymbol \theta ) \triangleq & 
 \int  \min_{j\in \mathcal N} \Big\{   \theta_0 a_{0j} +  \sum^N_{i=1} \theta_i \exp\big\{\tau \beta_{i0}+ \boldsymbol{e}_i \mathbb{R}  (t  \boldsymbol  x+ \mathbb{P}  \boldsymbol h    ) \big \} a_{ij}  \Big\}   \boldsymbol g^{*(\tau-t)}(\boldsymbol h)  d \boldsymbol h, \ \tau >t, \\
 \mathcal{L}^t_t(\boldsymbol x; \boldsymbol \theta ) \triangleq & \min_{ j\in \mathcal N}\{ \theta_0 a_{0j}+ \sum^N_{i=1} \theta_i \exp [ t (\beta_{i0}+ \boldsymbol{e}_i \mathbb{R}  \boldsymbol x) ]   a_{ij} \big\}, \ \tau =t. \nonumber
\end{align}
\end{definition}
The function $\mathcal{L}^\tau_t(\boldsymbol x; \boldsymbol \theta )$ can be interpreted as the (scaled) expected termination cost of the decision rule that keeps observing until time $\tau$ and then make the best diagnoses, given that the diagnostic statistic is $\boldsymbol x$ at time $t$. The following lemma shows that this function is decreasing in $\tau$ and can lead to both lower and upper bounds for the scaled value function. 

 \begin{lemma}
\label{lemma: bounds on the value functions}
The following inequalities hold for all $\boldsymbol x$, $t$, and $\tau \geqslant t$:
\begin{enumerate}
\item   $  \mathcal{L}^\tau_t(\boldsymbol x; \boldsymbol \theta ) \geqslant  \mathcal{L}^{\tau+1}_t(\boldsymbol x; \boldsymbol \theta )$.
\item $ \min_{ t \leqslant \tau  \leqslant T } \{  (\tau-t) [\theta_0 c_0+ \sum^N_{i=1} \theta_i \exp\{t(\beta_{i0}+  \boldsymbol{e}_i \mathbb{R}  \boldsymbol x_t) \} c_i ]+ \mathcal{L}^\tau_t(\boldsymbol x_t; \boldsymbol \theta )  \}  \geqslant \mathcal{V}_t(\boldsymbol x_t; \boldsymbol \theta ) \geqslant \mathcal{L}^T_t(\boldsymbol x_t; \boldsymbol \theta )$.
\end{enumerate}
\end{lemma}

  \proof{Proof.}
To begin, we denote the unnormalized belief at period $t$ as
$$
\varphi_{t}(\boldsymbol x_{t}, \boldsymbol \theta)=\big[\theta_0, \theta_1 \exp\big\{t (\beta_{10}+ \boldsymbol{e}_1 \mathbb{R}  \boldsymbol x_{t}) \big \} , \ldots, \theta_N \exp\big\{t (\beta_{N0}+ \boldsymbol{e}_N \mathbb{R}  \boldsymbol x_{t}) \big \}   \big].
$$
With this notation, we can rewrite the function $\mathcal{L}^\tau_t(\boldsymbol x; \boldsymbol \theta )$ as follows 
\begin{align}
\mathcal{L}^\tau_t(\boldsymbol x; \boldsymbol \theta ) 
=& \int  \min_{j\in \mathcal N} \Big\{ \varphi_{\tau} \Big(  \frac{t  \boldsymbol  x+ \mathbb{P}  \boldsymbol h    }{\tau} , \boldsymbol \theta \Big) \boldsymbol a_j  \Big\}   \boldsymbol g^{*(\tau-t)}(\boldsymbol h)  d \boldsymbol h \nonumber\\
=& \int_\mathcal{Y} \ldots  \int_\mathcal{Y} \min_{j\in \mathcal N} \Big\{  \varphi_\tau \Big(\frac{  t \boldsymbol  x+ \mathbb{P}  \sum^\tau_{m=t+1}   \boldsymbol h(y_m)  }{\tau}, \boldsymbol \theta \Big)   \boldsymbol a_j    \Big\}    \prod^{\tau}_{m=t+1} f_0(y_m) d y_{t+1} \ldots d y_\tau . \nonumber
\end{align}
The first equality is obtained from the definition of $\varphi_{t}(\boldsymbol x_{t}, \boldsymbol \theta)$. The second equality holds because the convolution power of densities represents the density of the sum of iid. random variables. In the subsequent part of proof, we will use the expression in the second line above to represent $\mathcal{L}^\tau_t(\boldsymbol x; \boldsymbol \theta ) $.

After obtaining $n$ more observations, $y_{t+1}, y_{t+2}, \ldots, y_{t+n}$, the unnormalized belief, $\varphi_{t}(\boldsymbol x_{t}, \boldsymbol \theta)$, will be updated to
\begin{align}
\varphi_{t}(\boldsymbol x_{t}, \boldsymbol \theta) \prod^{t+n}_{m=t+1} G(y_m) 
 = & \big[\theta_0 \prod^{t+n}_{m=t+1} f_0(y_m),   \ldots, \theta_N \exp\big\{t (\beta_{N0}+ \boldsymbol{e}_N \mathbb{R}  \boldsymbol x_{t}) \big \}  \prod^{t+n}_{m=t+1} f_N(y_m) ]  \nonumber\\
  = &\prod^{t+n}_{m=t+1} f_0(y_m) \big[\theta_0 ,  \ldots, \theta_N \exp\big\{t (\beta_{N0}+ \boldsymbol{e}_N \mathbb{R}  \boldsymbol x_{t}) + n \beta_{N0} + \boldsymbol e_N \mathbb{RP} \sum^{t+n}_{m=t+1}   \boldsymbol h(y_m)  \} ]   \nonumber\\
    = &\prod^{t+n}_{m=t+1} f_0(y_m) \Big[\theta_0 ,  \ldots, \theta_N \exp\Big\{ (t+n) \Big[ \beta_{N0} + \boldsymbol e_N \mathbb{R}  \frac{  t \boldsymbol  x_{t}+ \mathbb{P} \sum^{t+n}_{m=t+1}   \boldsymbol h(y_m) }{t+n} \Big] \Big \} \Big ]   \nonumber\\
    =&\prod^{t+n}_{m=t+1} f_0(y_m)  \varphi_{t+n} \Big(\frac{  t \boldsymbol  x_{t}+ \mathbb{P}  \sum^{t+n}_{m=t+1}   \boldsymbol h(y_m)  }{t+n}, \boldsymbol \theta \Big). \nonumber
\end{align}
Therefore,
\begin{align}
\label{eq: unnormalized belief is martingale}
 &\int_\mathcal{Y} \ldots  \int_\mathcal{Y} \varphi_{t+n} \Big(\frac{  t \boldsymbol  x_{t}+ \mathbb{P}  \sum^{t+n}_{m=t+1}   \boldsymbol h(y_m)  }{t+n}, \boldsymbol \theta \Big)  \prod^{t+n}_{m=t+1} f_0(y_m) d y_{t+1} \ldots d y_{t+n}  \nonumber\\
=  &\int_\mathcal{Y} \ldots  \int_\mathcal{Y}   \varphi_{t}(\boldsymbol x_{t}, \boldsymbol \theta) \prod^{t+n}_{m=t+1} G(y_m)  d y_{t+1} \ldots d y_{t+n}  \nonumber\\
= &  \varphi_{t}(\boldsymbol x_{t}, \boldsymbol \theta). 
\end{align}

Let $\boldsymbol a_j  =[a_{0j}, a_{1j}, \ldots, a_{Nj}]^\mathsf{T}$. It is clear that $ \min_{j\in \mathcal N} \{  \varphi_{t}(\boldsymbol x_{t}, \boldsymbol \theta) \boldsymbol a_j \}$ is concave in $ \varphi_{t}(\boldsymbol x_{t}, \boldsymbol \theta)$. Then, by  Jensen's inequality and  \eqref{eq: unnormalized belief is martingale} we have
\begin{align}
\label{eq: 667-jkkekwjkuiisjk}
\min_{j\in \mathcal N} \{    \varphi_{t}(\boldsymbol x_{t}, \boldsymbol \theta)   \boldsymbol a_j  \}  
=&\min_{j\in \mathcal N} \Big\{  \int_\mathcal{Y}  \varphi_{t+1} \Big(\frac{  t \boldsymbol  x_{t}+ \mathbb{P}     \boldsymbol h(y_{t+1})  }{t+1}, \boldsymbol \theta \Big) \boldsymbol a_j   f_0(y_{t+1}) d y_{t+1}       \Big\}  \nonumber\\
\geqslant & \int_\mathcal{Y}  \min_{j\in \mathcal N} \Big\{  \varphi_{t+1} \Big(\frac{  t \boldsymbol  x_t+ \mathbb{P}    \boldsymbol h(y_{t+1})  }{t+1}, \boldsymbol \theta \Big)  \boldsymbol a_j    \Big\}   f_0(y_{t+1}) d y_{t+1}    ,
\end{align}
for all $t$. Repeating \eqref{eq: 667-jkkekwjkuiisjk} for $t+1$, we have
\begin{align}
\label{eq: 710-jkkekwjk773882uiisjk}
& \min_{j\in \mathcal N}\Big \{    \varphi_{t+1} \Big(\frac{  t \boldsymbol  x_t+ \mathbb{P}    \boldsymbol h(y_{t+1})  }{t+1}, \boldsymbol \theta \Big)   \boldsymbol a_j \Big \}  \nonumber\\
\geqslant & \int_\mathcal{Y}  \min_{j\in \mathcal N} \Big\{  \varphi_{t+2} \Big(\frac{   t \boldsymbol  x_t+ \mathbb{P}    \boldsymbol h(y_{t+1})  + \mathbb{P}    \boldsymbol h(y_{t+2})  }{t+2}, \boldsymbol \theta \Big)  \boldsymbol a_j    \Big\}   f_0(y_{t+2}) d y_{t+2}    .
\end{align}
Substitute \eqref{eq: 710-jkkekwjk773882uiisjk} into the right-hand side of \eqref{eq: 667-jkkekwjkuiisjk}, we obtain 
\begin{align}
\min_{j\in \mathcal N} \{    \varphi_{t}(\boldsymbol x_{t}, \boldsymbol \theta)   \boldsymbol a_j  \}  
\geqslant & \int_\mathcal{Y} \int_\mathcal{Y}  \min_{j\in \mathcal N} \Big\{  \varphi_{t+2} \Big(\frac{   t \boldsymbol  x_t+ \mathbb{P}    \boldsymbol h(y_{t+1})  + \mathbb{P}    \boldsymbol h(y_{t+2})  }{t+2}, \boldsymbol \theta \Big)  \boldsymbol a_j    \Big\}   f_0(y_{t+1}) f_0(y_{t+2}) d y_{t+1}     d y_{t+2}   .\nonumber
\end{align}
Similarly, repeating the above step to $t+2, \ldots, t+n$ yields the following inequality
\begin{align}
\label{eq: 722-jjuyiqi714441ugdgsggs}
&\min_{j\in \mathcal N} \{    \varphi_{t}(\boldsymbol x_{t}, \boldsymbol \theta)   \boldsymbol a_j  \}  \nonumber\\
\geqslant & \int_\mathcal{Y} \ldots  \int_\mathcal{Y} \min_{j\in \mathcal N} \Big\{  \varphi_{t+n} \Big(\frac{  t \boldsymbol  x_{t}+ \mathbb{P}  \sum^{t+n}_{m=t+1}   \boldsymbol h(y_m)  }{t+n}, \boldsymbol \theta \Big)   \boldsymbol a_j    \Big\}    \prod^{t+n}_{m=t+1} f_0(y_m) d y_{t+1} \ldots d y_{t+n}  ,
\end{align}
which is an important inequality that will be used to establish the lower bound.

\subsection*{Part 1}
By letting $t=\tau, n=1$, and $\boldsymbol x_t= [  t \boldsymbol x_t + \mathbb{P}  \sum^{\tau}_{m=t+1}   \boldsymbol h(y_m)  ]/\tau$ in \eqref{eq: unnormalized belief is martingale}, we obtain 
\begin{align}
 \int_\mathcal{Y}   \varphi_{\tau+1} \Big(\frac{  t \boldsymbol  x+ \mathbb{P}  \sum^{\tau+1}_{m=t+1}   \boldsymbol h(y_m)  }{\tau+1}, \boldsymbol \theta \Big)     f_0(y_{\tau+1}) d y_{\tau+1} =  \varphi_{\tau} \Big(\frac{  t \boldsymbol  x+ \mathbb{P}  \sum^{\tau}_{m=t+1}   \boldsymbol h(y_m)  }{\tau}, \boldsymbol \theta \Big). \nonumber
\end{align}
Following Jensen's inequality we have
\begin{align}
 \int_\mathcal{Y}  \min_{j\in \mathcal N} \Big\{  \varphi_{\tau+1} \Big(\frac{  t \boldsymbol  x+ \mathbb{P}  \sum^{\tau+1}_{m=t+1}   \boldsymbol h(y_m)  }{\tau+1}, \boldsymbol \theta \Big) \boldsymbol a_j    \Big\}      f_0(y_{\tau+1}) d y_{\tau+1} 
 \leqslant \min_{j=\in \mathcal N} \Big\{   \varphi_{\tau} \Big(\frac{  t \boldsymbol  x+ \mathbb{P}  \sum^{\tau}_{m=t+1}   \boldsymbol h(y_m)  }{\tau}, \boldsymbol \theta \Big)  \boldsymbol a_j    \Big\}  . \nonumber
\end{align}
Integrating on both sides gives
\begin{align}
\mathcal{L}^{\tau+1}_t(\boldsymbol x; \boldsymbol \theta ) 
=& \int_\mathcal{Y} \ldots  \int_\mathcal{Y} \min_{j\in \mathcal N} \Big\{  \varphi_{\tau+1} \Big(\frac{  t \boldsymbol  x+ \mathbb{P}  \sum^{\tau+1}_{m=t+1}   \boldsymbol h(y_m)  }{\tau+1}, \boldsymbol \theta \Big)   \boldsymbol a_j    \Big\}    \prod^{\tau+1}_{m=t+1} f_0(y_m) d y_{t+1} \ldots d y_\tau d y_{\tau+1} . \nonumber\\
\leqslant & \int_\mathcal{Y} \ldots  \int_\mathcal{Y} \min_{j\in \mathcal N} \Big\{  \varphi_\tau \Big(\frac{  t \boldsymbol  x+ \mathbb{P}  \sum^\tau_{m=t+1}   \boldsymbol h(y_m)  }{\tau}, \boldsymbol \theta \Big)   \boldsymbol a_j    \Big\}    \prod^{\tau}_{m=t+1} f_0(y_m) d y_{t+1} \ldots d y_\tau \nonumber\\
=&\mathcal{L}^\tau_t(\boldsymbol x; \boldsymbol \theta ),  \nonumber
\end{align}
thereby proving part 1 of the lemma.

\subsection*{Part 2 (Lower bound)}
We now prove the lower bound by induction. Consider $t=T-1$. Since $c_i \geqslant 0$ for all $i$, it is easy to obtain a lower bound for $\mathcal{V}^w_{T-1}$ by letting $c_i=0$ for all $i$, namely,
\begin{align}
\label{eq: lowerbound of Vw-657iiwiqj}
\mathcal{V}^w_{T-1} (\boldsymbol x_{T-1}; \boldsymbol \theta ) \geqslant & \int_\mathcal{Y} \mathcal{V}_T \Big( \frac{  (T-1) \boldsymbol  x_{T-1}+ \mathbb{P} \boldsymbol h(y_T) }{T} ; \boldsymbol \theta  \Big)  f_0(y_T) d y_T  \nonumber\\
=&\int_\mathcal{Y} \min_{j\in \mathcal N} \Big\{ \varphi_{T} \Big(  \frac{(T-1)\boldsymbol x_{T-1} +\mathbb{P} \boldsymbol h(y_T) }{T} , \boldsymbol \theta \Big) \boldsymbol a_j  \Big\}   f_0(y_T) d y_T  .  
\end{align}
Therefore, 
\begin{align}
&\mathcal{V}_{T-1}(\boldsymbol x_{T-1}; \boldsymbol \theta ) \nonumber\\
& \geqslant \min\Big\{   \min_{j\in \mathcal N}\big\{   \mathcal{V}^j_{T-1}(\boldsymbol x_{T-1}; \boldsymbol \theta )  \big\},\int_\mathcal{Y} \min_{j\in \mathcal N} \Big\{ \varphi_{T} \Big(  \frac{(T-1)\boldsymbol x_{T-1} +\mathbb{P} \boldsymbol h(y_T) }{T} , \boldsymbol \theta \Big) \boldsymbol a_j  \Big\}   f_0(y_T) d y_T    \Big\} \nonumber\\
& = \min\Big\{   \min_{j\in \mathcal N}\big\{  \varphi_{T-1}(\boldsymbol x_{T-1}, \boldsymbol \theta)    \boldsymbol a_j  \big\},\int_\mathcal{Y} \min_{j\in \mathcal N} \Big\{ \varphi_{T} \Big(  \frac{(T-1)\boldsymbol x_{T-1} +\mathbb{P} \boldsymbol h(y_T) }{T} , \boldsymbol \theta \Big) \boldsymbol a_j  \Big\}   f_0(y_T) d y_T    \Big\} \nonumber\\
&= \int_\mathcal{Y} \min_{j\in \mathcal N} \Big\{ \varphi_{T} \Big(  \frac{(T-1)\boldsymbol x_{T-1} +\mathbb{P} \boldsymbol h(y_T) }{T} , \boldsymbol \theta \Big) \boldsymbol a_j  \Big\}   f_0(y_T) d y_T \nonumber
\end{align}
The inequality above follows from the optimality equation in Proposition~\ref{prop: unnormalizedValueFunction_convex} and the lower bound of~$\mathcal{V}^w_{T-1} (\boldsymbol x_{T-1}; \boldsymbol \theta )$ in \eqref{eq: lowerbound of Vw-657iiwiqj}. The first equality follows from the definition of $\mathcal{V}^j_{T-1}$. The second equality is adapted from \eqref{eq: 722-jjuyiqi714441ugdgsggs} by letting $t=T-1$ and $n=1$. 

Next, suppose the following induction hypothesis hold 
\begin{align}
\mathcal{V}_t(\boldsymbol x_t; \boldsymbol \theta ) 
\geqslant & \int_\mathcal{Y} \ldots  \int_\mathcal{Y} \min_{j\in \mathcal N} \Big\{  \varphi_T \Big(\frac{  t \boldsymbol  x_{t}+ \mathbb{P}  \sum^T_{m=t+1}   \boldsymbol h(y_m)  }{T}, \boldsymbol \theta \Big)   \boldsymbol a_j    \Big\}    \prod^{T}_{m=t+1} f_0(y_m) d y_{t+1} \ldots d y_{T}  . \nonumber
\end{align}
Then, we obtain the following lower bound for the value function of waiting
\begin{align}
\label{eq: 700-jdjkiqi92jks90021i}
&\mathcal{V}^w_{t-1}(\boldsymbol x_{t-1}; \boldsymbol \theta )  \nonumber\\
\geqslant & \int_\mathcal{Y} \mathcal{V}_t \Big( \frac{ (t-1) \boldsymbol  x_{t-1}+ \mathbb{P} \boldsymbol h(y_{t}) }{t} ; \boldsymbol \theta  \Big)  f_0(y_t) d y_t  \nonumber\\
\geqslant & \int_\mathcal{Y} \ldots  \int_\mathcal{Y} \min_{j\in \mathcal N} \Big\{  \varphi_T \Big(\frac{  (t-1) \boldsymbol  x_{t-1}+ \mathbb{P}  \sum^T_{m=t}   \boldsymbol h(y_m)  }{T}, \boldsymbol \theta \Big)   \boldsymbol a_j    \Big\}    \prod^{T}_{m=t} f_0(y_m) d y_{t} \ldots d y_{T} . 
\end{align}
The first inequality is obtained by setting all $c_i$'s to zero, and the second inequality follows from the induction hypothesis. With this lower bound, the value function can also be bounded as
\begin{align}
&\mathcal{V}_{t-1}(\boldsymbol x_{t-1}; \boldsymbol \theta )  \nonumber\\
\geqslant &\min\Big\{   \min_{j\in \mathcal N}\big\{  \varphi_{t-1}(\boldsymbol x_{t-1}, \boldsymbol \theta)    \boldsymbol a_j  \big\}, \int_\mathcal{Y} \ldots  \int_\mathcal{Y} \min_{j\in \mathcal N} \Big\{  \varphi_T \Big(\frac{  (t-1) \boldsymbol  x_{t-1}+ \mathbb{P}  \sum^T_{m=t}   \boldsymbol h(y_m)  }{T}, \boldsymbol \theta \Big)   \boldsymbol a_j    \Big\}  \nonumber\\
& \hspace{2cm} \times   \prod^{T}_{m=t} f_0(y_m) d y_{t} \ldots d y_{T}     \Big\} \nonumber\\
 =& \int_\mathcal{Y} \ldots  \int_\mathcal{Y} \min_{j\in \mathcal N} \Big\{  \varphi_T \Big(\frac{  (t-1) \boldsymbol  x_{t-1}+ \mathbb{P}  \sum^T_{m=t}   \boldsymbol h(y_m)  }{T}, \boldsymbol \theta \Big)   \boldsymbol a_j    \Big\}    \prod^{T}_{m=t} f_0(y_m) d y_{t} \ldots d y_{T} . \nonumber
\end{align}
The inequality follows from the optimality equations as well as from \eqref{eq: 700-jdjkiqi92jks90021i}, and the equality holds because of \eqref{eq: 722-jjuyiqi714441ugdgsggs}.

\subsection*{Part 2 (Upper bound)}
Consider an open-loop policy that chooses waiting at period $t, t+1, \ldots, \tau-1$ and then stopping at period~$\tau$. The expected cost of this policy is 
\begin{align}
\bar{\mathcal{V}}^\tau_t(\boldsymbol x_t; \boldsymbol \theta ) \triangleq
& (\tau-t) [\theta_0 c_0+ \sum^N_{i=1} \theta_i \exp\{t(\beta_{i0}+  \boldsymbol{e}_i \mathbb{R}  \boldsymbol x_t) \} c_i ] \nonumber\\
& + \int_\mathcal{Y} \ldots  \int_\mathcal{Y} \min_{j\in \mathcal N} \Big\{  \varphi_\tau \Big(\frac{  t \boldsymbol  x_{t}+ \mathbb{P}  \sum^\tau_{m=t+1}   \boldsymbol h(y_m)  }{\tau}, \boldsymbol \theta \Big)   \boldsymbol a_j    \Big\}    \prod^{\tau}_{m=t+1} f_0(y_m) d y_{t+1} \ldots d y_\tau  . \nonumber\\
=& (\tau-t) [\theta_0 c_0+ \sum^N_{i=1} \theta_i \exp\{t(\beta_{i0}+  \boldsymbol{e}_i \mathbb{R}  \boldsymbol x_t) \} c_i ]+ \mathcal{L}^\tau_t(\boldsymbol x_t; \boldsymbol \theta ) .\nonumber
\end{align}
The expected cost of the optimal open-loop policy is  
$$
\min_{ t \leqslant \tau  \leqslant T } \{  \bar{\mathcal{V}}^\tau_t(\boldsymbol x_t; \boldsymbol \theta ) \} 
=\min_{ t \leqslant \tau  \leqslant T } \{   (\tau-t) [\theta_0 c_0+ \sum^N_{i=1} \theta_i \exp\{t(\beta_{i0}+  \boldsymbol{e}_i \mathbb{R}  \boldsymbol x_t) \} c_i ]+ \mathcal{L}^\tau_t(\boldsymbol x_t; \boldsymbol \theta )   \},
$$
which is an upper bound of the optimal cost.  \Halmos
  \endproof

\subsection{Subsets of the optimal control regions}

Using the bounds in Lemma~\ref{lemma: bounds on the value functions}, the following proposition identifies sufficient conditions for certain action to be optimal. It essentially characterizes a subset of the corresponding control region in $\Omega_t$.   

 \begin{proposition}
\label{prop: bounds on the control regions}
\mbox{}
\begin{enumerate}
\item It is optimal to wait if 
\begin{align}
\label{eq: sufficient condition for waiting to dominate stopping}
& \min_{ t \leqslant \tau  \leqslant T } \{  (\tau-t) [\theta_0 c_0+ \sum^N_{i=1} \theta_i \exp\{t(\beta_{i0}+  \boldsymbol{e}_i \mathbb{R}  \boldsymbol x_t) \} c_i ]+ \mathcal{L}^\tau_t(\boldsymbol x_t; \boldsymbol \theta )  \}  
 <  \mathcal{L}^t_t(\boldsymbol x_t; \boldsymbol \theta ).
\end{align} 
\item It is optimal to stop and diagnose the type $j$ if $\mathcal{V}^j_t(\boldsymbol x_t; \boldsymbol \theta ) \leqslant \mathcal{V}^k_t(\boldsymbol x_t; \boldsymbol \theta )$, for all $k\neq j$, and
\begin{align}
\label{eq: sufficient condition for stopping to dominate waiting}
& \theta_0 c_0+ \sum^N_{i=1} \theta_i \exp\{t(\beta_{i0}+  \boldsymbol{e}_i \mathbb{R}  \boldsymbol x_t) \} c_i 
\geqslant  \mathcal{V}^j_t(\boldsymbol x_t; \boldsymbol \theta )-\mathcal{L}^T_t (  \boldsymbol  x_t ; \boldsymbol \theta  ) .
\end{align}  
\end{enumerate}
\end{proposition}

  \proof{Proof.}
  We prove the first part by contradiction. Suppose it is optimal to stop, then, by the optimality equations, we must have $\mathcal{V}_t(\boldsymbol x_t; \boldsymbol \theta ) = \min_{j\in \mathcal N}\{\mathcal{V}^j_t(\boldsymbol x_t; \boldsymbol \theta ) \}$. Note that $\mathcal{L}^t_t(\boldsymbol x_t; \boldsymbol \theta )=\min_{j\in \mathcal N}\{\mathcal{V}^j_t(\boldsymbol x_t; \boldsymbol \theta ) \}$, therefore 
  \begin{align}
  \label{eq:676-ejwi83hhhe}
\mathcal{L}^t_t(\boldsymbol x_t; \boldsymbol \theta ) =  \mathcal{V}_t(\boldsymbol x_t; \boldsymbol \theta ) \leqslant  \min_{ t \leqslant \tau  \leqslant T } \{  (\tau-t) [\theta_0 c_0+ \sum^N_{i=1} \theta_i \exp\{t(\beta_{i0}+  \boldsymbol{e}_i \mathbb{R}  \boldsymbol x_t) \} c_i ]+ \mathcal{L}^\tau_t(\boldsymbol x_t; \boldsymbol \theta )  \} ,
  \end{align}
 where the inequality follows from Lemma~\ref{lemma: bounds on the value functions}.  Since \eqref{eq:676-ejwi83hhhe} contradicts with \eqref{eq: sufficient condition for waiting to dominate stopping}, it is therefore never optimal to stop in this case. 
  
 We now prove the second part. When the inequality \eqref{eq: sufficient condition for stopping to dominate waiting} holds, we have
  \begin{align}
 \mathcal{V}^j_t(\boldsymbol x_t; \boldsymbol \theta ) =  & \theta_0 a_{0j}+ \sum^N_{i=1} \theta_i \exp\big\{t (\beta_{i0}+ \boldsymbol{e}_i \mathbb{R}  \boldsymbol x_t) \big \}   a_{ij} \nonumber\\
\leqslant& \theta_0 c_0+ \sum^N_{i=1} \theta_i \exp\{t(\beta_{i0}+  \boldsymbol{e}_i \mathbb{R}  \boldsymbol x_t) \} c_i+\mathcal{L}^T_t (  \boldsymbol  x_t ; \boldsymbol \theta  )   \nonumber\\
=& \theta_0 c_0+ \sum^N_{i=1} \theta_i \exp\{t(\beta_{i0}+  \boldsymbol{e}_i \mathbb{R}  \boldsymbol x_t) \} c_i+\int_\mathcal{Y} \mathcal{L}^T_{t+1} \Big( \frac{  t \boldsymbol  x_t+ \mathbb{P} \boldsymbol h(y_{t+1}) }{t+1} ; \boldsymbol \theta  \Big)  f_0(y_{t+1}) d y_{t+1}   \nonumber\\
\leqslant& \theta_0 c_0+ \sum^N_{i=1} \theta_i \exp\{t(\beta_{i0}+  \boldsymbol{e}_i \mathbb{R}  \boldsymbol x_t) \} c_i+\int_\mathcal{Y} \mathcal{V}_{t+1} \Big( \frac{  t \boldsymbol  x_t+ \mathbb{P} \boldsymbol h(y_{t+1}) }{t+1} ; \boldsymbol \theta  \Big)  f_0(y_{t+1}) d y_{t+1}  \nonumber\\
=& \mathcal{V}^w_t(\boldsymbol x_t; \boldsymbol \theta ) .\nonumber
\end{align}  
The first equality comes from the definition of $ \mathcal{V}^j_t(\boldsymbol x_t; \boldsymbol \theta )$ and the first inequality is \eqref{eq: sufficient condition for stopping to dominate waiting}. The second equality follows from the definition of $\mathcal{L}^T_{t+1}$ and the second inequality holds because $ \mathcal{V}_{t+1}(\boldsymbol x; \boldsymbol \theta ) \geqslant \mathcal{L}^T_{t+1}(\boldsymbol x; \boldsymbol \theta )$, as shown in Lemma~\ref{lemma: bounds on the value functions}. The last equality is by the definition of $\mathcal{V}^w_t(\boldsymbol x_t; \boldsymbol \theta )$. By this, we have shown that stopping and diagnosing the system in state $j$ is less costly than waiting. Further, when $\mathcal{V}^j_t(\boldsymbol x_t; \boldsymbol \theta ) \leqslant \mathcal{V}^k_t(\boldsymbol x_t; \boldsymbol \theta )$ for all $k\neq j$, diagnosing the system in state $j$ incurs the least cost among other stopping alternatives. It is therefore optimal to stop and diagnose the system in type~$j$.~\Halmos
  \endproof

\subsection{$\omega$-convexity and computational benefits}
\label{sec: Omega convexity over time}

\begin{figure}
\centering \includegraphics[width=6.5in]{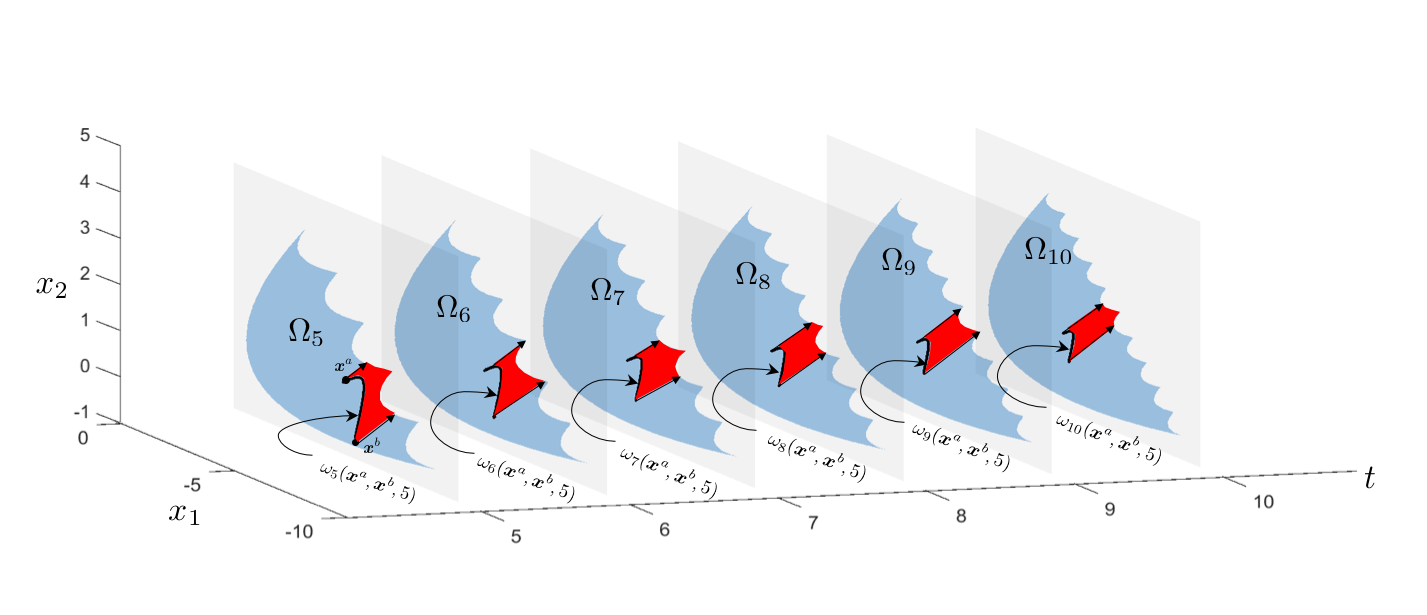} 
\caption{An illustration of $\omega$-convexity propagating through time.}
\label{Fig: OmegaConvexitySlices} 
\end{figure}

\subsection*{Additional interpretation of $\omega$-convexity}

The curve $\{\boldsymbol x \big|  \mathcal{T}^s(\boldsymbol x, \boldsymbol \theta )  =\rho \mathcal{T}^t (\boldsymbol x^a, \boldsymbol \theta) +  (1-\rho) \mathcal{T}^t (\boldsymbol x^b, \boldsymbol \theta) , \rho\in(0,1)  \}$ connecting $\boldsymbol x^a$ and $\boldsymbol x^b$ may intersect  not only with $\Omega_t$ but also with $\Omega_s, s\geqslant t$. In this case, the $\omega$-convexity can ``propagate through time". An illustration is provided in Figure~\ref{Fig: OmegaConvexitySlices}. Suppose we have identified two points, $\boldsymbol x^a, \boldsymbol x^b$, from a stopping region at period $t=5$, then the curve connecting them may intersect with $\Omega_6, \Omega_7,...$ and generate a series of intersection curves $\omega_s(\boldsymbol x^a, \boldsymbol x^b, 5), s=6,7,...$ that belong to the same stopping region in period $t=6, 7,...$ which, by Theorem~\ref{prop: w(x1,x2) belongs to Omegai}, lead to a series of subset regions (highlighted in red). In other words, two points from the same stopping region can characterize a series of subsets of future stopping regions.

\subsection*{Computational benefits}

The $\omega$-convexity can be exploited to reduce the amount of computation. Specifically, once we have identified some sample states from a stopping region, we can use the $\omega$-convexity to identify a set of additional states from the same stopping region. As such, we can reduce the total number of states involved in the value iteration.

In each period of the value iteration, we need to compute the value function $J_t(\boldsymbol x_t; \boldsymbol \theta )$ for all (discretized) states $\boldsymbol x_t$ in $\Omega_t$. To exploit the $\omega$-convexity,  we  first draw $n$ (random) samples $\{\boldsymbol x^{(1)}, \boldsymbol x^{(2)}, \ldots, \boldsymbol x^{(n)} \}$ from $\Omega_t$ and compute the values of $J_t(\boldsymbol x_t; \boldsymbol \theta )$ as well as the optimal actions for these sample states. Then, for each class $i\in\mathcal{N}$, we select the sample states falling on the corresponding stopping region $\Omega_{it}$, denoted by $\{\boldsymbol x^{(1)}_i, \boldsymbol x^{(2)}_i, \ldots, \boldsymbol x^{(n_i)}_i \}$. Following Theorem~\ref{prop: w(x1,x2) belongs to Omegai}, we can use $\{\boldsymbol x^{(1)}_i, \boldsymbol x^{(2)}_i, \ldots, \boldsymbol x^{(n_i)}_i \}$ to characterize a subset of $\Omega_{it}$, which can then be excluded from the remaining computation. Thus, the state space that requires computation is reduced. We can then (randomly) select another set of states from the reduced state space and repeat the above procedure, until the optimal actions for all states in $\Omega_t$ are identified. 

To continue the value iteration, we need to assign the value of $J_t(\boldsymbol x_t; \boldsymbol \theta )$ to each state in $\Omega_t$. For states in the stopping region $ j \in \mathcal N$, the value function is given in closed-form,  
$
J^j_t(\boldsymbol x_t; \boldsymbol \theta ) =\sum^N_{i=0}  \mathcal{T}^t_{i}(\boldsymbol x_t, \boldsymbol \theta) a_{ij}. 
$
These states requires no iterative computation, so that the computational resources can be focused on states in the waiting region. Therefore, it is expected that the computational savings increase with the size of the stopping regions. 

We conduct a series of numerical experiments to measure the computational savings from the $\omega$-convexity. The results are displayed in Table~\ref{Personalized Med: Computational saving from structural results}, in which the parameters are from the personalized-medicine example in~\S\ref{sec: PD example}, and the goal is to quickly identify the subtypes of Parkinson's diseases~(PD). We observe that the computational savings are significant when the delay cost $c$ is large (namely, in the quick-diagnosis regimes), and increase in the horizon length. 

\begin{table}\footnotesize
\centering
\smallskip
\caption{Computational savings from exploiting the $\boldsymbol\omega$-convexity (PD diagnosis example).} 
\label{Personalized Med: Computational saving from structural results}
\begin{threeparttable}
\begin{tabular}{c c c c c c c c c c}
\toprule
 & c=0.001 &c=0.005 & c=0.01 & $c=0.025$ &  c=0.05 & c=0.10 & c=0.15 & c=0.18  \\
\toprule
T=5 & $16.212\%$  & $18.344\%$ & $20.282\%$& $31.044\%$  & $48.560\%$ & $83.870\%$ & $98.400\%$ & $99.998\%$\\
T=10 & $19.709\%$ &	$26.689\%$ &	$32.462\%$ &	$44.781\%$ &	$63.947\%$ &	$90.215\%$ &	$98.789\%$ &	$99.971\%$ \\
T=15 & $23.558\%$ &	$	33.055\%$ &	$	40.975\%$ &	$	54.815\%$ &	$	72.110\%$ &	$	93.166\%$ &	$	99.092\%$ &	$	99.959\%$\\
T=20 & $29.093\%$  & $40.820\%$ & $48.051\%$ & $61.504\%$ & $77.656\%$  & $94.608\%$ & $99.257\%$ & $99.974\%$\\
T=25 & $34.486\%$	& $46.783\%$ & $53.666\%$ & $67.451\%$ & $81.241\%$ & $95.553\%$ & $99.436\%$ & $99.976\%$ \\
T=30 & $39.369\%$  &$51.185\%$  &$58.313\%$  &$71.884\%$  &$84.030\%$  &$96.271\%$  &$99.537\%$  &$99.974\%$\\
T=40 & $48.163\%$  &$58.778\%$  &$65.731\%$  &$77.778\%$  &$87.734\%$  &$97.278\%$  &$99.636\%$  &$99.983\%$\\
T=50 & $54.240\%$  &$65.153\%$  &$71.077\%$  &$81.489\%$  &$89.908\%$  &$97.778\%$  &$99.694\%$  &$99.985\%$\\
T=75 & $65.493\%$  &$74.157\%$  &$79.377\%$  &$87.292\%$  &$93.189\%$  &$98.532\%$  &$99.795\%$  &$99.992\%$\\
T=100 & $72.864\%$  &$80.424\%$  &$84.413\%$  &$90.463\%$  &$94.868\%$  &$98.872\%$  &$99.848\%$  &$99.992\%$\\
\bottomrule
\end{tabular}
\end{threeparttable}
 \end{table}

\subsection{Control-limit representation of a stopping region}
\label{ECsec: connected middle stopping interval}

\begin{proposition}[Control-limit representation]
\label{proposition: connected middle stopping interval}
For $N=2$ and $r=1$, suppose $a_{ii} \leqslant a_{ij}$ for all $i,j=0,1,2$  and, without loss of generality, $\mathbb{R}=[R_1, R_2]^\mathsf{T}$, where $R_1<R_2$. For each period $k$, there exist two thresholds, $\underline{x}^\star_k$  and  $\bar{x}^\star_k$,  such that it is optimal to diagnose type~$1$ if and only if $\underline{x}^\star_k  \leqslant x_k \leqslant \bar{x}^\star_k$.
\end{proposition}  

\emph{Proof.} 
The set of beliefs reachable in $k$ periods is
$$
\mathcal{F}^{\boldsymbol\theta}_k =\{(\pi_{0k}(x), \pi_{1k}(x), 1-\pi_{0k}(x)-\pi_{1k}(x)  )\in S^2 | x \in \Omega_k \}, 
$$
in which 
\begin{align} 
\label{eq: 1112-uhufidsjkekwna}
\pi_{0k} (x)=& \Big( \sum^2_{i=1}\frac{\theta_i}{\theta_0} \exp\big\{
k (\beta_{i0} +  \boldsymbol e_i \mathbb{R}  x  ) \big\} +1 \Big)^{-1}, \nonumber  \\
\pi_{1k} (x)=& \frac{\theta_1 \exp\big\{ k (\beta_{10} +  \boldsymbol e_1 \mathbb{R}  x  ) \big\} }
 {\sum^2_{i=1} \theta_i \exp\big\{ k (\beta_{i0} +  \boldsymbol e_i \mathbb{R}  x  ) \big\} +\theta_0}.
 \end{align}
From now on, we focus on the projection of the belief track on the sub-simplex $S^2_2 \triangleq \{\Pi: \pi_0+\pi_1 \leqslant 1, \pi_2=0 \}$. The projected belief track is defined as
$
\mathcal{F}^2_k =\{(\pi_{0k}(x), \pi_{1k}(x) )\in S^2_2 | x \in \Omega_k \}, 
$
where the coordinates are still specified by \eqref{eq: 1112-uhufidsjkekwna}. For notational simplicity, we write $\pi_{0k} (x), \pi_{1k} (x)$ as $\pi_{0k}, \pi_{1k}$ and $\mathbb{R}=[R_1, R_2]^\mathsf{T}$. Then, \eqref{eq: 1112-uhufidsjkekwna} can be rewritten as
\begin{align} 
\label{eq: 1122-7721899222}
k R_1 x  & = \ln \Big( \frac{\theta_0\pi_{1k}}{\theta_1 \pi_{0k}} \Big)-k \beta_{10},     \\
\label{eq: 1124-9982h191}
\frac{1}{\pi_{0k}}-1 &= \sum^2_{i=1}\frac{\theta_i}{\theta_0} \exp \{k (R_i  x   + \beta_{i0} ) \} .
 \end{align}
 Substituting \eqref{eq: 1122-7721899222} into \eqref{eq: 1124-9982h191}, we obtain
 \begin{align} 
 \label{eq:1127:287ieuuisi112}
\frac{1}{\pi_{0k}} &=\sum^2_{i=1}\frac{\theta_i}{\theta_0}  \Big( \frac{\theta_0\pi_{1k}}{\theta_1\pi_{0k}} \Big)^{\frac{R_i }{R_1}}+C,
 \end{align}
 where 
 $$
 C=1+\sum^2_{i=1}\frac{\theta_i}{\theta_0} \exp\big\{  k \big( \beta_{i0}- \frac{R_i }{R_1} \beta_{10} \big) \big\}
 $$ 
 is a constant independent of $\pi_{0k}$ and $\pi_{1k}$.

 \begin{figure}
\centering \includegraphics[width=2.5in]{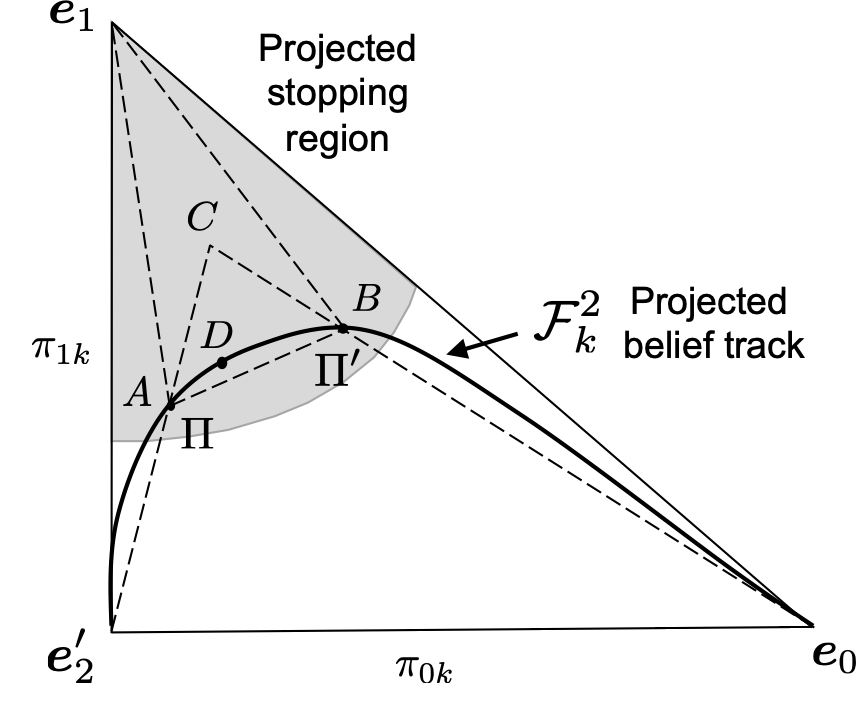} 
\caption{Projection of the stopping region and a belief track on the sub-simplex $S^2_2$.}
\label{projectedBeliefTrack} 
\end{figure}

Equation~\eqref{eq:1127:287ieuuisi112} specifies the relation between $\pi_{1k}$ and $\pi_{0k}$ on the projected curve. For a given $\pi_{0k}$, it is easy to observe that the right-hand side of~\eqref{eq:1127:287ieuuisi112} is increasing in $\pi_{1k}$. Therefore, there is a unique~$\pi_{1k}$ corresponding to each value of $\pi_{0k}$, as illustrated in Figure~\ref{projectedBeliefTrack}. With a slight abuse of notation, we use the function $\pi_{1k}(\pi_{0k})$ to represent the projected curve.  Implicit differentiation of \eqref{eq:1127:287ieuuisi112} with respect to $\pi_{0k}$ leads to 
   \begin{align} 
-1 &=\sum^2_{i=1} \frac{\theta_i R_i }{\theta_1  R_1}  \Big( \frac{\theta_0 \pi_{1k}}{\theta_1 \pi_{0k}} \Big)^{\frac{R_i }{R_1}-1} \Big( \frac{d \pi_{1k}}{d \pi_{0k}}  \pi_{0k}-\pi_{1k} \Big). \nonumber
 \end{align}
The second-order implicit differentiation gives
   \begin{align} 
0&=\sum^2_{i=1} \frac{\theta_iR_i }{\theta_1 R_1} \frac{(R_i -R_1)}{R_1} \Big( \frac{\theta_0 \pi_{1k}}{\theta_1 \pi_{0k}} \Big)^{\frac{R_i }{R_1}-2} \frac{\theta_0}{\theta_1}  \frac{(\frac{d \pi_{1k}}{d \pi_{0k}}  \pi_{0k}- \pi_{1k})^2}{(\pi_{0k})^2} 
+\Big[ \sum^2_{i=1} \frac{\theta_i R_i }{\theta_1 R_1}  \Big( \frac{\theta_0 \pi_{1k}}{\theta_1 \pi_{0k}} \Big)^{\frac{R_i }{R_1}-1} \pi_{0k} \Big] \frac{d^2 \pi_{1k}}{d (\pi_{0k})^2}  \nonumber\\
&= \frac{\theta_2 R_2 }{\theta_1 R_1} \frac{(R_2  -R_1)}{R_1} \Big( \frac{\theta_0 \pi_{1k}}{\theta_1 \pi_{0k}} \Big)^{\frac{R_2 }{R_1}-2} \frac{\theta_0}{\theta_1}  \frac{(\frac{d \pi_{1k}}{d \pi_{0k}}  \pi_{0k}- \pi_{1k})^2}{(\pi_{0k})^2} 
+\Big[ \sum^2_{i=1} \frac{\theta_i R_i }{\theta_1 R_1}  \Big( \frac{\theta_0 \pi_{1k}}{\theta_1 \pi_{0k}} \Big)^{\frac{R_i }{R_1}-1} \pi_{0k} \Big] \frac{d^2 \pi_{1k}}{d (\pi_{0k})^2}  . \nonumber
 \end{align}
 Note that $\pi_{0k} >0$ and $\pi_{1k}>0$. When $R_2>R_1$, we observe from the above equation that
 $$
\Big[ \sum^2_{i=1} \frac{\theta_i R_i }{\theta_1 R_1}  \Big( \frac{\theta_0 \pi_{1k}}{\theta_1 \pi_{0k}} \Big)^{\frac{R_i }{R_1}-1} \pi_{0k} \Big] \frac{d^2 \pi_{1k}}{d (\pi_{0k})^2}  <0,
 $$
 which implies $\frac{d^2 \pi_{1k}}{d (\pi_{0k})^2}  <0$, suggesting that the function $\pi_{1k}(\pi_{0k})$ is concave in $\pi_{0k}$.

By~\eqref{eq:1127:287ieuuisi112} we have
 $
 \lim_{x\to-\infty} \pi_{0k} (x)=1,   \lim_{x\to-\infty} \pi_{1k} (x)=0;   \lim_{x\to \infty} \pi_{0k} (x)=0,    \lim_{x\to\infty} \pi^k_2 (x)=1.
 $
Therefore, when $x$ is unbounded, the vertex $\boldsymbol e_0$ and $\boldsymbol e_2$ lies on the belief track $\mathcal{F}_k$. The projected belief track passes through~$\boldsymbol e_0$ and $\boldsymbol e_2$'s projection, denoted by $\boldsymbol e'_2$, as illustrated in Figure~\ref{projectedBeliefTrack}. 
 
Since the stopping region $\Gamma_{1t}$ is convex (see the proof of Theorem~\ref{prop: w(x1,x2) belongs to Omegai}), its projection on the sub-simplex $S^2_2$, namely, $\{ (\pi_0,\pi_1) \in S^2_2: V_{t}(\pi_0,\pi_1,1-\pi_0-\pi_1) =S^1_t(\pi_0,\pi_1,1-\pi_0-\pi_1) \}$ is also convex. The projected stopping region is shown as the shaded region in Figure~\ref{projectedBeliefTrack}. 
 
Suppose belief states $\Pi=(\pi_{0k} (x), \pi_{1k} (x), 1-\pi_{0k} (x)-\pi_{1k}(x))$ and $\Pi'=(\pi_{0k} (x'), \pi_{1k} (x'), 1-\pi_{0k} (x')-\pi_{1k}(x'))$ are both in the stopping region $\Gamma_{1t}$. The projection of $\Pi$ on $S^2_2$ is represented by point $A$ in Figure~\ref{projectedBeliefTrack} and the projection of $\Pi'$ is represented by point $B$. We will show that, for $\rho\in[0,1]$, the belief state $(\pi_{0k} (\rho x+(1-\rho)x'), \pi_{1k} (\rho x+(1-\rho)x'), 1-\pi_{0k} (\rho x+(1-\rho)x')-\pi_{1k}(\rho x+(1-\rho)x'))$, represented by point $D$, also lies in the stopping region $\Gamma_{1t}$. Toward this end, we make use of some geometric properties of the belief track and the stopping region. Recall that the projected belief track $\mathcal{F}^2_k$ passes through $\boldsymbol e'_2, A, B$ and $\boldsymbol e_0$ (see Figure~\ref{projectedBeliefTrack}). Since $\mathcal{F}^2_k$ is concave, point $D$ must stay within the triangle $ABC$. Note further that $A, B$ and $\boldsymbol e_1$ all lie on the convex stopping region $\Gamma_{1t}$. Therefore, $D\in ABC \subset AB\boldsymbol e_1 \subset \Gamma_{1t}$, thereby completing the proof. \Halmos

\subsection{The counter-intuitive properties of the stopping regions}
\label{sec: Counter-intuitive Properties of the Stopping Regions}
\begin{figure}
\centering \includegraphics[width=6.0in]{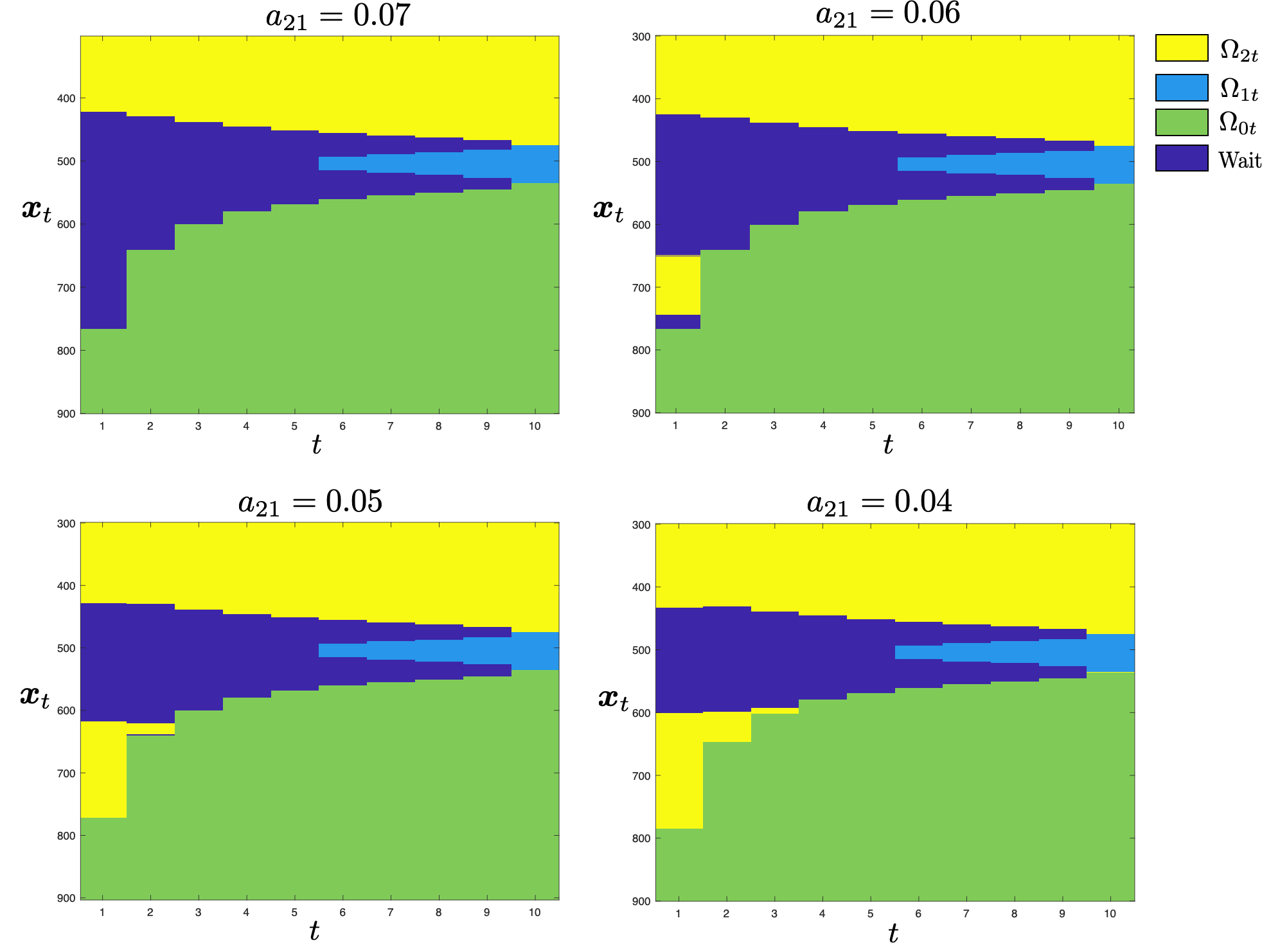} 
\caption{Exploration of the noncontiguous behavior of the stopping regions. Parameters are $N=2$, $c=0.03$, $\boldsymbol\theta=(1/3,1/3,1/3)$, $(\mu_0, \mu_1, \mu_2)=(-2, 0, 1.5)$, $\sigma^2_0=\sigma^2_1=\sigma^2_2=2.4$. The termination costs are zero-one except that $a_{12}=5$ and $a_{21}$ is varied from 0.07 to 0.04.}
\label{Fig: ExclaveExplore} 
\end{figure}

We have seen from Figure~\ref{Fig: DisconnectedOmega}--(a) in the main text that the stopping region can be noncontiguous, which occurs when the termination costs are asymmetric (namely, class-specific). In Figure~\ref{Fig: ExclaveExplore}, we change the degree of asymmetry by varying the misdiagnosis cost $a_{21}$. As $a_{21}$ decreases from 0.07 to 0.04, the cost asymmetry increases and the exclave stopping region of $\Omega_{2t}$ enlarges, suggesting that the asymmetry in termination cost is an important driver of the noncontiguous behavior of the stopping regions.

We have also seen in Figure~\ref{Fig: DisconnectedOmega}--(b) that the optimal policy may be non-monotone in time. That is, if the same observation keeps reappearing, the precision of inference will increase and hence the expected misdiagnosis cost will decrease. However, the marginal value of information does not necessarily decrease with the sample size. In other words, the marginal value of information may actually increase as more samples are collected. In this situation, the expected cost of waiting may become lower than stopping, and, hence, the optimal action may change from ``stop" back to ``wait".  

\begin{figure}
\centering \includegraphics[width=6.2in]{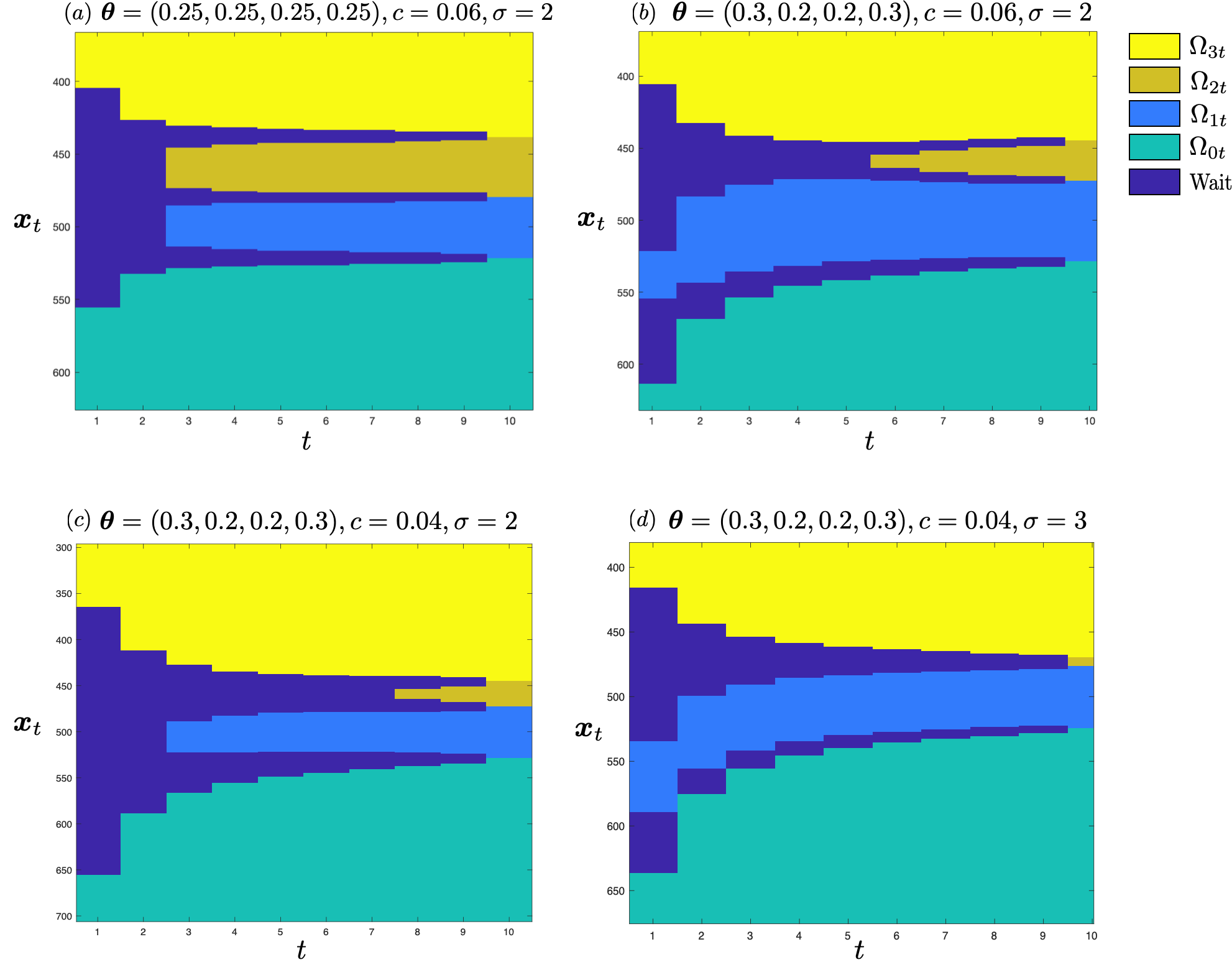} 
\caption{Exploration of the time-monotonicity of the optimal policy under a fixed $\boldsymbol x$ (zero-one termination cost; $T=10$; $\mu_0=-1, \mu_1=0, \mu_2=1, \mu_3=2$; $N=3$). }
\label{Fig: NonMonotoneInTime} 
\end{figure}

In Figure~\ref{Fig: NonMonotoneInTime}, we explore the non-monotonicity of the optimal policy in time (when the diagnostic statistic, $\boldsymbol x$, remains fixed). Under the uninformative prior, the optimal policy exhibits time-monotonicity in Figure~\ref{Fig: NonMonotoneInTime}--(a), but the non-monotonicity appears when the prior becomes informative, as shown in panel (b). Under the same prior, a lower delay cost reduces the non-monotonicity in panel (c), but it reappears when the observations become noisier, as shown in Figure~\ref{Fig: NonMonotoneInTime}--(d). To summarize, the non-monotonicity is associated with informative priors, higher delay costs, and noisier observations. These situations are largely in the non-asymptotic regimes, where the information from a single observation may substantially influence decisions.

\subsection{A sufficient condition for finite critical stopping time} 
\label{ECsubsec: sufficient condition for critical stopping time}

The following lemma gives a sufficient condition for the existence of a finite critical stopping time. 

\begin{lemma}
\label{theorem: cutoff horizon}
If there exists a time $t_c$ such that 
\begin{align}
\label{eq: sufficient condition for stopping to dominate waiting-2}
& \theta_0 c_0+ \sum^N_{i=1} \theta_i \exp\{t_c (\beta_{i0}+  \boldsymbol{e}_i \mathbb{R}  \boldsymbol x) \} c_i + \mathcal{L}^T_{t_c} (  \boldsymbol  x; \boldsymbol \theta  ) 
\geqslant  \mathcal{L}^{t_c}_{t_c} (\boldsymbol x; \boldsymbol \theta ), 
\end{align}  
for all $\boldsymbol x \in\Omega_{t_c}$, then the decision process must terminate no later than $t_c$. 
\end{lemma}

\proof{Proof.}
Note that
$
\mathcal{L}^{t_c}_{t_c}(\boldsymbol x; \boldsymbol \theta ) =  \min_{ j\in \mathcal N}\{ \mathcal{V}^j_{t_c}(\boldsymbol x; \boldsymbol \theta )  \big\}.
$
Let  $j^*= \argmin_{ j\in \mathcal N}\{ \mathcal{V}^j_{t_c}(\boldsymbol x; \boldsymbol \theta )  \big\}$, namely, $\mathcal{V}^{j^*}_{t_c}(\boldsymbol x; \boldsymbol \theta ) \leqslant \mathcal{V}^k_{t_c}(\boldsymbol x; \boldsymbol \theta )$, for all $k\neq j^*$, which satisfies the first condition in part 2 of Proposition~\ref{prop: bounds on the control regions}. In addition, since $\mathcal{L}^{t_c}_{t_c}(\boldsymbol x; \boldsymbol \theta ) = \mathcal{V}^{j^*}_{t_c}(\boldsymbol x; \boldsymbol \theta ) $, \eqref{eq: sufficient condition for stopping to dominate waiting-2} can be written as  
$
 \theta_0 c_0+ \sum^N_{i=1} \theta_i \exp\{t_c (\beta_{i0}+  \boldsymbol{e}_i \mathbb{R}  \boldsymbol x) \} c_i  
\geqslant \mathcal{V}^{j^*}_{t_c}(\boldsymbol x; \boldsymbol \theta ) - \mathcal{L}^T_{t_c} (  \boldsymbol  x; \boldsymbol \theta  ), 
$
which satisfies the second condition \eqref{eq: sufficient condition for stopping to dominate waiting} in Proposition~\ref{prop: bounds on the control regions}. Therefore, it is optimal to stop and diagnose type~$j^*$. When the condition  \eqref{eq: sufficient condition for stopping to dominate waiting-2} is satisfied for all $\boldsymbol x \in\Omega_{t_c}$, then the decision process must terminate at time $t_c$.  \Halmos
\endproof

\section{Proof of Theorem~\ref{corollary: cutoff horizon in MSPRT}}

\proof{Proof.}
Under the zero-one cost structure and uninformative prior, we have
$$
\frac{\mathcal{L}^t_t (\boldsymbol x; \boldsymbol \theta )}{ \alpha_t(\boldsymbol  x; \boldsymbol \theta)} = 1-\frac{\exp[ \max_j \{t(\beta_{j0}+\boldsymbol e_j \mathbb{R} \boldsymbol x) \} ]}{\sum^N_{i=0} \exp[  t(\beta_{i0}+\boldsymbol e_i \mathbb{R} \boldsymbol x)  ]}.
$$
Let $j^*(\boldsymbol x)$ denote the optimal action for $\boldsymbol x$,  which implies that 
\begin{align}
\label{eq: 1368-hiudhfiuhwieouhiuwe}
\beta_{i0}+\boldsymbol e_i \mathbb{R} \boldsymbol x - \beta_{j^*(\boldsymbol x)0}-\boldsymbol e_{j^*(\boldsymbol x)} \mathbb{R} \boldsymbol x \leqslant  0, 
\end{align}
for all $i\neq j^*$. Define set $\Delta_{xt}=\{ \boldsymbol x \in \Omega_t |  \beta_{i0}-\beta_{j^*(\boldsymbol x)0}=(\boldsymbol e_{j^*(\boldsymbol x)}-\boldsymbol e_i ) \mathbb{R} \boldsymbol x, \text{ for some } i \neq  j^*(\boldsymbol x) \in\mathcal{N}\}$, which represents the boundaries  on which the type $i$ and type $j^*$ have the same probability. According to the definition of $n_s$, there are at most $n_s$ types sharing the same density.

 \begin{figure}
\centering \includegraphics[width=5.0in]{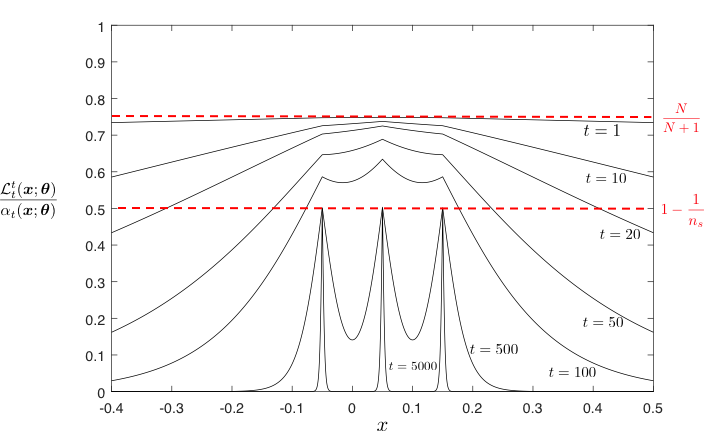} 
\caption{An illustration of the proof of Theorem~\ref{corollary: cutoff horizon in MSPRT} for $r=1, N=3$ and $n_s=2$.}
\label{Appendix_Proof_LAdT} 
\end{figure}

We first prove that $\frac{\mathcal{L}^t_t (\boldsymbol x; \boldsymbol \theta )}{ \alpha_t(\boldsymbol  x; \boldsymbol \theta)}$ is decreasing in $t$ and $\lim_{t\to \infty}  \frac{\mathcal{L}^t_t (\boldsymbol x; \boldsymbol \theta )}{ \alpha_t(\boldsymbol  x; \boldsymbol \theta)}=0$ for $ \boldsymbol x \notin  \Delta_{xt} $. Note that $\frac{\mathcal{L}^t_t (\boldsymbol x; \boldsymbol \theta )}{ \alpha_t(\boldsymbol  x; \boldsymbol \theta)}$ can be written as follows
$$
\frac{\mathcal{L}^t_t (\boldsymbol x; \boldsymbol \theta )}{ \alpha_t(\boldsymbol  x; \boldsymbol \theta)} = 1-\frac{1}{1+ \sum_{i \neq j^*_t (\boldsymbol x)} \exp[  t(\beta_{i0}+\boldsymbol e_i \mathbb{R} \boldsymbol x -  \beta_{j^*_t(\boldsymbol x)0}-\boldsymbol e_{j^*_t(\boldsymbol x)} \mathbb{R} \boldsymbol x)  ]}.
$$
For $\boldsymbol x \notin \Delta_{xt}$, the inequality~\eqref{eq: 1368-hiudhfiuhwieouhiuwe} is strict for all $i$. Therefore, $ \frac{\mathcal{L}^t_t (\boldsymbol x; \boldsymbol \theta )}{ \alpha_t(\boldsymbol  x; \boldsymbol \theta)}$  is decreasing in $t$ and converges to $0$ as~$t\to\infty$. We divide both sides of \eqref{eq: sufficient condition for stopping to dominate waiting-2} by the normalization term $ \alpha_t(\boldsymbol  x; \boldsymbol \theta)$ and obtain
\begin{align}
&c  + \frac{  \mathcal{L}^T_t (  \boldsymbol  x; \boldsymbol \theta  )  }{ \alpha_t(\boldsymbol  x; \boldsymbol \theta)   }  \geqslant  \frac{  \mathcal{L}^t_t(\boldsymbol x; \boldsymbol \theta )   } { \alpha_t(\boldsymbol  x; \boldsymbol \theta) } . \nonumber
\end{align}  
Since $\frac{\mathcal{L}^t_t (\boldsymbol x; \boldsymbol \theta )}{ \alpha_t(\boldsymbol  x; \boldsymbol \theta)}$ is decreasing in $t$ and $\lim_{t\to \infty}  \frac{\mathcal{L}^t_t (\boldsymbol x; \boldsymbol \theta )}{ \alpha_t(\boldsymbol  x; \boldsymbol \theta)}=0$,  for  $\boldsymbol x \notin \Delta_{xt}$, there must exist a finite threshold $t'<\infty$ such that $c \geqslant  \frac{\mathcal{L}^t_t (\boldsymbol x; \boldsymbol \theta )}{ \alpha_t(\boldsymbol  x; \boldsymbol \theta)}$ for all $t>t'$. In addition, since $\frac{ \mathcal{L}^T_t (  \boldsymbol  x; \boldsymbol \theta  )}{\alpha_t(\boldsymbol  x; \boldsymbol \theta)}>0$, it follows immediately that $ c   \geqslant  \frac{\mathcal{L}^t_t (\boldsymbol x; \boldsymbol \theta )}{ \alpha_t(\boldsymbol  x; \boldsymbol \theta)} -  \frac{ \mathcal{L}^T_t (  \boldsymbol  x; \boldsymbol \theta  )}{\alpha_t(\boldsymbol  x; \boldsymbol \theta)}$, for all $t>t'$, which satisfies the condition~\eqref{eq: sufficient condition for stopping to dominate waiting-2} in Lemma~\ref{theorem: cutoff horizon}. Therefore, the decision process must terminate in finite time, for $\boldsymbol x \notin \Delta_{xt}$. 
 
Next, we consider $\boldsymbol x \in \Delta_{xt}$. By the definition of $n_s$, we observe that 
$$
\lim_{t\to\infty} \sum_{i \neq j^*_t (\boldsymbol x)} \exp[  t(\beta_{i0}+\boldsymbol e_i \mathbb{R} \boldsymbol x -  \beta_{j^*_t(\boldsymbol x)0}-\boldsymbol e_{j^*_t(\boldsymbol x)} \mathbb{R} \boldsymbol x)  ] \leqslant  n_s-1,
$$
and hence $$
\lim_{t\to\infty}\frac{\mathcal{L}^t_t (\boldsymbol x; \boldsymbol \theta )}{ \alpha_t(\boldsymbol  x; \boldsymbol \theta)} \leqslant 1-\frac{1}{n_s},
$$
for all $\boldsymbol x \in \Delta_{xt}$. An illustration is given in Figure~\ref{Appendix_Proof_LAdT}. Since $ \frac{\mathcal{L}^t_t (\boldsymbol x; \boldsymbol \theta )}{ \alpha_t(\boldsymbol  x; \boldsymbol \theta)}$  is decreasing in $t$, when $c>1-1/n_s$, there must exist a finite threshold $t'<\infty$ such that $c \geqslant  \frac{\mathcal{L}^t_t (\boldsymbol x; \boldsymbol \theta )}{ \alpha_t(\boldsymbol  x; \boldsymbol \theta)}$ for all $t>t'$ and $\boldsymbol x \in \Delta_{xt}$. By the same argument as given earlier, the decision process must terminate in finite time. 

Finally, we prove that it is optimal to stop at $t=0$ if $c\geqslant N/(N+1)$.  This is obvious because when $t=0$, we have 
$$
\frac{\mathcal{L}^0_0 (\boldsymbol x; \boldsymbol \theta )}{ \alpha_0(\boldsymbol  x; \boldsymbol \theta)} = \frac{N}{ N+1};
$$
it is therefore optimal to stop when $c\geqslant N/(N+1)$. \Halmos
\endproof

\section{Details of the Medical Application in \S\ref{sec: PD example}}
\label{EC: Details of the medical application}

The observation variable, MoCA, follows a normal distribution $y\sim f_i=N(\mu_i, \sigma_i)$ that depends on the patient subtype~$i$. Specifically, 
$
\mu_0=27.98, \sigma_0=1.86,  \mu_1=27.09, \sigma_1=2.4, \mu_2=24.62, \sigma_2=4.06. 
$
Therefore, we have three simple hypothese: 
$$
H_0: y\sim N(27.98, 1.86), \ \ H_1: y\sim N(27.09, 2.4), \ \ H_2: y\sim N(24.62, 4.06),
$$
which correspond to subtype 1, 2, 3, respectively.  Using the expression of normal distribution in Table\ref{list_distributions-ETM_EF}, we obtain the tilting parameters as follows 
\begin{align}
\boldsymbol\beta^\mathsf{T}_1 &= \Big(\frac{\mu_1}{\sigma^2_1}-\frac{\mu_0}{\sigma^2_0}, \frac{1}{2\sigma^2_0}-\frac{1}{2\sigma^2_1} \Big) =  ( -3.3845, \    0.0577), \ \beta_{10}= 49.1874,\nonumber\\
\boldsymbol\beta^\mathsf{T}_2 &= \Big(\frac{\mu_2}{\sigma^2_2}-\frac{\mu_0}{\sigma^2_0}, \frac{1}{2\sigma^2_0}-\frac{1}{2\sigma^2_2} \Big) =  (-6.5940, \    0.1142), \ \beta_{20}= 93.9792. \nonumber
\end{align}
Next, we describe how to specify the prior. The prevalence of subtype 1 is 43.13\%, subtype 2 is 22.96\%, and subtype 3 is 33.91\%. Therefore, before knowing the patient's age and gender, we form an original prior 
$
(\gamma_0, \gamma_1, \gamma_2)=(0.4312, 0.2296, 0.3391). 
$
Consider a 65-year-old male patient. Given the patient's gender and age, we use Bayes' rule to obtain
$$
\theta_0=\Pr(s=0 |\text{Male, 65 yo})= \frac{\gamma_0 \Pr(\text{Male} |s=0) \Pr(\text{65 yo} |s=0)}{ \sum^2_{i=0} \gamma_i \Pr(\text{Male} | s=i) \Pr(\text{65 yo} | s=i)}.
$$
In the above, we assume that the age and gender are conditionally independent since their correlation is not reported in~\cite{zhang2019data}. According to Table \ref{Personalized Med: Class Parameter Table}, 
$$
\Pr(\text{Male} |s=0)=0.636, \Pr(\text{Male} |s=1)=0.588, \Pr(\text{Male} |s=2)=0.689. 
$$
$\Pr(\text{65 yo} |s=0)$ is the probability density of $N(62.66, 9.55)$ evaluated at $65$, $\Pr(\text{65 yo} |s=1)$ is the probability density of $N(64.61, 9.20)$ evaluated at $65$, and $\Pr(\text{65 yo} |s=2)$ is the probability density of $N(69.16, 8.82)$ at $65$. Using these inputs, we obtain $(\theta_0, \theta_1, \theta_2)=(0.4208, 0.2214, 0.3579)$, which will be used as the prior for sequential diagnosis.

\section{Proof of Proposition~\ref{prop: Bound on performance loss}}
\label{EC proof: bound on performance loss}
\proof{Proof.}

Consider the original optimality equation~\eqref{eq: finite-beliefstate-OE} for $t=0$, 
\begin{align}
\label{ECeq:888-hjdh}
V_0(\boldsymbol\theta)&=\min\Big\{\sum^N_{i=0} \theta_i a_{i0}, \sum^N_{i=0} \theta_i a_{i1}, \ldots, \sum^N_{i=0} \theta_i a_{iN}, V^w_0(\boldsymbol\theta)  \Big\}, 
\end{align}
where $V^w_0(\boldsymbol\theta)= \sum^N_{i=0} \theta_i c_{i}+\int_{\mathcal{Y}} J_1(\mathbb{P}  \boldsymbol h(y) ; \boldsymbol \theta )  \boldsymbol\theta F(y) dy$. Since $V_0(\boldsymbol\theta)=J^*(\boldsymbol\theta)$ by definition, our goal is to prove that $J^{lb}(\boldsymbol\theta) \leqslant  V_0(\boldsymbol\theta) $. The key inequality comes from Lemma~\ref{lemma: bounds on the value functions}, which leads to
\begin{align}
\label{eq: EC-892-jjks}
J_1(\mathbb{P}  \boldsymbol h(y); \boldsymbol \theta ) =\frac{\mathcal{V}_1   \big(\mathbb{P}  \boldsymbol h(y); \boldsymbol\theta   \big)}{\alpha_1( \mathbb{P}  \boldsymbol h(y) ;\boldsymbol\theta )}  \geqslant \frac{\mathcal{L}^T_1(\mathbb{P}  \boldsymbol h(y); \boldsymbol \theta ) } {\alpha_1( \mathbb{P}  \boldsymbol h(y) ;\boldsymbol\theta )},
\end{align}
where $
\alpha_1( \mathbb{P}  \boldsymbol h(y);\boldsymbol\theta )  \triangleq  \theta_0+ \sum^N_{i=1}  \theta_i  \exp \{ \beta_{i0}+ \boldsymbol{e}_i \mathbb{B} \boldsymbol h(y) \},
$ 
and 
\begin{align}
\label{ECEq: 900-hkdsu}
\mathcal{L}^T_1(\mathbb{P}  \boldsymbol h(y); \boldsymbol \theta ) \triangleq & 
 \int  \min_{j\in \mathcal N} \Big\{   \theta_0 a_{0j} +  \sum^N_{i=1} \theta_i \exp\big\{T \beta_{i0}+ \boldsymbol{e}_i \mathbb{B}  (\boldsymbol h(y) +  \boldsymbol h    ) \big \} a_{ij}  \Big\}   \boldsymbol g^{*(T-1)}(\boldsymbol h)  d \boldsymbol h. \nonumber
\end{align}
The inequality~\eqref{eq: EC-892-jjks} leads to a lower bound of $V^w_0(\boldsymbol\theta)$ given below
\begin{align}
V^w_0(\boldsymbol\theta) 
\geqslant & \sum^N_{i=0} \theta_i c_{i} + \int_{\mathcal{Y}} \frac{\mathcal{L}^T_1(\mathbb{P}  \boldsymbol h(y); \boldsymbol \theta ) } {\alpha_1( \mathbb{P}  \boldsymbol h(y) ;\boldsymbol\theta )}  \boldsymbol\theta F(y) dy \nonumber\\
=&  \sum^N_{i=0} \theta_i c_{i} + \int_{\mathcal{Y}}  \int  \min_{j\in \mathcal N} \Big\{ \frac{  \theta_0 a_{0j} +  \sum^N_{i=1} \theta_i \exp\big\{T \beta_{i0}+ \boldsymbol{e}_i \mathbb{B}  (\boldsymbol h(y) +  \boldsymbol h    ) \big \} a_{ij} } {\theta_0+ \sum^N_{i=1}  \theta_i  \exp \{ \beta_{i0}+ \boldsymbol{e}_i \mathbb{B} \boldsymbol h(y) \}}  \Big\}   \boldsymbol g^{*(T-1)}(\boldsymbol h)  d \boldsymbol h  \nonumber\\
&\hspace{1in} \times  \{ \theta_0 f_0(y) + \theta_1 f_1(y)+ \cdots + \theta_N f_N(y) \} dy . \nonumber\\
=&  \sum^N_{i=0} \theta_i c_{i} + \int_{\mathcal{Y}}  \int  \min_{j\in \mathcal N} \Big\{ \frac{  \theta_0 a_{0j} +  \sum^N_{i=1} \theta_i \exp\big\{T \beta_{i0}+ \boldsymbol{e}_i \mathbb{B}  (\boldsymbol h(y) +  \boldsymbol h    ) \big \} a_{ij} } {\theta_0+ \sum^N_{i=1}  \theta_i  \exp \{ \beta_{i0}+ \boldsymbol{e}_i \mathbb{B} \boldsymbol h(y) \}}  \Big\}   \boldsymbol g^{*(T-1)}(\boldsymbol h)  d \boldsymbol h  \nonumber\\
&\hspace{1in} \times \{ \theta_0  + \sum^N_{i=1}  \theta_i  \exp \{ \beta_{i0}+ \boldsymbol{e}_i \mathbb{B} \boldsymbol h(y) \} \} f_0(y)dy  \nonumber\\
=&  \sum^N_{i=0} \theta_i c_{i} + \int_{\mathcal{Y}}  \int  \min_{j\in \mathcal N} \Big\{  \theta_0 a_{0j} +  \sum^N_{i=1} \theta_i \exp\big\{T \beta_{i0}+ \boldsymbol{e}_i \mathbb{B}  (\boldsymbol h(y) +  \boldsymbol h    ) \big \} a_{ij}   \Big\}   \boldsymbol g^{*(T-1)}(\boldsymbol h)  d \boldsymbol h  f_0(y)dy  \nonumber\\
=& \sum^N_{i=0} \theta_i c_{i} + \int  \min_{j\in \mathcal N} \Big\{   \theta_0 a_{0j} +  \sum^N_{i=1} \theta_i \exp\big\{T \beta_{i0}+ \boldsymbol{e}_i \mathbb{B}   \boldsymbol h     \big \} a_{ij}  \Big\}   \boldsymbol g^{*T}(\boldsymbol h)  d \boldsymbol h \nonumber\\
=& \sum^N_{i=0} \theta_i c_{i} + \mathcal{L}^T_0(\boldsymbol \theta ). 
\end{align}
The first equality follows from the definitions of $\alpha_1( \mathbb{P}  \boldsymbol h(y);\boldsymbol\theta )$, $\mathcal{L}^T_1(\mathbb{P}  \boldsymbol h(y); \boldsymbol \theta )$, and $F(y)$. The second equality follows from the ETM assumption, $f_i(y)=f_0(y) \exp\{ \beta_{i0}+ \boldsymbol\beta^\mathsf{T}_i \boldsymbol h(y)  \}$. The third equality is simply the reduction of fraction, and the fourth equality is based on the definition of convolution power. Finally, note that $J^{lb}(\boldsymbol\theta)$ is obtained by replacing $V^w_0(\boldsymbol\theta)$ in \eqref{ECeq:888-hjdh} with its lower bound in the last line of~\eqref{ECEq: 900-hkdsu}. Hence, we have $J^{lb}(\boldsymbol\theta) \leqslant V_0(\boldsymbol\theta)$, or equivalently, $J^{lb}(\boldsymbol\theta) \leqslant J^*(\boldsymbol\theta)$, thereby proving $ J^{\pi_{abr}}(\boldsymbol\theta)-J^*(\boldsymbol\theta) \leqslant J^{\pi_{abr}}(\boldsymbol\theta)-J^{lb}(\boldsymbol\theta)$. \Halmos

%

\endproof

\section{Proof of Proposition~\ref{prop: Approx low-rank approximation characterize a subset of the waiting region}}
\proof{Proof.}
Since the dimension reduction loses information, the corresponding expected cost-to-go is higher than the optimal cost. That is, $\hat J^w_t \geqslant J^w_t$. Therefore, if $\hat J^w_t =\min\{J^0_t,\ldots, J^N_t, \hat J^w_t \}$, then we also have $J^w_t =\min\{J^0_t,\ldots, J^N_t, J^w_t \}$, suggesting that it is optimal to wait. The inequalities follow from the arguments after the proposition in the main body. \Halmos
\endproof

\section{Proof of Proposition~\ref{prop: LRA interpretation-MVN-Equal covariance}}

\proof{Proof.}

The solution to the problem~\eqref{eq: Rayleigh quotient} is obtained by an eigendecomposition of~$ \mathbb{S}^* $. The maximization is achieved by $v_1$, the first eigenvector of $ \mathbb{S}^*$. The second eigenvector $v_2$ is orthogonal to $v_1$ and maximizes~$v^\mathsf{T}_2 \mathbb{S}^* v_2 /{v^\mathsf{T}_2 v_2}$, and so on. Next, we find an eigendecomposition of $ \mathbb{S}^*$. Toward this end, we first define  
\begin{align}
\mathbb{B}_d \triangleq
\begin{bmatrix}
&(\boldsymbol\mu^\mathsf{T}_1-\boldsymbol\mu^\mathsf{T}_0) \boldsymbol\Sigma^{-1} \\
&\vdots   \\
&(\boldsymbol\mu^\mathsf{T}_N-\boldsymbol\mu^\mathsf{T}_0) \boldsymbol\Sigma^{-1} \\
\end{bmatrix}
,
 \nonumber
\end{align}
with singular value decomposition $\mathbb{B}_d=U_d \Sigma_d V^\mathsf{T}_d$. A rank decomposition of $\mathbb{B}_d$ is 
\begin{align}
\mathbb{B}_d=U_d \Sigma_d V^\mathsf{T}_d
=
\begin{bmatrix}
U_r &U_{N-r} 
\end{bmatrix}
\begin{bmatrix}
&\Sigma_r & \boldsymbol 0 \\
&\boldsymbol 0 & \boldsymbol 0
\end{bmatrix}
\begin{bmatrix}
& V^\mathsf{T}_r \\
&V^\mathsf{T}_{p-r}
\end{bmatrix}
=U_r ( \Sigma_r V^\mathsf{T}_r)=\mathbb{R}_d \mathbb{P}_d, \nonumber
\end{align}
where $\mathbb{R}_d$ is the first~$r$ columns of $U_d$, and $\mathbb{P}_d=\Sigma_r V^\mathsf{T}_r$ is the first $r$ rows of~$\Sigma_d V^\mathsf{T}_d$. 
Then, the diagnostic parameter matrix can be written as 
\begin{align}
\mathbb{B}=
\begin{bmatrix}
&(\boldsymbol\mu^\mathsf{T}_1-\boldsymbol\mu^\mathsf{T}_0) \boldsymbol\Sigma^{-1},   &0  &\cdots  &0 \\
&\vdots   &\vdots   &\ddots &\vdots\\
&(\boldsymbol\mu^\mathsf{T}_N-\boldsymbol\mu^\mathsf{T}_0) \boldsymbol\Sigma^{-1},   &0  &\cdots  &0 \\
\end{bmatrix}
=
\mathbb{B}_d
\begin{bmatrix}
&\mathbb{I}, &\boldsymbol 0   \\
\end{bmatrix}
=\mathbb{R}_d \mathbb{P}_d 
\begin{bmatrix}
&\mathbb{I}, &\boldsymbol 0   \\
\end{bmatrix}
=\mathbb{RP}
,
 \nonumber
\end{align}
where $\mathbb{I}$ is the $d\times d$ identity matrix, and $\boldsymbol 0 $ is the $d\times d(d+1)/2$ zero matrix. The reconstruction matrix is $\mathbb{R}=\mathbb{R}_d$, and the projection matrix is $\mathbb{P}=\mathbb{P}_d[\mathbb{I}, \boldsymbol 0 ]$, in which  the first $d$ columns are~$\mathbb{P}_d$. Therefore, $\mathbb{P} \boldsymbol h(y) = \mathbb{P}_d y$. Next, we write
\begin{align}
\mathbb{S}^* 
=   \sum^N_{k=1}  ( \boldsymbol\Sigma^{-1} )^\mathsf{T} (\boldsymbol\mu_k-\boldsymbol\mu_0)(\boldsymbol\mu_k-\boldsymbol\mu_0)^\mathsf{T}   \boldsymbol\Sigma^{-1} 
=   \mathbb{B}^\mathsf{T}_d   \mathbb{B}_d,  \nonumber
\end{align}
following the definition of $\mathbb{B}_d$. Since $\mathbb{B}_d$ has a singular value decomposition $\mathbb{B}_d=U_d \Sigma_d V^\mathsf{T}_d$, and $U^\mathsf{T}_d$ is a unitary matrix (i.e., $U^\mathsf{T}_d U_d=\mathbb{I}$), we obtain 
\begin{align}
\mathbb{S}^* 
=&  V_d \Sigma_d  U^\mathsf{T}_d U_d \Sigma_d V^\mathsf{T}_d 
= V_d  \Sigma^2_d  V^\mathsf{T}_d,  \nonumber
\end{align}
which is an eigendecomposition of $\mathbb{S}^* $. The eigenvectors of $\mathbb{S}^*$ are stored as columns of $V_d$. Note that the columns of $V_d$ represent the same directions as the rows of~$\Sigma_d V^\mathsf{T}_d$ (because $\Sigma_d$ is diagonal), and since $\mathbb{P}_d$ consists of the first~$r$ rows of~$\Sigma_d V^\mathsf{T}_d$, the rows of $\mathbb{P}_d$ must represent the same directions as the eigenvectors of~$\mathbb{S}^*$.  \Halmos
\endproof

%
%
%
\section{Computational Benefits}
We compare the computational requirements of the belief-reconstruction approach with the standard POMDP algorithm that discretizes the entire belief space.  When $N$ is large but $r$ is small, the belief reconstruction algorithm is considerably more efficient than the standard algorithm. The amount of computational saving depends on $N$ and the number of discretization points in the state space. Suppose each dimension of the state space (the belief space or diagnostic statistic space) is discretized into $n$ points, the standard algorithm would require $n^N/N!$ states in the belief space whereas the belief reconstruction requires only $n^r$ states. That is, the standard algorithm requires $n^{(N-r)}/N!$ times more memory and computations than belief reconstruction. 

Consider $n=200$ discretization points, we compare the computation times and the memory uses of the two algorithms in Table~\ref{Standard POMDP vs Belief reconstruction time}-\ref{Standard POMDP vs Belief reconstruction memory size}. The numerical experiments are performed on a laptop with 2.4 GHz Intel Core i5 CPU and 8 GB memory (the results for $N=5,6$ are based on estimation). The standard algorithm suffers from the curse of dimensionality in the state space and becomes difficult to implement when $N$ exceeds 4. On the other hand, computation time of the belief reconstruction algorithm is mostly determined by $r$ and almost independent of $N$. The per-period computing time is less than $0.05$ second for $r=1$, less than 5 second for $r=2$ and around 5 minute for $r=3$. For normal observations with intrinsic dimension $r=2$, the belief-reconstruction algorithm is over 6,000 times faster than the standard algorithm when $N=3$, over 300,000 faster when $N=4$, and over 400 million times faster when $N=6$. The acceleration factor could be even higher when the number of discretization points, $n$, gets larger. 

The results of the belief reconstruction algorithm in Table~\ref{Standard POMDP vs Belief reconstruction time}-\ref{Standard POMDP vs Belief reconstruction memory size} can be further reduced by exploiting the structural properties of the optimal policy; the amount of reduction would depend on the model parameters. Such reduction is not implemented in the comparison above so that the acceleration factors are not parameter-specific and can be used as the general worse-case estimates.  

\begin{table}\footnotesize
\centering
\caption{Comparing the (per-period) run times between the standard and proposed algorithm.  }
\label{Standard POMDP vs Belief reconstruction time}
\begin{tabular}{ccccccccc}
  \toprule
\multicolumn{3}{c}{ ($n=200$)} & $N=1$ & $N=2$ & $N=3$ & $N=4$ & $N=5$ & $N=6$  \\
  \midrule
       & \multicolumn{2}{c}{Standard approach (sec.) }      	&0.04568		&4.5684		&304.56	&15228 	&609120  &20304000	\\
         \midrule
        & \multicolumn{2}{c}{Belief reconstruction $r=1$ (sec.)}      &0.04568	& 0.04569	& 0.04570	& 0.04571	& 0.04572 & 0.04572  \\
       & \multicolumn{2}{c}{Acceleration factor}         &1		&100		&6664		&3.332$\times 10^5$	&1.3329$\times 10^7$	 &4.4429 $\times 10^8$	\\
        \midrule
        & \multicolumn{2}{c}{Belief reconstruction $r=2$ (sec.)}      &--	& 4.5614	& 4.5621	& 4.5624& 4.5650& 4.5677 \\
       & \multicolumn{2}{c}{Acceleration factor}         &--		&1		&66.66		&3.33$\times 10^3$	&1.33$\times 10^5$	 &4.44 $\times 10^6$	\\
       \midrule
        & \multicolumn{2}{c}{Belief reconstruction $r=3$ (sec.)}      &--	& --	& 304.11	& 304.14& 304.51& 304.76 \\
       & \multicolumn{2}{c}{Acceleration factor}         &--		&--		&1		& 50.07	&2.00$\times 10^3$	 &6.68 $\times 10^4$	\\
  \bottomrule
\end{tabular}
\end{table}

\begin{table}\footnotesize
\centering
\caption{Comparing the (per-period) memory size between the standard and proposed algorithm.  }
\label{Standard POMDP vs Belief reconstruction memory size}
\begin{tabular}{ccccccccc}
  \toprule
\multicolumn{3}{c}{($n=200$)} & $N=1$ & $N=2$ & $N=3$ & $N=4$ & $N=5$ & $N=6$  \\
  \midrule
       & \multicolumn{2}{c}{Standard approach }      	&	1.6kB	&	160kB	& 10.66MB	& 4.26GB	& 170.61GB &	5.69 TB \\
        & \multicolumn{2}{c}{Belief reconstruction ($r=1$)}      &1.6kB& 1.6kB	& 1.6kB& 1.6kB	& 1.6kB & 1.6kB  \\
                & \multicolumn{2}{c}{Belief reconstruction ($r=2$)}      &--& 160kB	& 160kB& 160kB	& 160kB & 160kB  \\
                                & \multicolumn{2}{c}{Belief reconstruction ($r=3$)}      &--& --	& 10.66MB& 10.66MB	& 10.66MB & 10.66MB  \\
  \bottomrule
\end{tabular}
\end{table}

\end{document}